\documentclass{aa}

\usepackage{graphicx}
\usepackage{txfonts}
\usepackage{color}
\usepackage{multirow}
\usepackage{hyperref}
\newcommand{\dd}{\mathrm{d}}
\makeatletter
\renewcommand*\aa@pageof{, page \thepage{} of \pageref*{LastPage}}
\makeatother
\hypersetup{colorlinks=true,allcolors=[rgb]{0,0,0.8}}

\begin{document}

   \title{The impact of baryons on the sparsity of simulated galaxy clusters from \textsc{The Three Hundred} Project}

   \subtitle{Astrophysical and cosmological implications}

   \author{P.~S. Corasaniti
          \inst{1,2}
          \and
          T.~R.~G. Richardson\inst{3}
          \and
          S. Ettori\inst{4,5}
          \and
          M. De Petris\inst{6}
          \and
          E.~Rasia\inst{7,8,9} 
          \and
          W.~Cui\inst{10,11}
          \and
          G.~Yepes\inst{10,11}
          \and
          G.~Gianfagna\inst{12}
          \and
          A.~M.~C. Le Bun\inst{1}
          \and
          Y.~Rasera\inst{1,13}
          }

   \institute{LUX, Observatoire de Paris, Université PSL, Sorbonne Université, CNRS, 92190 Meudon, France\\
              \email{Pier-Stefano.Corasaniti@obspm.fr}
         \and
             Sorbonne Universit\'e, CNRS, UMR 7095, Institut d'Astrophysique de Paris, 98 bis bd Arago, 75014 Paris, France
        \and
        Donostia International Physics Center (DIPC), Paseo Manuel de Lardizabal, 4, 20018, Donostia-San Sebasti\'an,
        Gipuzkoa, Spain\\
             \email{tamara.richardson@dipc.org}
        \and
        INAF, Osservatorio di Astrofisica e Scienza dello Spazio, via Piero Gobetti 93/3, I-40129 Bologna, Italy
        \and
        INFN, Sezione di Bologna, viale Berti Pichat 6/2, I-40127 Bologna, Italy
        \and
        Dipartimento di Fisica, Sapienza Università di Roma, Piazzale Aldo Moro 5, I-00185 Rome, Italy
        \and
        INAF-Osservatorio Astronomico di Trieste, via Tiepolo 11, I-34131, Trieste, Italy
        \and
        IFPU-Institute for Fundamental Physics of the Universe, via Beirut 2, 34151, Trieste, Italy
        \and
        Department of Physics, University of Michigan, 450 Church St, Ann Arbor, MI 48109, USA
        \and     
        Departamento de Física Te\'orica, M\'odulo 15, Facultad de Ciencias, Universidad Autónoma de Madrid, 28049 Madrid, Spain   
        \and
        Centro de Investigaci\'on Avanzada en F\'isica Fundamental (CIAFF), Facultad de Ciencias, Universidad Aut\'onoma de Madrid, 
        28049, Madrid, Spain
        \and
        INAF - Istituto di Astrofisica e Planetologia Spaziali, via Fosso del Cavaliere 100, I-00133 Rome, Italy
        \and
        Universit\'e Paris Cit\'e, F-75006 Paris
             }

  \abstract 
   {Measurements of the sparsity of galaxy clusters can be used to probe the cosmological information encoded in the host dark matter halo profile, and infer constraints on the cosmological model parameters. Key to the success of these analyses is the control of potential sources of systematic uncertainty. As an example, the presence of baryons can alter the cluster sparsity with respect to predictions from N-body simulations. Similarly, a radial dependent mass bias, as in the case of cluster masses inferred under the hydrostatic equilibrium (HE) hypothesis, can affect sparsity estimates.}
   {First, we examine the imprint of baryonic processes on the sparsity statistics. Then, we investigate the relation between cluster sparsities and gas mass fraction. Finally, we perform a study of the impact of HE mass bias on sparsity measurements and the implication on cosmological parameter inference analyses.}
   {We use catalogues of simulated galaxy clusters from {\sc The Three Hundred} project and run a comparative analysis of the sparsity of clusters from N-body/hydro simulations implementing different feedback model scenarios.}
   {Sparsities which probe the mass profile across a large radial range are affected by the presence of baryons in a way that is particularly sensitive to astrophysical feedback, whereas those probing exclusively external cluster regions are less affected. In the former case, we find the sparsities to be moderately correlated with measurements of the gas fraction in the inner cluster regions. We infer constraints on $S_8$ using synthetic average sparsity measurements generated to evaluate the impact of baryons, selection effects and HE bias. In the case of multiple sparsities ($s_{200,500}$, $s_{200,2500}$ and $s_{500,2500}$), these lead to highly bias results. Hence, we calibrate linear bias models that enable us to correct for these effects and recover unbiased constraints that are significantly tighter than those inferred from the analysis of $s_{200,500}$ only.}
   {}


   \maketitle

\section{Introduction}
There is a widespread consensus that observations of galaxy clusters can provide constraints on cosmological model parameters through a variety of observable properties \citep[see e.g.][]{2011ARA&A..49..409A,2012ARA&A..50..353K}. The availability of large homogeneous galaxy cluster samples has enabled in recent years the realisation of cosmological data analyses through measurements of the cluster abundance \citep[see e.g.][]{PlanckSZ2014,PlanckSZ2016,Pacaud2018,Bocquet2019, 2022A&A...659A..88L,To2021}, spatial clustering \citep{2014A&A...571A..21P,Maruli2021} as well as estimates of the cluster gas mass fraction \citep[see e.g.][]{2009A&A...501...61E,2014MNRAS.440.2077M,2022MNRAS.510..131M,2022arXiv220412823W}. Cosmological information is also encoded in the spatial distribution of mass within clusters. This is because galaxy clusters are hosted in massive dark matter haloes, which result of the hierarchical bottom-up gravitational assembly process of cosmic structure formation \citep[see e.g.][]{2002ApJ...568...52W,2013MNRAS.432.1103L,2020MNRAS.498.4450W,2022MNRAS.513.4951R}. Such a process depends on the global properties of the universe, such as the cosmic matter density, the state of cosmic expansion and the amplitude of initial matter density fluctuations, thus leaving a cosmological imprint on the radial mass profile of clusters.

The possibility of testing cosmology through measurements of the distribution of mass within clusters has primarily been investigated in the context of a parametric description of the host dark matter halo density profile. This is motivated from the analysis of N-body simulations which have shown that the spherically averaged halo density profile is well-fitted by a two-parameter formula, the Navarro-Frenk-White (NFW) profile \citep{NFW1997}, which is fully specified by the halo mass $M$, and the concentration parameter $c$. 

Numerical studies have shown that the concentration relates to the halo formation history and the halo dynamical state \citep[see e.g.][]{2002ApJ...568...52W,2007MNRAS.381.1450N,2009ApJ...707..354Z,2012MNRAS.427.1322L,2013MNRAS.432.1103L,2017MNRAS.466.3834L,2019MNRAS.485.1906R}. However, due to the stochastic nature of the gravitational collapse that drives the halo assembly process, the concentration behaves as a random variate characterised by a large scatter and a median which varies as function of halo mass \citep[see e.g.][]{2001MNRAS.321..559B,2003ApJ...597L...9Z,2007MNRAS.378...55M,2012MNRAS.423.3018P,Ishiyama2021}. Quite importantly, the amplitude of such a relation has been found to depend on the underlying cosmological model \citep[see e.g.][]{2001MNRAS.321..559B,2004A&A...416..853D,2009ApJ...707..354Z,2012MNRAS.422..185G}, thus opening the possibility to infer cosmological parameter constraints from measurements of the concentration-mass relation. Nevertheless, this has proven to be challenging \citep[see e.g.][]{2010A&A...524A..68E}, since systematic uncertainties, such as selection effects \citep{2015MNRAS.449.2024S} and astrophysical process altering the matter distribution in galaxy clusters \citep[see e.g.][]{2010MNRAS.406..434M,2011MNRAS.416.2539K,2013ApJ...776...39R} can significantly impact the determination of the concentration parameter of galaxy clusters.

Alternatively, \citet{Balmes2014} have proposed a non-parametric approach based on measurements of ratios of spherical halo masses estimated at radii enclosing different overdensities. These ratios, dubbed sparsities, probe the level of sparseness of the halo mass distribution, while retaining cosmological information encoded in the halo mass profile \citep[][]{Balmes2014,2020PhRvD.102d3501C}. In particular, this can be retrieved through measurements of the average sparsity of a sample of clusters at different redshifts \citep{Corasaniti2018,2022arXiv221203233R}. Furthermore, the combination of measurements of average sparsities for multiple overdensity pairs can provide stronger constraints by extracting the cosmological signal imprinted across the entire halo mass profile, beyond what is captured by the NFW's concentration \citep{2022MNRAS.516..437C}. Recently, \citet{2022arXiv221203233R} have also shown that constraints can also be obtained from individual cluster sparsity measurements, rather than ensemble average estimates, provided the availability of predictions of the sparsity probability distribution function from N-body simulations.

Cosmological parameter constraints have been inferred from measurements of the average sparsity of X-ray \citep{Corasaniti2018} and weak gravitational lensing \citep{Corasaniti2021} mass estimates of cluster samples. These studies have shown that current sparsity estimates are primarily limited by the large statistical uncertainties affecting the mass measurements of observed galaxy clusters. Nonetheless, in the upcoming years these are expected to improve thanks to a new generation of cluster survey programs such as the CHEX-MATE project\footnote{\href{http://xmm-heritage.oas.inaf.it}{http://xmm-heritage.oas.inaf.it}} \citep{2021A&A...650A.104C}, or the Euclid mission \citep[see e.g.][]{2016MNRAS.459.1764S,2019A&A...627A..23E}. Hence, in preparation of the future cosmological analyses, it is of primary importance to study potential sources of systematic errors that can affect the cosmological model parameter inference. For instance, the presence of baryons may alter the predictions of the sparsity with respect to that inferred from DM-only simulations. Similarly, a radial (aperture) dependent mass bias can affect the determination of the cluster sparsity and induce a sparsity bias. As an example, deviations from the hydrostatic equilibrium (HE) hypothesis may result in a radial dependent bias of the cluster masses obtained from the analysis of X-ray and Sunyaev-Zel'dovich (SZ) observations of galaxy clusters, thus leading to biased sparsity estimates.

Assessing these systematics requires us to investigate the statistical properties of sparsity using hydrodynamical simulations. Here, we perform a thorough analysis using the simulated cluster catalogues from {\sc The Three Hundred} project \citep{2018MNRAS.480.2898C}, hereafter The300. Our goal is twofold. On the one hand, we aim to study the extent to which baryons (gas and stars) alter the mass profile of clusters, as probed by sparsity. To this purpose, we evaluate the deviation of sparsities of clusters from hydrodynamical simulations against those estimated from DM-only simulations. This enables us to evaluate the sensitivity of different sparsities to the effect of astrophysical processes parameterised by the sub-grid models implemented in The300. On the other hand, we investigate the effect of mass and selection bias on the cosmological parameter inference of a synthetic sample of sparsity data generated
using the HE mass estimates obtained by \citet{2023MNRAS.518.4238G} from the analysis of The300.

The article is organised as follows: in Section~\ref{sec:method} we introduce the sparsity statistics and describe the simulation dataset used in our analysis. In Section~\ref{sec:spars_baryon_analysis}, we present the results of numerical analyses devoted to the assessment of the statistical representativeness of the sparsity samples, the evaluation of the impact of baryons on the halo sparsities and the role of baryonic feedback models. In Section~\ref{sec:astro_implications}, we discuss the astrophysical implications of such effects, while in Section~\ref{sec:hydro_bias} we present the analysis of the impact of the HE mass bias on the cluster sparsities. In Section~\ref{sec:cosmo}, we discuss the cosmological implications of the HE mass bias and selection effects and evaluate their effects on cosmological parameter inference analyses. Finally, in Section~\ref{sec:conclu} we present our conclusions.

\section{Methodology}\label{sec:method}

\subsection{Dark Matter Halo Sparsity}
Dark matter halo sparsity is defined as
\citep{Balmes2014}:
\begin{equation}
s_{\rm \Delta_1,\Delta_2}\equiv\frac{M_{\Delta_1}}{M_{\Delta_2}},\label{eq:sparsity}
\end{equation}
where $M_{\Delta_1}$ and $M_{\Delta_2}$ are the masses enclosed in spheres of radii $r_{\Delta_1}$ and $r_{\Delta_2}$, that contain the overdensity $\Delta_1$ and $\Delta_2$ respectively, where $\Delta_1<\Delta_2$. Notice that the sparsity can be written as a fractional mass estimate, i.e. $s_{\Delta_1,\Delta_2}=\Delta{M}_{12}/M_{\Delta_2}+1$, where $\Delta{M}_{12}=M_{\Delta_1}-M_{\Delta_2}$ is the mass enclosed within the radial shell comprised between the outer radius $r_{\Delta_1}$ and the inner radius $r_{\Delta_2}$. Hence, depending on the choice of $\Delta_1$ and $\Delta_2$ the sparsity provides a proxy of the fractional halo mass profile in different halo regions\footnote{Alternatively, the halo sparsity can be seen as an indirect proxy of the logarithmic steepness of the radial halo mass profile, since
 \begin{equation}\label{slopelogM}
 \frac{\Delta{\log{M}}}{\Delta{\log{r}}}\equiv \frac{\log{M_{\Delta_1}}-\log{M_{\Delta_2}}}{\log{r_{\Delta_1}}-\log{r_{\Delta_2}}}=\frac{3\log{s_{\Delta_1,\Delta_2}}}{\log{({\Delta_2}/{\Delta_1})}+\log{s_{\Delta_1,\Delta_2}}}.
 \end{equation}}. Hereafter, we will always refer to masses at overdensities given in units of the critical density.

\begin{figure*}[ht]
    \centering
    \includegraphics[width = 1\linewidth]{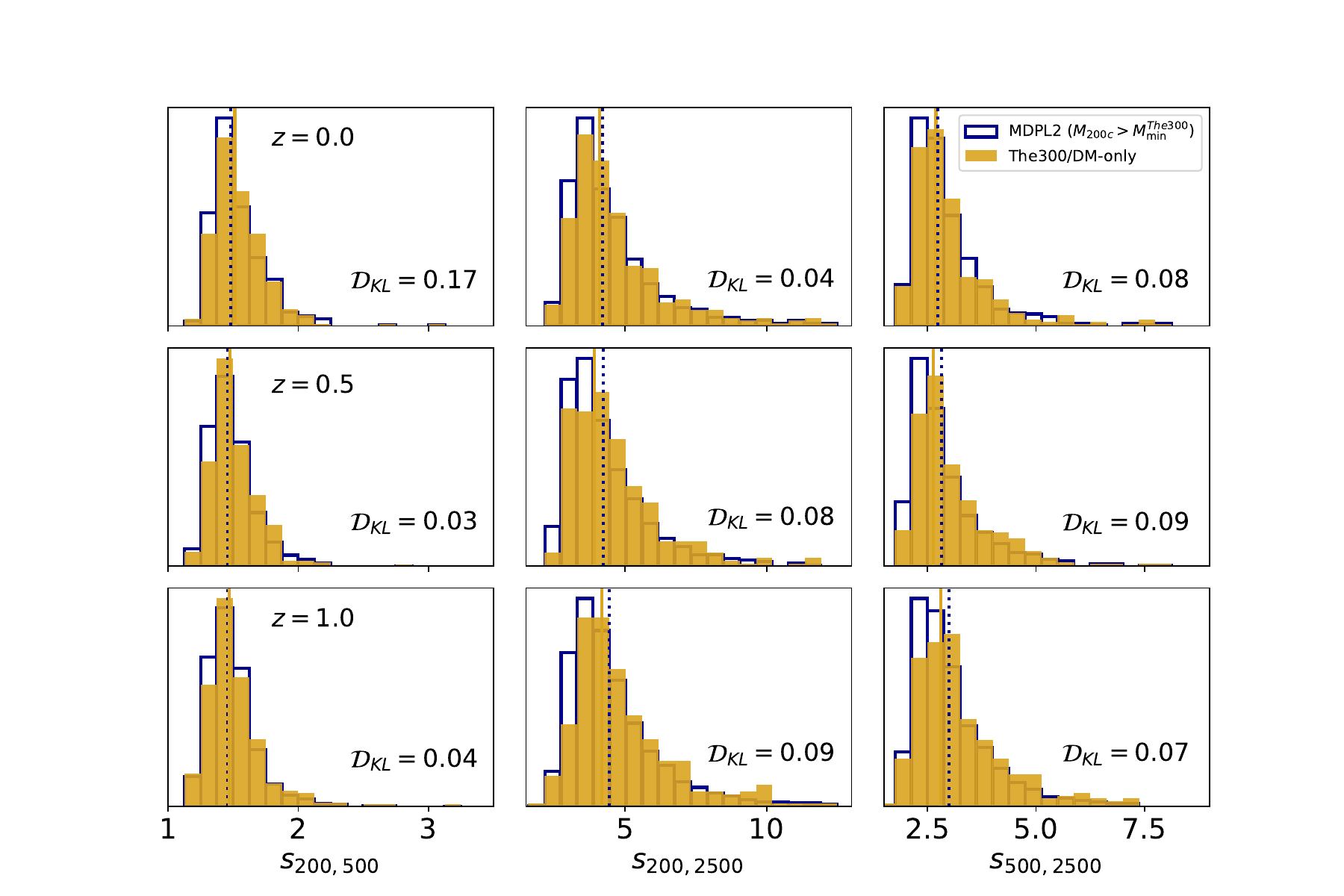}
    \caption{Normalised histograms of the sparsity $s_{200,500}$ (left panels), $s_{200,2500}$ (central panels) and $s_{500,2500}$ (right panels) obtained from the analysis of the MDPL2 sub-sample (dark blue bars) and The300/DM-only catalog (goldenrod filled bars) at $z=0.0$ (top panels), $0.5$ (middle panels) and $1.0$ (bottom panels), where the MDPL2 sub-sample consists of haloes with mass larger than the minimum mass of the DM-only catalog at the redshift considered. The vertical lines in each panel correspond to the median sparsities for the MDPL2 sub-sample (dark blue dotted line) and the The300/DM-only sample (goldenrod solid line). \label{fig:spars_dist}}
\end{figure*}

\subsection{Numerical Simulation Dataset}\label{subsec:dataset}
Here, we describe the numerical simulations we have used in our analyses. Specifically, we have used the simulations realised in the context of The Three Hundred project collaboration\footnote{\url{http://the300-project.org}} to investigate the impact of baryons on the sparsity of halos relies. We have also used the halo catalogs from the \textsc{uchuu} simulation suite \citep{Ishiyama2021} to calibrate the halo mass function used in the cosmological parameter inference analyses and the evaluation of selection bias effects.
 
\subsubsection{The Three Hundred}
We use catalogues of simulated galaxy clusters from The300 \citep{2018MNRAS.480.2898C}. These have been generated from zoom simulations of 324 spherical regions of $15\,h^{-1}$ Mpc radius centred on the most massive haloes of the MultiDark-Planck2 (MDPL2) simulation \citep{2016MNRAS.457.4340K} detected at redshift $z=0$ using the \textsc{rockstar}\footnote{\url{https://bitbucket.org/gfcstanford/rockstar/src/main/}} halo finder \citep{2013ApJ...762..109B}. These regions have been re-simulated assuming purely gravitational physics (DM-only) and including baryon hydrodynamics with different sub-grid models. Cluster catalogues were then generated using the \textsc{AHF}\footnote{\url{http://popia.ft.uam.es/AHF}} \citep{2009ApJS..182..608K} halo finder. 

The simulations assume a flat $\Lambda$CDM model with parameters set to the values inferred from the Planck data analysis \citep{2016A&A...594A..13P}: total matter density $\Omega_m=0.307$, baryon density $\Omega_b=0.048$, reduced Hubble constant $h=0.678$, amplitude of linear matter density fluctuations on $8\,h^{-1}$ Mpc scale $\sigma_8=0.823$, and scalar spectral index $n_s=0.96$.

The \textsc{gizmo-simba} zoom simulations \citep[see][for more details]{2022MNRAS.514..977C} have been realised with the \textsc{gizmo} code \citep{2015MNRAS.450...53H} implementing the sub-grid model from the \textsc{simba} simulation \citep{2019MNRAS.486.2827D}. The \textsc{gadget-music} simulations were carried out with the \textsc{gadget}-3 Tree-PM gravity solver \citep[an updated version of \textsc{gadget}-2 code by][]{2005MNRAS.364.1105S} implementing a classic entropy-conserving formulation smoothed-particle hydrodynamics (SPH) scheme with the baryonic physics described in \citet{2013MNRAS.429..323S}, while the \textsc{gadget-x} simulations were realised using an improved version of the SPH scheme by \citet{2016MNRAS.455.2110B} with the baryon physics models discussed in \citet{2015ApJ...813L..17R}. Notice that both \textsc{gizmo-simba} and \textsc{gadget-x} simulations include AGN feedback, though with different model characteristics, while the \textsc{gadget-music} runs do not. Because of this the simulated clusters from the \textsc{gizmo-simba} and \textsc{gadget-x} catalogues have properties that are better in agreement with observations \citep[see][]{2018MNRAS.480.2898C,2022MNRAS.514..977C}.

In the analysis presented here, we consider the most massive cluster of each of the 324 re-simulated regions of the \textsc{gadget-x}, \textsc{gadget-music}, \textsc{gizmo-simba}, and DM-only simulations, for which we compute the sparsity for several pairs of overdensities. In particular, we compute the mass $M_{\Delta}$ at $\Delta=200,500,1000,1500,2000$ and $2500$ for each of the clusters in the numerical catalogues. Then, we estimate the sparsity for different combination of overdensities as defined by Eq.~(\ref{eq:sparsity}). Note that in the case of the catalogues from the hydrodynamical simulations, the mass $M_{\Delta}$ is the total cluster mass, i.e. the sum of the mass in the dark matter, gas mass and stellar component within $R_{\Delta}$. 

In order to distinguish the effect of the baryons from that of the dynamical state of clusters, we perform a statistical analysis of the sparsity on a subsample of relaxed objects from each of the numerical catalogues listed above. More specifically, we select systems that object-by-object are characterised by the virial energy ratio $\eta$, the centre-of-mass offset $\Delta_r$ and the fraction of mass in sub-haloes $f_s$ (all evaluated within $R_{200}$) within the following bounds: $0.85<\eta<1.15$, $\Delta_r<0.04$ and $f_s<0.1$ both for the hydrodynamical case and their DM-only counterpart. Notice that the bounds adopted here are identical to those assumed in \citet{2018MNRAS.480.2898C}, although our selection is more restrictive since for each cluster in the relaxed subsample the above conditions need to be fulfilled simultaneously in both the hydrodynamical and DM-only simulations. It is worth mentioning that such a characterisation results in a binary selection of clusters. This differs from the approach adopted in \citet{2020MNRAS.492.6074H}, where the authors introduce a parameter, $\chi_{DS}$, which links the values of $\eta$, $\Delta_r$ and $f_s$, with the intent of obtaining a continuous measure of the dynamical state. Furthermore, as pointed out in \citet{2021MNRAS.504.5383D} these indicators may vary depending on the overdensity at which they are evaluated. Quite remarkably, their analysis of The300 shows that the fraction of relaxed clusters selected with criteria adopted in \citet{2018MNRAS.480.2898C} remains unaltered whether the dynamical indicators are evaluated at $R_{200}$ or $R_{500}$. This is due the aperture dependence of $\eta$ and results in more stringent criterion. Indeed, our selection is more restrictive than that assumed in \citet{2018MNRAS.480.2898C}, which enables us to more easily distinguish the impact of baryons on the halo sparsity with respect to that of the dynamical state.

Finally, in order to assess the impact of the HE mass bias on sparsity measurements, we use the three-dimensional HE mass estimates from the analysis of the \textsc{gadget-x} clusters presented in \citet{2023MNRAS.518.4238G}. These have been estimated for the most massive cluster of each of the zoomed regions of the \textsc{gadget-x} simulation that contain no low-resolution particles within the virial radius of the cluster.

\subsubsection{\textsc{Uchuu} Simulation}\label{sec:uchuu}
We use the halo catalogues from the \textsc{uchuu} N-body simulation suite \citep{Ishiyama2021} of ($2h^{-1}\textrm{Gpc}$)$^3$ volume with 12800$^3$ particles realised with the code GreeM \citep{2009PASJ...61.1319I,2012arXiv1211.4406I}. 
As in the case of The300, the simulation assume a flat $\Lambda$CDM model with parameters set to the {\it Planck}-CMB 2015 analysis (Planck Collaboration et al. 2016a): $\Omega_m= 0.3089$, $\Omega_b = 0.0486$, $h = 0.6774$, $n_s = 0.9667$, and $\sigma_8 = 0.8159$. Notice that these differ at sub-percent level from those assumed in the The300. Halos 

Haloes have been detected with the \textsc{rockstar} halo finder \citep{2013ApJ...762..109B} and spherical masses at different overdensities have been computed for all detected halos in 24 redshift snapshots in the range $0\le z \le 2$.

\begin{figure*}[ht]
    \centering
    \includegraphics[width = 1\linewidth]{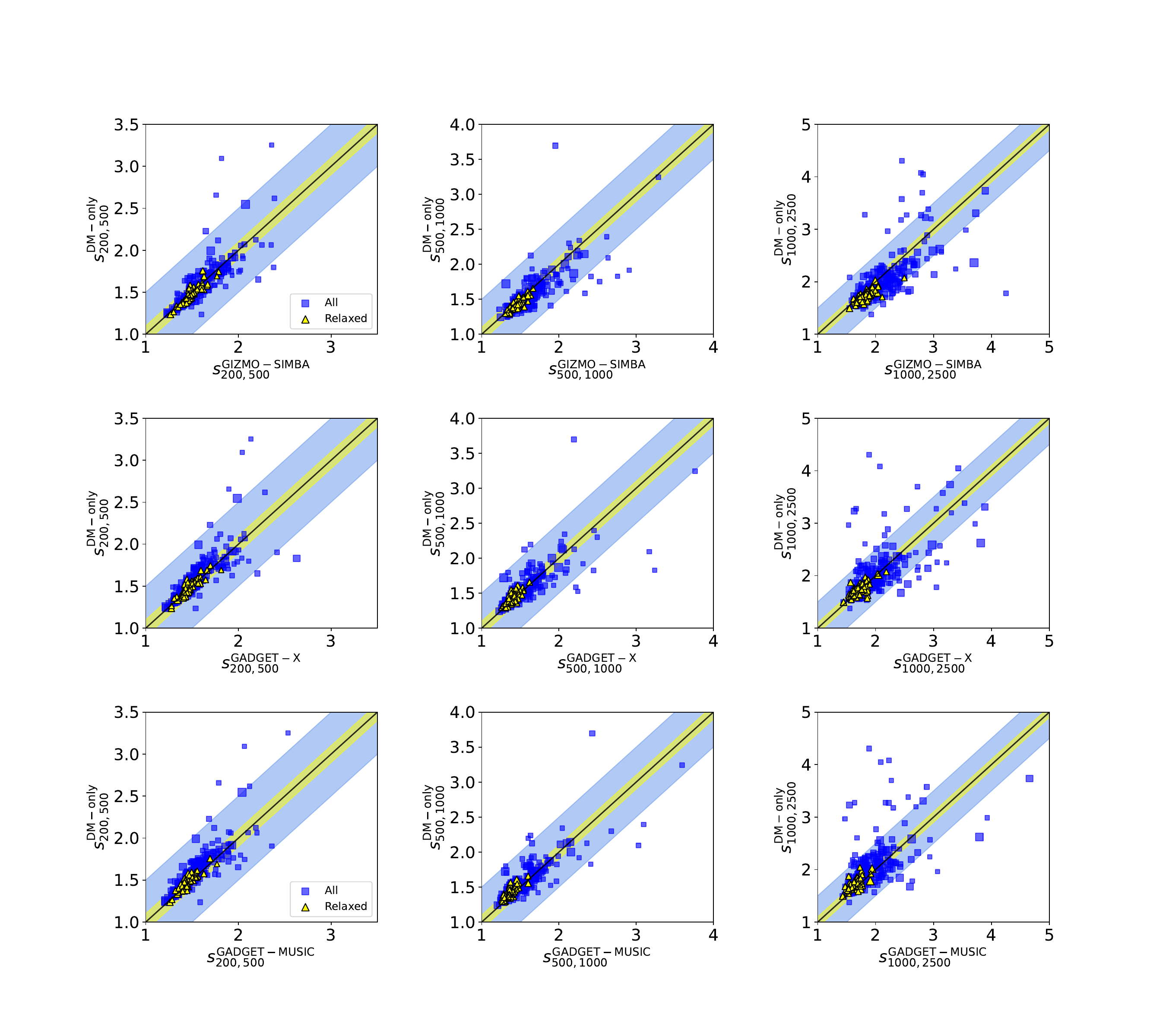}
    \caption{DM-only vs hydro simulation estimated sparsities $s_{200,500}$ (left panels), $s_{500,1000}$ (central panels) and $s_{1000,2500}$ (right panels) for the clusters in the \textsc{gizmo-simba} (top panels), \textsc{gadget-x} (central panels) and \textsc{gadget-music} (bottom panels) catalogues at $z=0$ respectively. The blue squares represent clusters from the full samples, while the yellow triangles corresponds to the relaxed systems as defined in Sec.~\ref{subsec:dataset}. The size of the data points is proportional to mass $M_{200}$ of each cluster from the DM-only simulation. The diagonal line represents the ideal case with no bias, while the shaded areas correspond to $10\%$ (yellow band) and $50\%$ (light-blue band) deviation respectively.}
    \label{fig:individual_sparsity}
\end{figure*}

\section{Numerical Data Analyses}\label{sec:spars_baryon_analysis}
\subsection{Statistical Representativeness}\label{subsec:spars_baryon_analysis}
The sample of simulated clusters from The300 project, corresponds to the 324 most massive haloes detected in the MDPL2 simulation at $z=0$ and followed throughout their evolution. Hence, the sample is mass-selected and volume-limited only for $z=0$, while the selection is unclear at higher redshifts. Given that we are interested in evaluating the impact of baryons on the sparsity statistics, it is worth assessing the representativeness of the estimated sparsities. Even though we can perform such a test only for the DM-only catalogue, this remains a useful check. For conciseness, we limit ourselves to the catalogues at $z=0.0,0.5$ and $1.0$, for which we can confront the distribution of sparsities inferred using the DM-only halo catalogue of The300 (The300/DM-only) against those obtained from the MDPL2 haloes with masses larger than the minimum halo mass of the The300/DM-only catalogue, that is $M_{200}\gtrsim 5\cdot 10^{14}\,h^{-1}$ M$_{\odot}$ at $z=0.0$, $\gtrsim 10^{14}\,h^{-1}$ M$_{\odot}$ at $z=0.5$ and $\gtrsim 6\cdot 10^{13}\,h^{-1}$ M$_{\odot}$ at $z=1.0$. This selection results in a reference MDPL2 halo subsample of $N_{\rm halo}= 612$ haloes at $z=0.0$, $N_{\rm halo}=8368$ at $z=0.5$ and $ N_{\rm halo}= 8162$ at $z=1.0$. 

As the publicly available MDPL2 halo catalogues provide estimates of the halo masses at overdensity $\Delta=200,500$ and $2500$, we only consider the following set of sparsities: $s_{200,500}$, $s_{200,2500}$ and $s_{500,2500}$. In Fig.~\ref{fig:spars_dist}, we plot the normalised histograms of $s_{200,500}$ (left panel), $s_{200,2500}$ (central panel) and $s_{500,2500}$ (right panel) for the MDPL2 subsample (blue bars) and The300/DM-only catalogues (goldenrod filled bars) at $z=0.000$ (top panels), $0.523$ (middle panels) and $1.031$ (bottom panels) respectively. The vertical lines in the panels correspond to the median values of the sparsity distribution for the MDPL2 subsample (dark blue dotted line) and the The300/DM-only sample (goldenrod solid line). We can see that independently of the redshift both samples span the same range of sparsity values. Moreover, we find that the distribution of sparsities are very similar to one another, with the median sparsities of the two samples having nearly identical values. This is confirmed by the estimation of the Kullback-Leibler (KL) divergence \citep{10.1214/aoms/1177729694}. In particular, by assuming the histogram of the MDPL2 subsample to be the target distribution, we find the KL divergence of The300/DM-only sample to be $\mathcal{D}_{KL}=0.17$ in the case of $s_{200,500}$,
$\mathcal{D}_{KL}=0.04$ for $s_{200,2500}$ and $\mathcal{D}_{KL}=0.08$ for $s_{500,2500}$ at $z=0$. Such an agreement might be expected given that The300 sample is mass-selected at $z=0$. However, we can see that this remains valid also at higher redshifts. In particular, at $z=0.5$ we have $\mathcal{D}_{KL}=0.03$ in the case of $s_{200,500}$,
$\mathcal{D}_{KL}=0.08$ for $s_{200,2500}$ and $\mathcal{D}_{KL}=0.09$ for $s_{500,2500}$,
and similar values at $z=1$. Hence, despite the difference in sample size and the fact that the MDPL2 haloes were detected using the \textsc{rockstar} halo finder, rather than \textsc{AHF}, the distributions of sparsities are compatible with one another. We can conclude that the sparsities from the The300/DM-only are representative of the population of very massive haloes from the larger volume MDPL2 simulation. Nonetheless, this does not ensure that the sparsities of the clusters from the zoom hydrodynamical simulations are also representative of the statistics of a large volume sample. Conservatively, it only guarantees that the distribution of sparsities of our reference DM-only sample is not affected by volume effects. 

A subtle point we would like to stress here is the fact that even if such representativeness may extend to the hydrodynamical cluster samples, by no means it entails that the selected clusters are representative of a cosmological selection. This is because the clusters considered here are not randomly drawn from the halo mass function distribution, rather that each of them is drawn from an extreme value distribution of the first to $324$th order most massive halo. This inevitably results in selection effects that need to be accounted when evaluating the cosmological information encoded in the sparsity from The300. As we will see in Sec.~\ref{sec:cosmo}, this is particularly important when assessing the impact of the HE mass bias on the cosmological parameter constraints inferred from sparsities measurements obtained using HE mass estimates from a subsample of the most massive clusters from the \textsc{gadget-x} simulations.

\subsection{The impact of baryons on the cluster sparsity}\label{impact_baryons}
Let us now focus on how baryons alter the mass profiles of simulated clusters as traced by sparsity. We can infer a first qualitative insight by comparing the sparsity of the haloes in the DM-only simulation against that of the corresponding clusters in the hydro simulations. In principle, we could choose the sparsity between any pair of overdensities, provided that $\Delta_1<\Delta_2$. Nonetheless, it is convenient to use overdensity values that are commonly adopted in observational studies. In particular, let us consider the sparsities $s_{200,500}$, $s_{500,1000}$ and $s_{1000,2500}$, which respectively probe the fractional mass profile in the external, intermediate and inner halo regions. 

In Fig.~\ref{fig:individual_sparsity} we plot the different sparsities in the case of the full sample of 324 clusters (blue squares) for the \textsc{gizmo-simba} (top panels), and \textsc{gadget-x} (central panels) and \textsc{gadget-music}(bottom panels) catalogues at $z=0$ against the sparsities of their DM-only counterparts. We also plot the comparison for the relaxed systems (yellow triangles) consisting of $62$ clusters from the \textsc{gizmo-simba} subsample, $68$ from \textsc{gadget-x} and $64$ from \textsc{gadget-music}. The diagonal line corresponds to the ideal case with the cluster sparsities being equal to that of the DM-only haloes, while the shaded areas corresponds to $10\%$ (yellow band) and $50\%$ (light-blue band) deviations respectively. The size of the data points in the plots is set to be proportional to the DM-only mass, $M_{200}$. 

First of all, we can see that in all the cases the sparsities are distributed along the diagonal line. The bulk of the clusters lies well within $50\%$ of the ideal case, and within $\sim 10\%$ in the case of relaxed systems. We can also notice that in the case of the full cluster sample, the scatter around the diagonal line increases for greater values of the cluster sparsity. This can be the results of multiple factors due to the interplay between the dynamical status of the clusters \citep{2022MNRAS.513.4951R}, the delay in the assembly of substructures between DM-only and hydrodynamical simulations \citep[see][]{2013MNRAS.429..323S} and the survival rate of substructure due to baryon physics. As far as the dependence on the halo mass is concerned, we do not find any particular trend and we refer the reader to Appendix \ref{sparsity_mass} for a more detailed analysis. 

\begin{figure}[t]
    \centering
    \includegraphics[width = 0.9\linewidth]{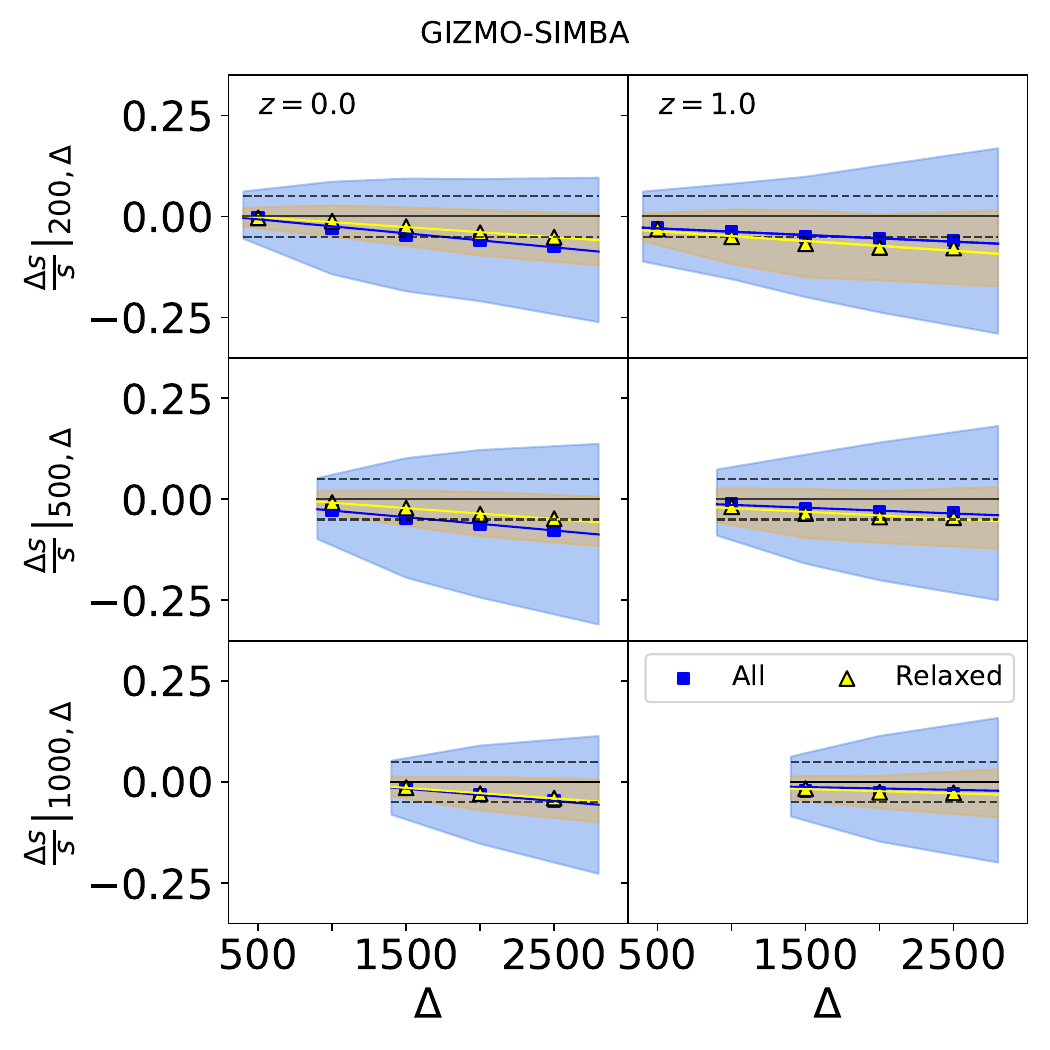}
    \caption{Sparsity relative difference $\Delta{s}/s\,|_{200,\Delta}$ (top panels), $\Delta{s}/s\,|_{500,\Delta}$ (central panels) and $\Delta{s}/s\,|_{1000,\Delta}$ (bottom panels) as function of $\Delta$ for the \textsc{gizmo-simba} clusters in the catalogues at $z=0$ (left panels) and $1$ (right panels). The blue squares (yellow triangles) represent the average for the full (relaxed) cluster sample, while the shaded areas correspond to the standard deviation. The black dashed lines in the plots mark the $5\%$ differences respectively, while the solid blue (yellow) lines correspond to the linear regression of the average values.}
    \label{fig:average_baryonic_bias_profile_GIZMO}
\end{figure}

\begin{figure}[ht]
    \centering
    \includegraphics[width = 0.9\linewidth]{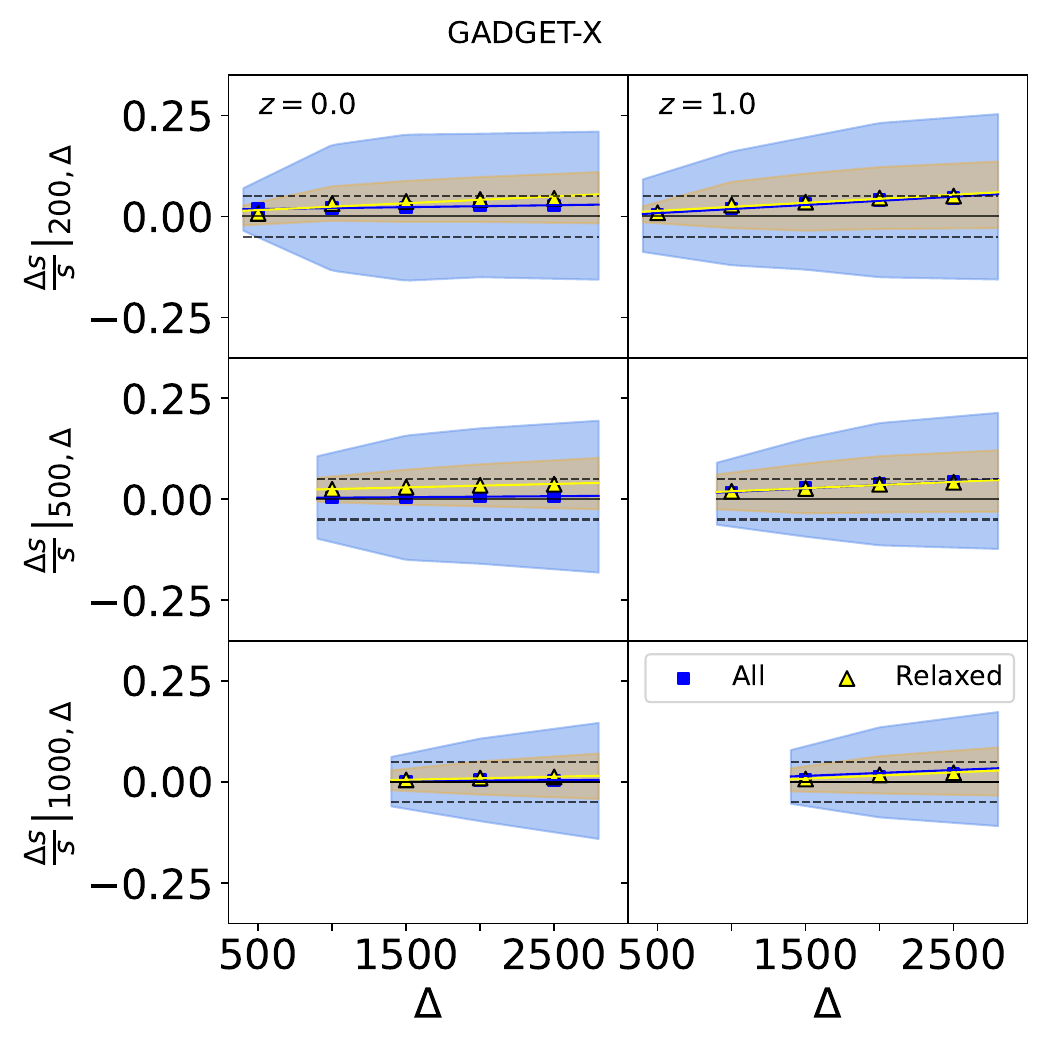}
    \caption{As in Fig.~\ref{fig:average_baryonic_bias_profile_GIZMO} for the \textsc{gadget-x} catalogues.}
    \label{fig:average_baryonic_bias_profile_GadgetX}
\end{figure} 
\begin{figure}[ht]
    \centering
    \includegraphics[width = 0.9\linewidth]{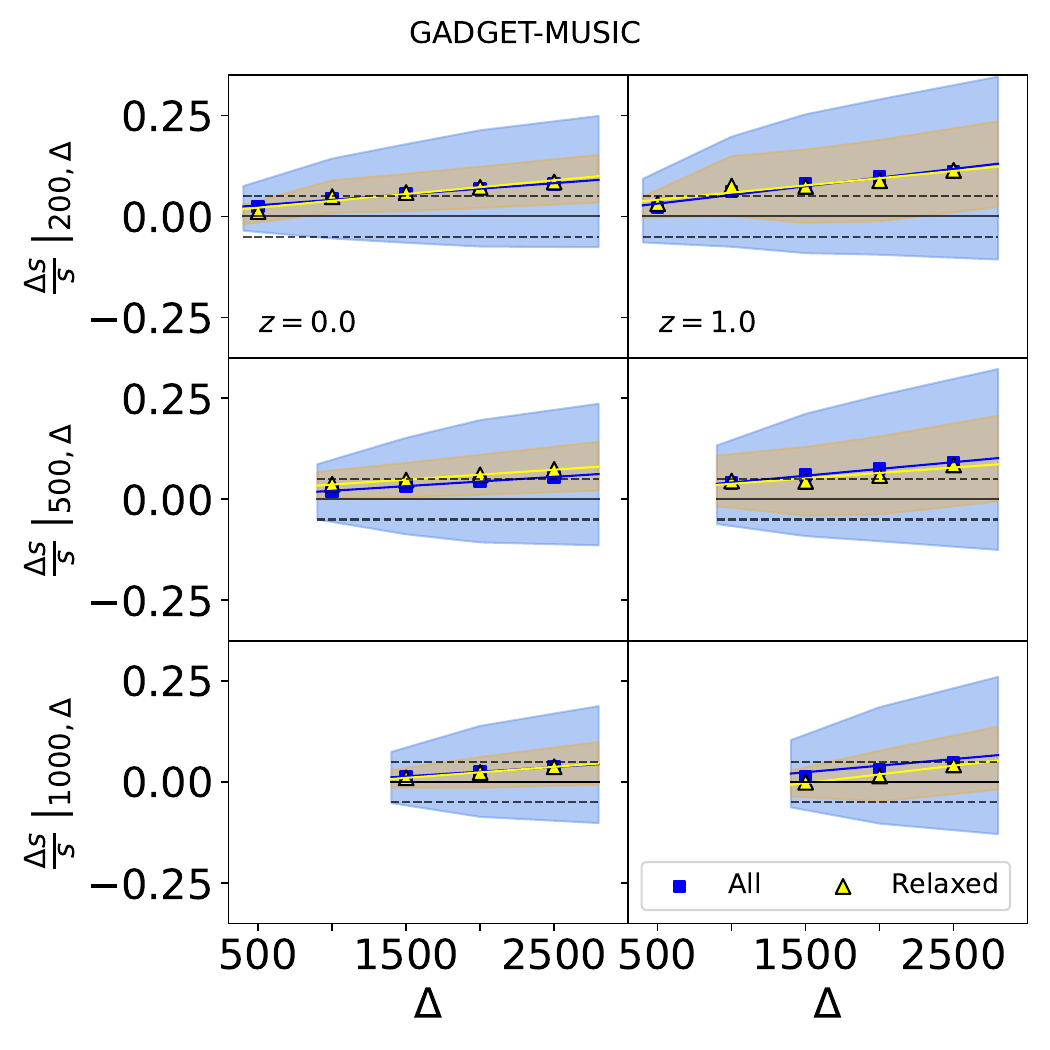}
    \caption{As in Fig.~\ref{fig:average_baryonic_bias_profile_GIZMO} for the \textsc{gadget-music} catalogues.}
    \label{fig:average_baryonic_bias_profile_GadgetMUSIC}
\end{figure} 

\subsubsection{Cluster sparsity: DM-only vs Hydrodynamical Simulations}\label{baryonic_effects}
We use the spherical masses at different overdensities computed for each of the clusters in the hydrodynamical and DM-only catalogues to estimate the sparsities $s_{\Delta_1,\Delta}$ as a function of $\Delta$ for $\Delta>\Delta_1$ with $\Delta_1=200,500$ and $1000$. Then, we compare the sparsity of each clusters in the hydrodynamical catalogue to that of its DM-only counterpart. More specifically, we compute for the $i$-th cluster the relative difference: 
\begin{equation}\label{eq:relative_difference_spars}
\frac{\Delta{s}}{s}\bigg\vert^{i-{\rm th}}_{\Delta_1,\Delta}= \frac{s^{i-{\rm th,\,DM-only}}_{\Delta_1,\Delta}-s^{i-{\rm th,\, hydro}}_{\Delta_1,\Delta}}{s^{i-{\rm th,\, DM-only}}_{\Delta_1,\Delta}}.
\end{equation}
We find the distribution of values of Eq.~(\ref{eq:relative_difference_spars}) to be well approximated by a Gaussian distribution. Hence, as summary statistics, we adopt the mean $\langle \Delta{s}/{s}\rangle$ and the standard deviation $\sigma_{\Delta{s}/s}$ estimated for each cluster sample. These are shown in Fig.~\ref{fig:average_baryonic_bias_profile_GIZMO} for the \textsc{gizmo-simba} clusters, in Fig.~\ref{fig:average_baryonic_bias_profile_GadgetX} for the \textsc{gadget-x} clusters and in Fig.~\ref{fig:average_baryonic_bias_profile_GadgetMUSIC} for the \textsc{gadget-music} clusters. The left (right) panels show the summary statistics for the catalogues at $z=0$ ($z=1$).

Firstly, we may notice that in all the cases the absolute value of the average sparsity relative difference $\langle \Delta{s}/s \rangle_{\Delta_1,\Delta}$ tends to linearly increase as function of $\Delta$, though the sign and amplitude of the variation depend on the simulated cluster sample. This is due to the different sub-grid models implemented in the different hydro simulations. In particular, we find that on average the \textsc{gizmo-simba} clusters have greater sparsities than their DM-only counterparts ($\langle \Delta{s}/s\rangle_{\Delta_1,\Delta}<0$), while \textsc{gadget-music} clusters exhibit smaller ones ($\langle \Delta{s}/s\rangle_{\Delta_1,\Delta}>0$). The \textsc{gadget-x} clusters represents an intermediary case, with smaller relative difference.  

Secondly, we may notice that the standard deviation of the sparsity relative difference also increases as function of $\Delta$, though the level of scatter is smaller for the sample of relaxed clusters. As already mentioned in Sec.~\ref{impact_baryons}, this indicative of the fact that deviations from the dynamical equilibrium largely contribute to the difference between the sparsity estimated from DM-only simulations and hydrodynamical one. 

Hereafter, we provide a more quantitative description of the observed trends. In the case of the \textsc{gizmo-simba} clusters at $z=0$, we find for the full sample that $-1\% \lesssim \langle\Delta{s}/s\rangle_{200,\Delta}\lesssim -7\%$ for $500\le \Delta\le 2500$, $-3\%\lesssim \langle \Delta{s}/s\rangle_{500,\Delta}\lesssim -8\%$ for $1000\le \Delta\le 2500$, and $-1\%\lesssim \Delta{s}/s\rangle_{1000,\Delta}\lesssim -5\%$ for $1500\le \Delta \le 2000$. In the case of the scatter, we find $7\%\lesssim \sigma_{\Delta{s}/s|_{200,\Delta}}\lesssim 17\%$, $9\%\lesssim \sigma_{\Delta{s}/s|_{500,\Delta}}\lesssim 21\%$, and
$7\%\lesssim \sigma_{\Delta{s}/s|_{1000,\Delta}}\lesssim 15\%$. A similar trend occurs at $z=1$. Notice that in the case of the relaxed clusters, we find similar values for the average relative difference both at $z=0$ and $1$, while the scatter results nearly a factor of two smaller than that found from the full sample. 

In the case of \textsc{gadget-x} clusters at $z=0$, we find for the full sample that $\langle \Delta{s}/s\rangle_{200,\Delta}$ remains approximately constant as function of $\Delta$ with a value $\sim 2\%$. On the other hand, we find the average relative difference to be at sub-percent level in the case of $\langle \Delta{s}/s\rangle_{500,\Delta}$ and $\langle \Delta{s}/s\rangle_{1000,\Delta}$. We obtain similar values in the case of the relaxed systems. As far as the scatter is concerned, it varies between $\sim 7\%$ and $\sim 18\%$ for the different sparsity configurations. At $z=1$ we observe a similar trend. Also in this case, the relaxed clusters exhibit similar values of the average relative difference of the sparsity configurations compared to the full case with nearly a factor of two smaller scatter. 

Finally, in the case of the \textsc{gadget-music} clusters at $z=0$, we find for the full sample that $\langle \Delta{s}/s\rangle_{200,\Delta}$ varies linearly between $\sim 2\%$ and $8\%$ for $500\le \Delta\le 2500$, while $\langle \Delta{s}/s\rangle_{500,\Delta}$ varies in the range $2-5\%$ for $1000\le\Delta\le 2500$ and $\langle \Delta{s}/s\rangle_{1000,\Delta}$ varies between $\sim 1\%$ and $\sim 4\%$ in the rage $1500\le\Delta\le 2500$. The standard deviation for the different sparsity configuration varies in the range $\sim 6\%$ to $\sim 15\%$. In the case of the relaxed clusters we find similar values for the average relative difference, while the scatter is nearly a factor of two smaller. Similar trends occur for the $z=1$ case.

\begin{figure}[t]
    \centering
    \includegraphics[width = 0.9\linewidth]{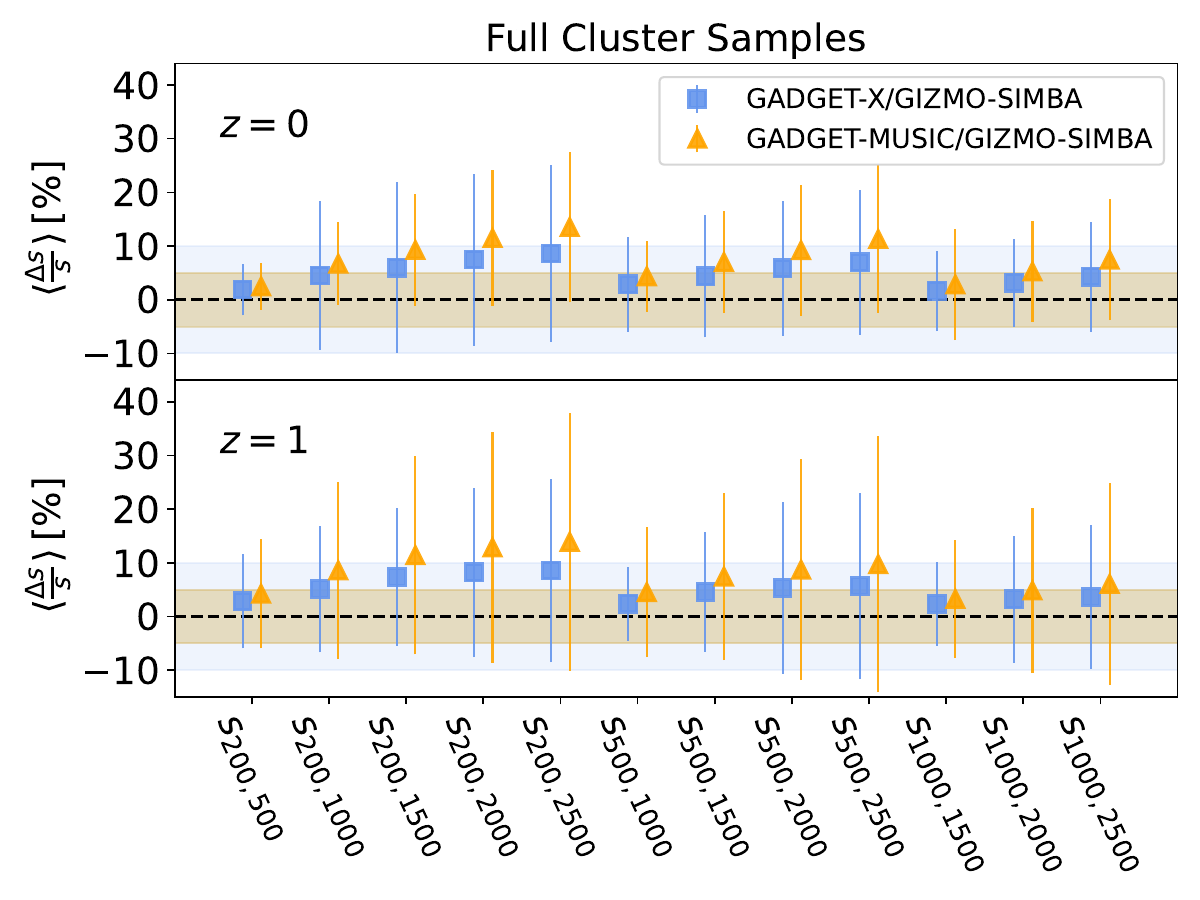}
    \caption{Average and scatter of the relative difference of the sparsity of the \textsc{gadget-x} (blue squares) and \textsc{gadget-music} (yellow triangles) clusters with respect to that of their \textsc{gizmo-simba} counterparts for different sparsity configurations at $z=0$ (top panel) and $1$ (bottom panel) respectively. The errorbars represent the standard deviations for each sparsity configuration, while the shaded areas correspond to $5\%$ and $10\%$ variation respectively. \label{fig:astro_dependence_full}}
    \centering
    \includegraphics[width = 0.9\linewidth]{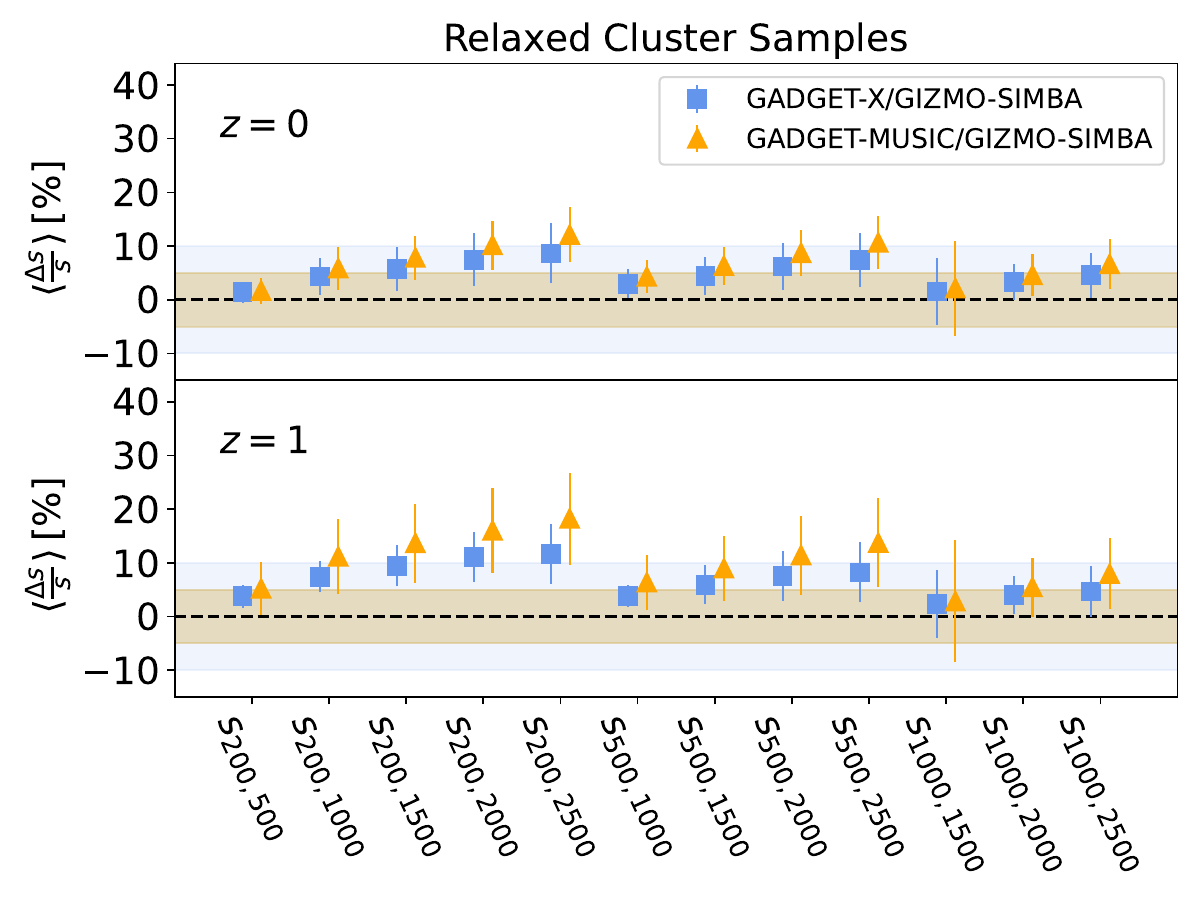}
    \caption{As in Fig.~\ref{fig:astro_dependence_full} for the samples of relaxed clusters.\label{fig:astro_dependence_relaxed}}
\end{figure}

\subsubsection{Baryon Feedback Model Dependence}
The results of the previous analysis show that the presence of baryons alters the average sparsity with respect to the DM-only results. Such differences vary across the cluster mass profile in a way that depends on the sub-grid physics model implemented in the simulations. This suggests that certain sparsity combinations are more sensitive to the astrophysical processes that alter the mass distribution within clusters, while others are less affected and thus better suited for cosmological studies. 

Here, we investigate this in more detail and we compute the average of the relative difference between the sparsities of the \textsc{gadget-x} and \textsc{gadget-music} clusters with respect to those from their \textsc{gizmo-simba} counterparts for different sparsity configurations, i.e. $\langle \Delta{s}/s\rangle\equiv\langle 1-s_{\Delta_1,\Delta_2}/s^{\textsc{gizmo-simba}}_{\Delta_1,\Delta_2} \rangle$. These are shown in Fig.~\ref{fig:astro_dependence_full} and Fig.~\ref{fig:astro_dependence_relaxed} for the full and relaxed samples at $z=0$ (top panels) and $z=1$ (bottom panels) respectively, where the errorbars correspond to the standard deviation. First of all, we observe similar trends for both the full and relaxed cluster samples. Secondly, we notice that in the case of the full cluster samples the scatter is much larger than that obtained from the analysis of the relaxed systems. Hence, to better appreciate the effect of the sub-grid physics implemented in the hydrodynamical simulations, let us focus on the results from the analysis of the relaxed samples shown in Fig.~\ref{fig:astro_dependence_relaxed}. 

We can see that relative differences are minimal for sparsity configurations probing the mass profile between nearby mass shells. For instance, in the case of $s_{200,500}$, which probes the external regions of the cluster profile, we find differences consistent with $0$ at $1\sigma$ for both \textsc{gadget-x} and \textsc{gadget-music} with respect to \textsc{gizmo-simba} clusters. Similarly, the differences in the case of $s_{500,1000}$ and $s_{1000,1500}$ never exceed the $5\%$ level. In contrast, these increase for large overdensity separations probing the slope of the mass profile between inner and outer regions. As an example, $s_{200,2500}$ shows the largest differences of $\sim 8\%$ ($\sim 12\%$) for \textsc{gadget-x} (\textsc{gadget-music}) at $z=0$ and $\sim 12\%$ ($\sim 18\%$) at $z=1$. 

The fact that at large overdensity separations the relaxed \textsc{gizmo-simba} clusters exhibit on average larger sparsities than their \textsc{gadget-x} counterparts is a direct consequence of the different AGN feedback models implemented in the two simulations. More specifically, the \textsc{gadget-x} simulation implements a thermal AGN feedback model, while the \textsc{gizmo-simba} simulations implement a kinetic one \citep[see][for more detail]{2022MNRAS.514..977C}. The latter is more effective at displacing the gas from the centre of the cluster, which reduces the amount of total mass in the core region of the cluster relative to the external one, thus resulting in higher sparsities. Hence, it is not surprising that the differences are even greater in the case of the \textsc{gadget-music} clusters, since the \textsc{gadget-music} simulations do not implement any AGN feedback at all. In the latter case, the dark matter mass in the centre of the cluster undergoes adiabatic contraction due to the collapse of baryons \citep{2004ApJ...616...16G,2011MNRAS.414..195T,2012MNRAS.422.3081M}. This leads to an increase of the total mass in the central region of the cluster relative to the outer one. Consequently, the \textsc{gadget-music} clusters have smaller sparsities compared to those of the simulations that account for AGN feedback.

Overall, our analysis suggests that sparsities such as $s_{200,500}$ and $s_{500,1000}$ are less sensitive to the effects of baryon feedback models, as such they can be a better proxy of the cosmological signal encoded in the cluster mass profile. On the other hand, sparsities probing the slope of the mass distribution across a large radial interval (from external to inner regions), such as $s_{200,2500}$ and $s_{500,2500}$, are more sensitive to the effects of baryonic processes, potentially providing a proxy for cluster astrophysics.

\section{Astrophysical Implications}\label{sec:astro_implications}
\subsection{Sparsity vs Gas Fraction}
In the previous section, we have seen that the presence of the baryons alters the mass profile of clusters as probed by sparsity with respect to the predictions of DM-only simulations in a way that depends on the astrophysical feedback model assumptions. In principle, we may expect these processes to induce correlations between the mass distribution and astrophysical properties of the clusters. To address this point, we now investigate the relation between the mass profile of clusters as probed by the sparsity and the baryon content as traced by the gas fraction. We do not report the analysis of stellar mass fractions as we have found no correlations with the cluster sparsities. 

In this analysis we only consider clusters from the \textsc{gadget-x} and \textsc{gizmo-simba} simulations, which are better in agreement with observations. For each of the clusters in the samples, we estimate the gas fraction within $R_{\Delta}$, $f^{\rm gas}_{R\Delta}=M^{\rm gas}_{\Delta}/M^{\rm tot}_{\Delta}$ for $\Delta=500$ and $2500$. We also consider the gas fraction in a shell comprised between $R_{2500}$ and $R_{500}$, defined as $f^{\rm gas}_{R2500-R500}=(M^{\rm gas}_{500}-M^{\rm gas}_{2500})/(M^{\rm tot}_{500}-M^{\rm tot}_{2500})$. For simplicity, we limit the analysis to sparsities $s_{200,500}$ and $s_{500,2500}$. 

\begin{figure}[t]
    \centering
    \includegraphics[width = 1.0\linewidth]{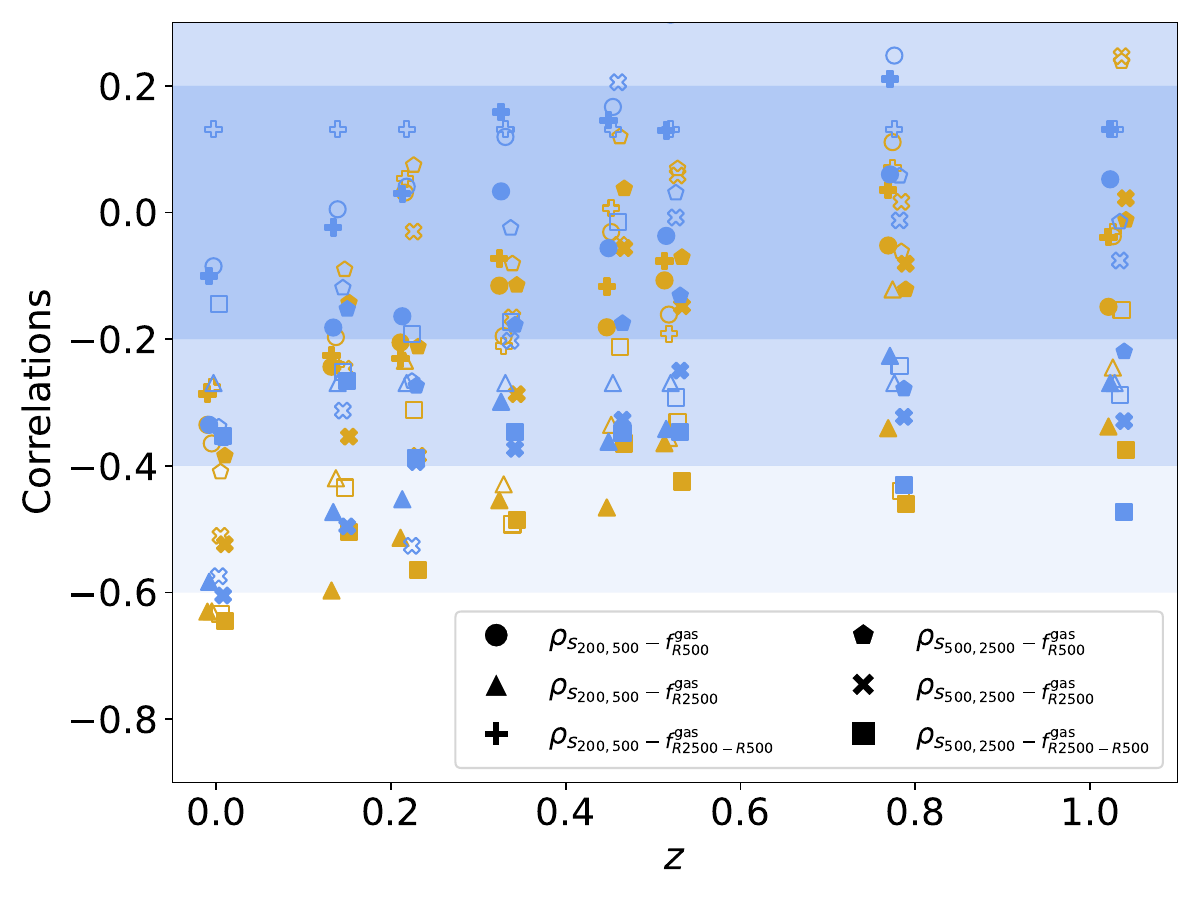}
    \caption{Pearson's correlation coefficients as function of redshift for the \textsc{gizmo-simba} (yellow markers) and \textsc{gadget-x} (blue markers) clusters from the full (filled markers) and relaxed (empty markers) samples. The shaded areas (dark to light blue) correspond to "very weak", "weak" and "moderate" correlations respectively.}
    \label{fig:sparsities_fgas_corr}
\end{figure}

We evaluate the Pearson' correlation coefficients, $\rho_{s_{\Delta_1,\Delta_2}-f^{\rm gas}_{R\Delta}}$, for the different sparsities and gas fraction configurations. These are shown in Fig.~\ref{fig:sparsities_fgas_corr} as function of redshift for the \textsc{gadget-x} (light blue markers) and \textsc{gizmo-simba} (yellow markers) in the case of the full (filled markers) and relaxed (empty markers) samples. The shaded areas corresponds to values of the Pearson coefficients indicative of correlations that are considered "very weak" (dark blue), "weak" (blue) and "moderate" (light-blue). 

We remark that for a given sparsity configuration the correlation coefficients from the \textsc{gizmo-simba} clusters are systematically larger in absolute value than those from the \textsc{gadget-x} ones. This is consistent with the fact that the AGN feedback model implemented in the \textsc{gizmo-simba} simulations is more efficient at displacing gas from the inner halo region than the thermal AGN model of the \textsc{gadget-x} simulations, thus inducing a stronger correlation between the distribution of gas and total mass. Not surprisingly in the case of the \textsc{gizmo-simba} clusters, we find that at $z=0$ $s_{200,500}$ and $f^{\rm gas}_{R2500}$, and $s_{500,2500}$ and $f^{\rm gas}_{R2500-R500}$ are anti-correlated with a "strong" level of correlation ($>60\%$). At higher redshifts, the correlation among these variates diminishes, though remaining above the $20\%$ level. In the case of the \textsc{gadget-x} clusters, the most anti-correlated variables are $s_{200,500}$ and $f^{\rm gas}_{R2500}$, and $s_{500,2500}$ and $f^{\rm gas}_{R2500}$ up to $z\sim 0.3$ with a correlation coefficient varying in the range $-0.6\lesssim \rho_{s-f^{\rm gas}}\lesssim -0.4$. At higher redshifts these variables become weakly correlated, while $s_{500,2500}$ and $f^{\rm gas}_{R2500-R500}$ show a moderate correlation. 

The relaxed clusters show correlations that are of slightly smaller amplitude than those inferred from the full cluster sample. Though, we should consider that for these samples, the estimation of the correlation coefficient is affected by a greater sample variance error, which is of order $\sim 20\%$ to be compared with the $\sim 3-5\%$ of the full cluster sample.

\begin{figure}[t]
    \centering
    \includegraphics[width = 1.0\linewidth]{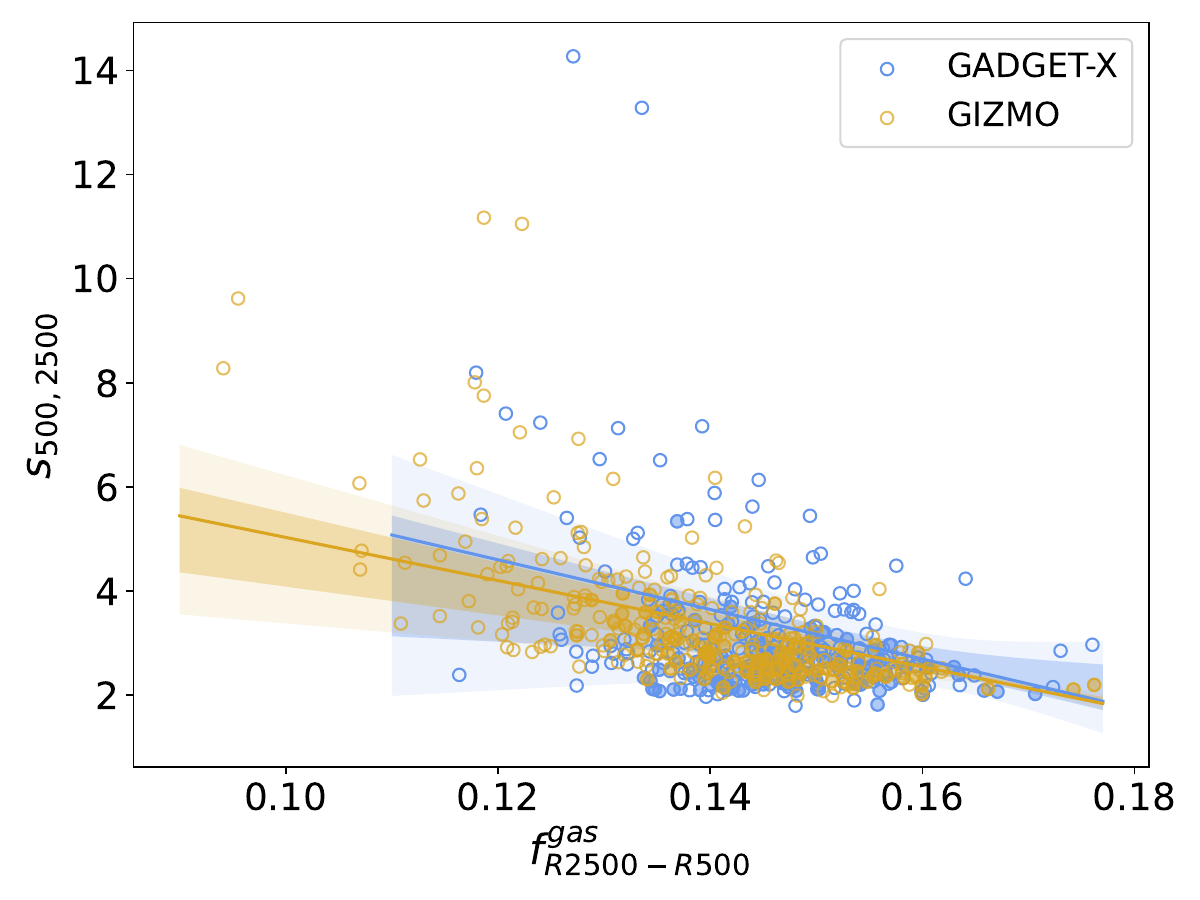}
    \caption{$s_{500,2500}$ vs $f^{\rm gas}_{R2500-R500}$ for the \textsc{gadget-x} (blue circles) and \textsc{gizmo-simba} (yellow circles) clusters at $z=0$ respectively. The relaxed systems are marked as filled circles. The solid lines correspond to the best-fit of the linear regression, while the shaded areas correspond to the $1$ and $2\sigma$ credibility regions.}
    \label{fig:s5002500_fgas}
\end{figure}

It is worth noticing that in all the cases $s_{200,500}$ and $s_{500,2500}$ appear to be weakly correlated with $f^{\rm gas}_{R500}$. And given the fact that these sparsities are weakly correlated sparsities (see Fig.~\ref{fig:corr} in Appendix~\ref{appendix:corr}), their measurements can be combined together with gas fraction estimates within $R_{500}$ to infer stronger cosmological constraints. In contrast, sparsities probing the slope of the cluster mass profile across a large radial range show a weak to a moderate correlation with gas fraction measurements that probe the gas distribution in inner cluster regions. 
The presence of such correlations can be tested through observations. As an example, in Fig.~\ref{fig:s5002500_fgas} we plot the values of $s_{500,2500}$ against the estimates of $f^{\rm gas}_{R2500-R500}$ for the \textsc{gizmo-simba} and \textsc{gadget-x} clusters at $z=0$. We can see that the lower the differential gas fraction the higher the sparsity. Again, this is a direct consequence of the AGN feedback models adopted in the simulations. Indeed, we may remark that $f^{\rm gas}_{R2500-R500}$ spans a lower range of values for the \textsc{gizmo-simba} clusters compared to the \textsc{gadget-x} ones. This is because the kinetic AGN model is more efficient at displacing the gas in the inner halo region compared to the thermal one.
This leads to lower values of the gas fraction, while at the same time it reduces the effect of the adiabatic contraction \citep{2004ApJ...616...16G} and therefore of the total mass in the inner cluster regions, which results in higher sparsity. 

The trends shown in Fig.~\ref{fig:s5002500_fgas} can modelled in terms of a simple linear regression. We perform a Bayesian analysis to infer the slope and intercept for the \textsc{gizmo-simba} and \textsc{gadget-x} clusters at different redshifts. In Table~\ref{tab:linreg} we quote the average, standard deviation and best-fit values of the linear regression coefficients. At fixed redshifts the slopes from the two cluster samples are statistically consistent with one another, though the uncertainties are smaller for the \textsc{gizmo-simba} clusters consistently with their stronger level of correlation for the variables considered. 

It would be interesting to perform such an analyses on a larger sample of simulated clusters, and eventually test such relations on observed cluster samples. We leave this to a future study.

\begin{table}
\centering
\caption{Summary results of the Bayesian linear regression analysis of $s_{500,2500}= m \cdot f_{\rm gas} + c$ for the full cluster sample from the \textsc{gizmo-simba} and \textsc{gadget-x} simulations at different redshifts. We quote average values of $m$ and $c$ with the associated $1\sigma$ uncertainties.\label{tab:linreg}}
\begin{tabular}{c c c c c}
\hline\hline
 & \multicolumn{2}{c}{\textsc{gizmo-simba}} 
 & \multicolumn{2}{c}{\textsc{gadget-x}} \\
\hline
z & $m$ & $c$ & $m$ & $c$ \\
\hline\hline
$0.0$ & $-36\pm{25}$ & $7\pm{3}$ & $-36\pm{28}$ & $7\pm{4}$ \\
$0.1$ & $-31\pm{14}$ & $7\pm{2}$ & $-29\pm{24}$ & $7\pm{4}$ \\
$0.2$ & $-32\pm{13}$ & $8\pm{2}$ & $-33\pm{31}$ & $8\pm{6}$ \\
$0.3$ & $-28\pm{14}$ & $6\pm{2}$ & $-31\pm{24}$ & $7\pm{4}$ \\
$0.5$ & $-21\pm{9}$ & $6\pm{1}$ & $-25\pm{25}$ & $6\pm{4}$ \\
$0.8$ & $-20\pm{15}$ & $6\pm{2}$ & $-27\pm{20}$ & $7\pm{3}$ \\
$1.0$ & $-20\pm{12}$ & $6\pm{2}$ & $-30\pm{19}$ & $7\pm{3}$ \\
\hline
\end{tabular}
\end{table}

\begin{figure}[t]
    \centering
    \includegraphics[width = 1.0\linewidth]{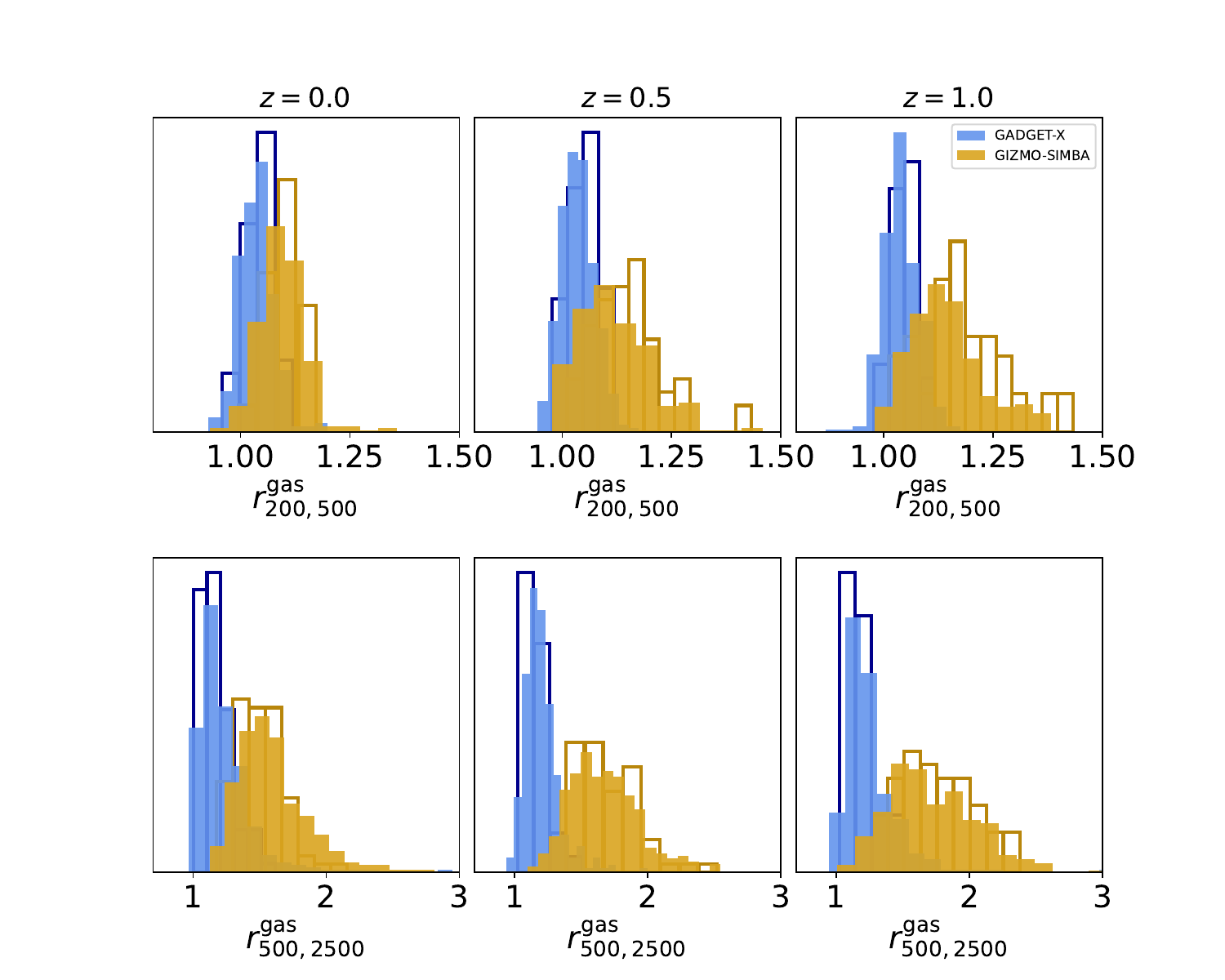}
    \caption{Normalised histograms of the gas mass fraction ratios $r^{\rm gas}_{200,500}$ (top panels) and $r^{\rm gas}_{500,2500}$ (bottom panels) for the \textsc{gadget-x} (blue filled bars) and \textsc{gizmo-simba} (yellow filled bars) clusters at $z=0$ (left panels), $0.5$ (middle panels) and $1$ (right panels) respectively. The histograms with empty bars correspond to the results from the relaxed cluster samples.}
    \label{fig:fgas_ratios}
\end{figure} 
 
\subsection{Sparsity and Gas Fraction Ratios}
Given that mass ratios provide an estimate of the logarithmic slope of the radial mass profile, we want to test to which extent ratios of the gas mass fraction estimated at two different overdensities directly probe the cluster sparsity. To this purpose we compute the gas mass fraction within $R_{\Delta}$ for $\Delta=200$, $500$ and $2500$ for each cluster in the \textsc{gadget-x} and \textsc{gizmo-simba} samples. Then, we estimate the gas mass fraction ratios $r^{\rm gas}_{200,500}=f^{\rm gas}_{R200}/f^{\rm gas}_{R500}$ and $r^{\rm gas}_{500,2500}=f^{\rm gas}_{R500}/f^{\rm gas}_{R2500}$. By definition $r^{\rm gas}_{\Delta_1,\Delta_2}\equiv s^{\rm gas}_{\Delta_1,\Delta_2}/s_{\Delta_1,\Delta_2}$, hence if the mass gas fraction profile is constant then $r^{\rm gas}_{\Delta_1,\Delta_2}=1$ and $s^{\rm gas}_{\Delta_1,\Delta_2}$ exactly trace the cluster sparsity $s_{\Delta_1,\Delta_2}$. 

In Fig.~\ref{fig:fgas_ratios} we plot the normalised histograms of $r^{\rm gas}_{200,500}$ (top panels) and $r^{\rm gas}_{500,2500}$ (bottom panels) for the \textsc{gadget-x} (blue filled bars) and \textsc{gizmo-simba} (yellow filled bars) clusters at $z=0$ (left panels), $0.5$ (middle panels) and $1.0$ (right panels) respectively. The histograms with empty bars correspond to the results from the analysis of the relaxed cluster samples. In Table~\ref{tab:fgas_ratio_stat}, we quote the summary statistics for the different cases in terms of median, average and percentiles of the gas mass fraction ratios.

We can see that the distributions are skewed toward values larger than unity\footnote{The fact that the sparsities are by definition strictly larger than unity does not imply that the gas mass fraction ratios have to be greater than one, since there is no physical reason to prevent $s^{\rm gas}_{\Delta_1,\Delta_2}<s_{\Delta_1,\Delta_2}$. Indeed, a few clusters in the sample have such values.}. This is a direct consequence of the fact that the gas fraction decreases at lower cluster radii. This is a well known result both from numerical simulations \citep[see e.g.][Rasia et al., in prep.]{1999ApJ...525..554F,2004MNRAS.351..237R,2006MNRAS.365.1021E,2013MNRAS.431.1487P,2013ApJ...777..123B,2023A&A...675A.188A} as well as observations of clusters \citep[see e.g.][]{2005A&A...437...31S,2010A&A...511A..85P,2013A&A...551A..23E,2016MNRAS.456.4020M}. In fact, even in non-radiative simulations, multiple astrophysical processes can deplete the gas mass from the inner cluster regions. Because of this, the gas mass fraction ratio is a biased tracer of the cluster sparsity, though the level of bias depends on the astrophysical processes that takes place in the intra-cluster medium. As an example, in the case of the \textsc{gadget-x} clusters we have an average bias of $\sim 4\%$ for $s_{200,500}$ in the redshift range $0<z<1$, while in the case of the \textsc{gizmo-simba} sample this is of the order of $\sim 9-14\%$. In the case of $s_{500,2500}$, the bias between the gas mass fraction ratio and the sparsity is of the order of $\approx 20\%$ for the \textsc{gadget-x} and $\approx 60-70\%$ for the \textsc{gizmo-simba} clusters respectively. We find results of the same order of magnitude in the case of the relaxed samples. The effect is larger on the \textsc{gizmo-simba} clusters than the \textsc{gadget-x} ones, consistently with the fact that the AGN feedback model is more efficient at displacing the gas from the inner cluster regions in the former case than the latter. Indeed, we may consider that the difference between the \textsc{gizmo-simba} and \textsc{gadget-x} distributions of the gas mass fraction ratios are a distinct signature of the feedback scenario implemented in the simulations, possibly providing a test against measurements of the gas fraction in clusters at different overdensities.

\begin{table}[t]
\centering
\caption{\label{tab:fgas_ratio_stat} Summary statistics of $r^{\rm gas}_{200,500}$ and $r^{\rm gas}_{500,2500}$ for the \textsc{gadget-x} and \textsc{gizmo-simba} clusters from the full (relaxed) samples. The different columns from left to right correspond to the redshift, the median (denoted as $\mu$), average, $16$-th and $84$-th percentiles.}
 \begin{tabular}{c c c c c}

 \multicolumn{5}{c}{\textsc{gadget-x}}\\ \\
 \hline\hline
 $z$ & $\mu(r^{\rm gas}_{200,500})$ & $\langle r^{\rm gas}_{200,500}\rangle$ & $p_{16}$ & $p_{84}$ \\
 \hline
 $0.0$ & $1.04$ ($1.04$) & $1.04$ ($1.04$)  & $0.99$ ($1.01$)  & $1.07$ ($1.07$) \\
 $0.5$ & $1.03$ ($1.05$) & $1.03$ ($1.05$) & $0.99$ ($1.02$) & $1.08$ ($1.08$) \\
 $1.0$ &$1.04$ ($1.04$) & $1.04$ ($1.04$) & $1.00$ ($1.01$) & $1.08$ ($1.07$) \\
 \hline\hline
  $z$ & $\mu(r^{\rm gas}_{500,2500})$ & $\langle r^{\rm gas}_{500,2500}\rangle$ & $p_{16}$ & $p_{84}$ \\
  \hline
 $0.0$ & $1.18$ ($1.15$) & $1.23$ ($1.17$) & $1.06$ ($1.06$) & $1.38$ ($1.27$)\\
 $0.5$ & $1.18$ ($1.14$) & $1.20$ ($1.16$) & $1.09$ ($1.08$) & $1.30$ ($1.24$) \\
 $1.0$ & $1.20$ ($1.16$) & $1.25$ ($1.19$) & $1.09$ ($1.10$) & $1.39$ ($1.28$) \\
 \hline\hline
 \\
 \multicolumn{5}{c}{\textsc{gizmo-simba}}\\ \\
 \hline\hline
  $z$ & $\mu(r^{\rm gas}_{200,500})$ & $\langle r^{\rm gas}_{200,500}\rangle$ & $p_{16}$ & $p_{84}$ \\
 \hline
 $0.0$ & $1.09$ ($1.10$) &  $1.09$ ($1.10$) & $1.04$ ($1.06$)& $1.14$ ($1.13$)\\
 $0.5$ & $1.11$ ($1.14$) & $1.12$ ($1.15$)& $1.04$ ($1.07$) & $1.19$ ($1.22$) \\
 $1.0$ & $1.13$ ($1.15$) & $1.14$ ($1.16$)& $1.06$ ($1.07$) & $1.22$ ($1.23$)\\
 \hline\hline
  $z$ & $\mu(r^{\rm gas}_{500,2500})$ & $\langle r^{\rm gas}_{500,2500}\rangle$ & $p_{16}$ & $p_{84}$ \\
  \hline
 $0.0$ & $1.56$ ($1.50$)& $1.60$ ($1.50$)& $1.36$ ($1.32$)& $1.85$ ($1.66$)\\
 $0.5$ & $1.63$ ($1.66$)& $1.67$ ($1.70$)& $1.42$ ($1.46$)& $1.91$ ($1.91$)\\
 $1.0$ & $1.67$ ($1.75$) & $1.72$ ($1.80$) & $1.40$ ($1.56$) & $2.08$ ($2.09$)\\
 \hline\hline
 \end{tabular}
\end{table}

\begin{figure*}[ht]
    \centering
    \includegraphics[width = 0.8\linewidth]{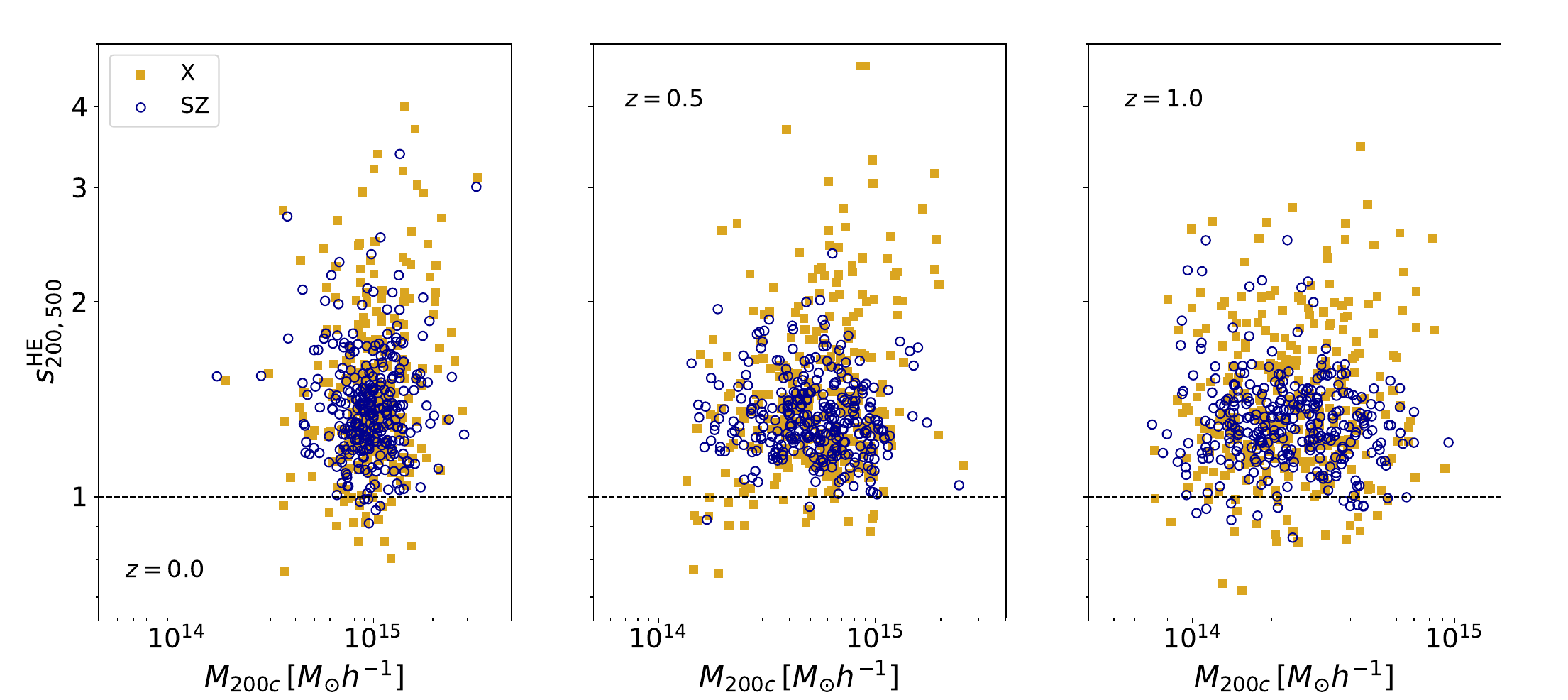}
    \caption{Sparsity $s_{200,500}$ estimated from the HE masses of the \textsc{gadget-x} clusters from \citet{2023MNRAS.518.4238G} at $z=0$ (left panel), $0.5$ (central panel) and $1.0$ (right panel) using the X-ray temperature and gas density profiles (filled goldenrod squares) and SZ pressure profiles (blue empty circles). Values of $s_{200,500}\le 1$ correspond to disturbed clusters for which the HE is not satisfied. Some of these are associated to systems with a vanishing or negative HE mass bias at $R_{500}$ and a significant bias at $R_{200}$.}
    \label{fig:he_sparsity_unphysical}
\end{figure*} 

\section{The impact of hydrostatic mass bias}
\label{sec:hydro_bias}

In the previous sections, we have investigated the impact of baryons on the mass profile of simulated clusters, as traced by the sparsity statistics. We have evaluated how the baryons modify the average sparsities with respect to expectations from DM-only simulations, and discussed how such effects vary depending on the radial interval probed by the sparsities and the underlying astrophysical feedback scenario implemented in the simulations. Estimating these effects is important given that cosmological model predictions of the sparsity statistics rely on DM-only simulations (see Sec.~\ref{subsec:cosmo_predictions} for predictions based on the relation between average sparsity halo mass functions). However, comparing with observations also requires identifying potential sources of bias that arise from specific aspects of the methodology adopted to measure the mass of clusters at different overdensities. 

Here, we consider the case of X-ray and SZ observations of galaxy clusters providing measurements of the cluster mass at different overdensities under the hydrostatic equilibrium hypothesis. Observationally, HE masses are known to provide a biased measurement of the cluster mass as estimated, for instance, by lensing observations \citep[see e.g.][for a review]{2019SSRv..215...25P}. Hydrodynamical simulations have been extensively used to investigate the amplitude and origin of the HE bias with respect to the true mass of simulated clusters \citep[see e.g.][]{2017MNRAS.466.4442L,2017MNRAS.465.3361H,2020MNRAS.491.1622P,2020A&A...634A.113A,2021MNRAS.506.2533B,2024MNRAS.534..251K,2025MNRAS.536.3784B}. These studies have mainly focused on determining the mass and redshift evolution of the HE bias at fixed overdensity. However, a key question for sparsity analyses is how this bias varies across the mass profile. In fact, a radial-dependent mass bias will inevitably induce a bias on the estimated sparsity and propagate as a systematic error in the cosmological model parameter inference analyses. To assess such an effect, we use the HE mass estimates obtained in \citet{2023MNRAS.518.4238G} from the spherically averaged properties of the intra-cluster medium of simulated clusters at different redshift snapshots from the \textsc{gadget-x} runs. Their study has focused on the sample of most massive clusters in the zoomed-in simulated regions that do not contain any low-resolution particles within the virial radius, thus resulting in a sample of $290$ clusters at $z=0$. For each of these clusters, the HE masses at $\Delta=200$, $500$, and $2500$ have been determined using parametric fitting functions of the spherically averaged 3D pressure profile (as in the case of SZ-like observations of clusters), and the 3D gas density and temperature profiles (as in the case of X-ray cluster observations). We have used the data to compute the following set of sparsities: $s^{\rm HE}_{200,500}$, $s^{\rm HE}_{200,2500}$ and $s^{\rm HE}_{500,2500}$. 

\begin{figure}[t]
    \centering
    \includegraphics[width = 1.05\linewidth]{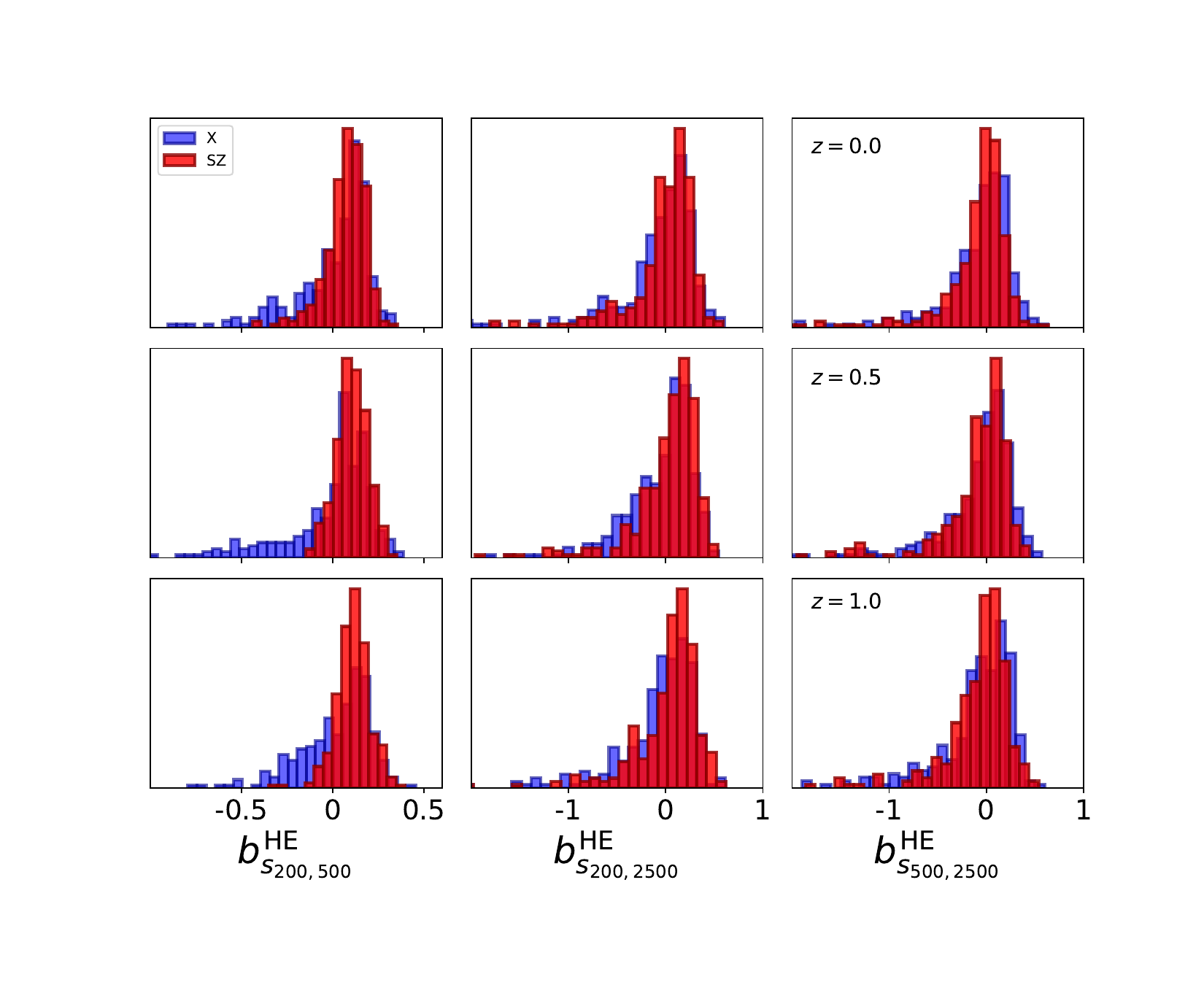}
    \caption{Probability density function of the HE sparsity bias $b_{s_{200,500}}$ (left columns), $b_{s_{200,2500}}$ (central columns) and $b_{s_{500,2500}}$ (right columns) for the X-ray (blue bars) and SZ (red bars) masses respectively at $z=0.07$ (top rows), $0.46$ (central rows) and $0.99$ (bottom rows).}
    \label{fig:he_bias_dist}
\end{figure} 

Firstly, we notice that the sparsities estimated using the X-ray masses have larger scatter than the SZ ones. Secondly, for each of the available redshifts, we found clusters with sparsities $s^{\rm HE}_{200,500}<1$, corresponding to $M^{\rm HE}_{200}<M^{\rm HE}_{500}$, while estimates of the sparsities $s^{\rm HE}_{200,2500}$ and $s^{\rm HE}_{500,2500}$ all lie above the physical threshold. These are shown in Fig.~\ref{fig:he_sparsity_unphysical}, where we plot $s_{200,500}$ from the X-ray (filled goldenrod squares) and SZ (empty blue circles) mass estimates against the true cluster mass $M_{200}$ at $z=0.0,0.5$ and $1.0$.
As we can see $5-10\%$ percent of the objects lie below the $s^{\rm HE}_{200,500}=1$ line and the number of such peculiar clusters is greater in the X-ray case than the SZ one. We verify that these cases are associated with disturbed systems that tend to have a negative bias (for which the HE mass is higher than the real mass) at  $R_{500}$ and a high positive bias at $R_{200}$. In real data, this case could be present and may be the result from extrapolation of the fitting functions over regions that are not constrained by the data \citep[see e.g. the case of HE masses of the HIFLUGCS clusters discussed in][]{2019MNRAS.487.4382C}. 

In the following, we exclude these clusters from the HE samples. Then, for each cluster we compare the HE estimated sparsities to those of the corresponding \textsc{gadget-x} cluster and evaluate the HE sparsity bias\footnote{Notice that Eq.~(\ref{eq:bias_he}) can be written in terms of the HE mass bias of a given cluster at overdensity $\Delta_1$ and $\Delta_2$ respectively:
\begin{equation}
b^{\rm HE}_{s_{\Delta_1,\Delta_2}}\equiv 1-\frac{1-b^{\rm HE}_{M_{\Delta_1}}}{1-b^{\rm HE}_{M_{\Delta_2}}},
\end{equation}
where $b^{\rm HE}_{M_{\Delta}}=1-M^{\rm HE}_{\Delta}/M^\textrm{true}_{\Delta}$.}:
\begin{equation}
b^{\rm HE}_{s_{\Delta_1,\Delta_2}}=1-\frac{s^{\rm HE}_{\Delta_1,\Delta_2}}{s^\textrm{true}_{\Delta_1,\Delta_2}},\label{eq:bias_he}
\end{equation}
where $s^{\rm true}_{\Delta_1,\Delta_2}$ is the sparsity obtained using the total 3D mass of the \textsc{gadget-x} cluster. In Fig.~\ref{fig:he_bias_dist}, we plot the probability density function of the HE sparsity bias for the different overdensity pairs (panels from left to right) at $z=0.0$ (top panels), $0.5$ (central panels) and $1.0$ (bottom panels) for the X-ray (blue bars) and SZ (red bars) mass estimates respectively. As we can see the distributions exhibit non-Gaussian tails, hence we use the mean, the median and the 16-th and 84-th percentiles as summary statistics.

\subsection{Mass \& Redshift Dependence}\label{he_redshift_mass_dependence}

\begin{figure}[t]
    \centering
    \includegraphics[width = 0.95\linewidth]{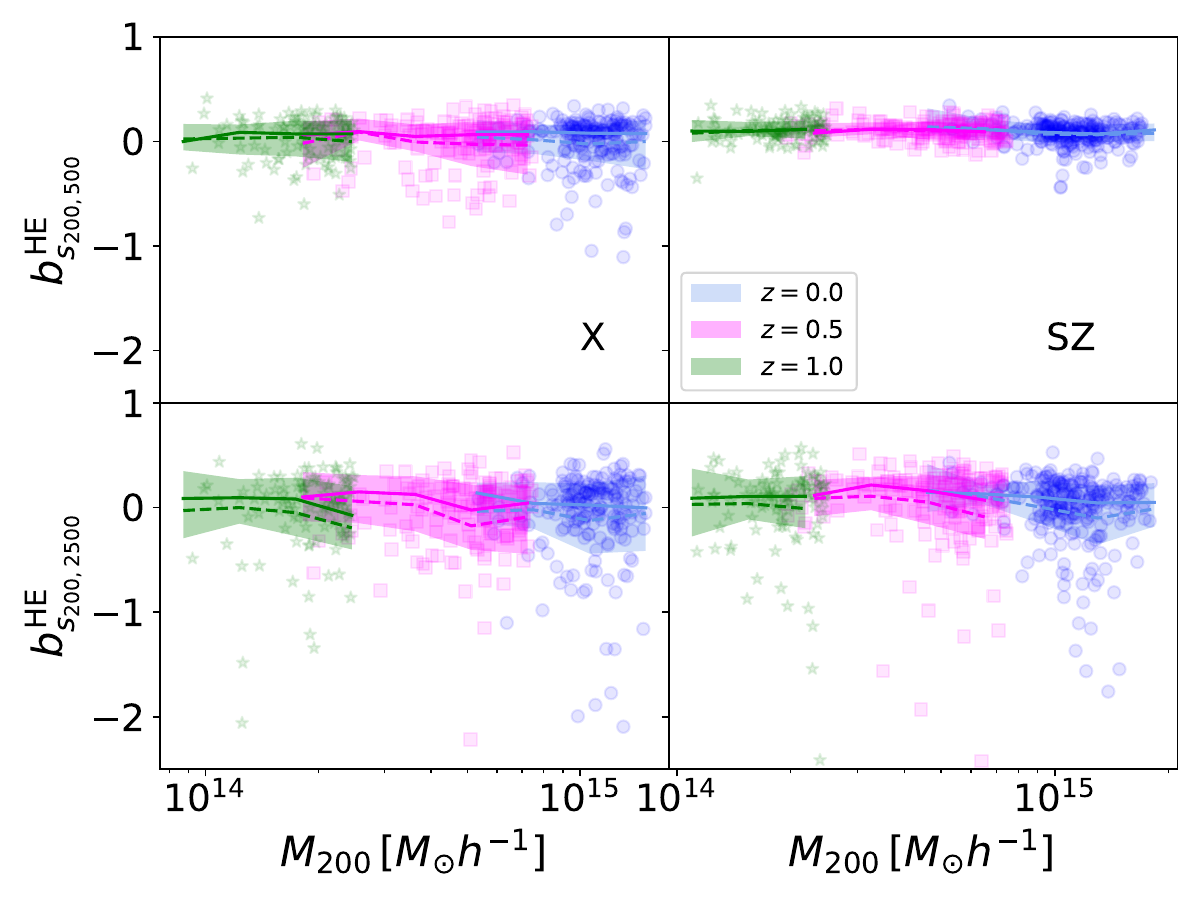}\\
    \includegraphics[width = 0.95\linewidth]{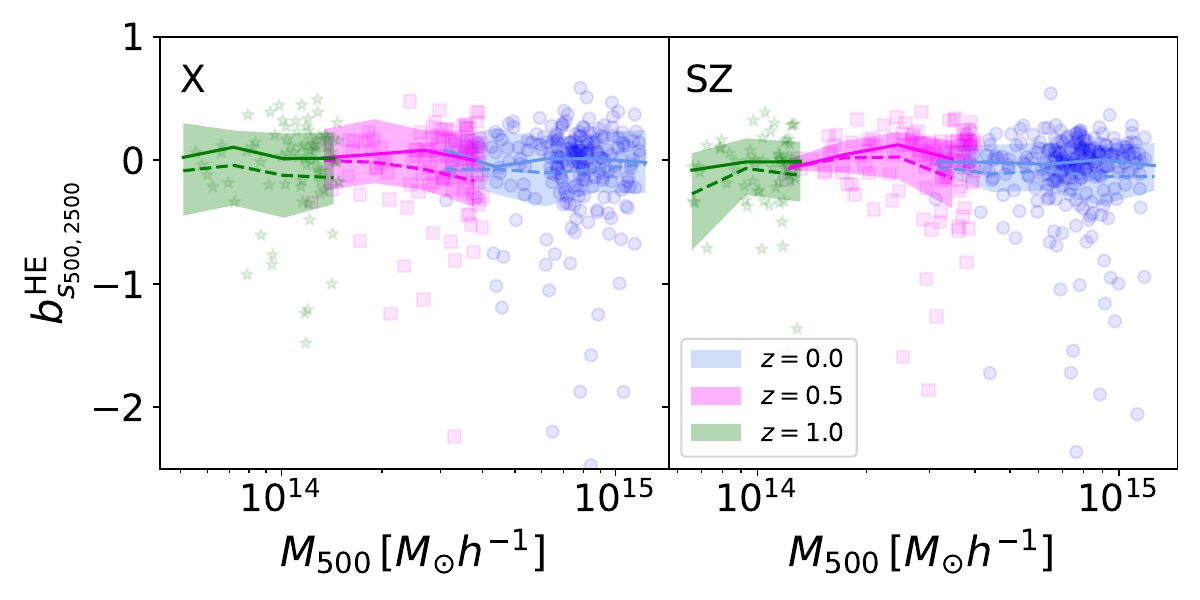}
    \caption{HE sparsity bias $b_{s_{200,500}}$ and $b_{s_{200,2500}}$ as function of $M_{200}$ (top panels) and $b_{s_{500,2500}}$ as function of $M_{500}$ (bottom panel) for the X (left panels) and SZ (right panels) masses of the full cluster sample at $z=0.0$ (blue circles), $0.5$ (magenta squares) and $1.0$ (green stars). The different lines correspond to the median (solid) and the mean (dashed), while the shaded area to the region comprised between the 16th and 84th percentile.}
    \label{fig:he_bias_mass}
\end{figure}

First, we investigate the dependence of the summary statistics on the mass of the simulated clusters. This is shown, in Fig.~\ref{fig:he_bias_mass}, where we plot the mean, the median and the 16-th and 84th percentiles of the HE bias for the three sparsity configuration considered as function of $M_{200}$ for the samples at $z=0.0$ (blue), $0.5$ (magenta) and $1.0$ (green) respectively. As we can see, the bias is largely independent of the mass enclosed within the outer radius probed by the sparsity. This is a direct consequence of the fact that the HE mass bias at different overdensity is independent of the cluster mass as found by several works in the literature in the high-mass range \citep[see e.g.][]{2017MNRAS.466.4442L,2017MNRAS.465.3361H,2020MNRAS.491.1622P,2020A&A...634A.113A,2021MNRAS.506.2533B}. We may also notice that there is no significant variation of the HE bias among the three different redshift samples, thus indicating a weak dependence on redshift. 

Indeed, concerning the redshift dependence we can see from Fig.~\ref{fig:he_bias_dist} that in all the cases the mean, median and percentiles of the HE sparsity bias remain approximately constant as a function of redshift. Moreover, we remark that the biases from the X-ray masses (left panels) exhibit a larger scatter than those from the SZ masses (right panels). This is especially the case for $b^{\rm SZ}_{s_{200,500}}$, that shows a level of scatter twice as small as that of $b^{\rm X}_{s_{200,500}}$. This is consistent with the different level of scatter of the HE mass bias at $\Delta=200$ and $500$ found in \citet{2023MNRAS.518.4238G} for the X-ray and SZ cases respectively (see their Fig~1). 

In Tab.~\ref{tab:bias_HE_full}, we quote the values of the mean and median of the different HE sparsity bias estimated by merging together the different redshift samples, although we point out that joining the samples may introduce some spurious correlations from the repetition of objects which we cannot quantify. A point which will try to assess when evaluating the impact of such biases on the cosmological parameter inference analyses.

\begin{figure}[h]
    \centering
    \includegraphics[width = 0.95\linewidth]{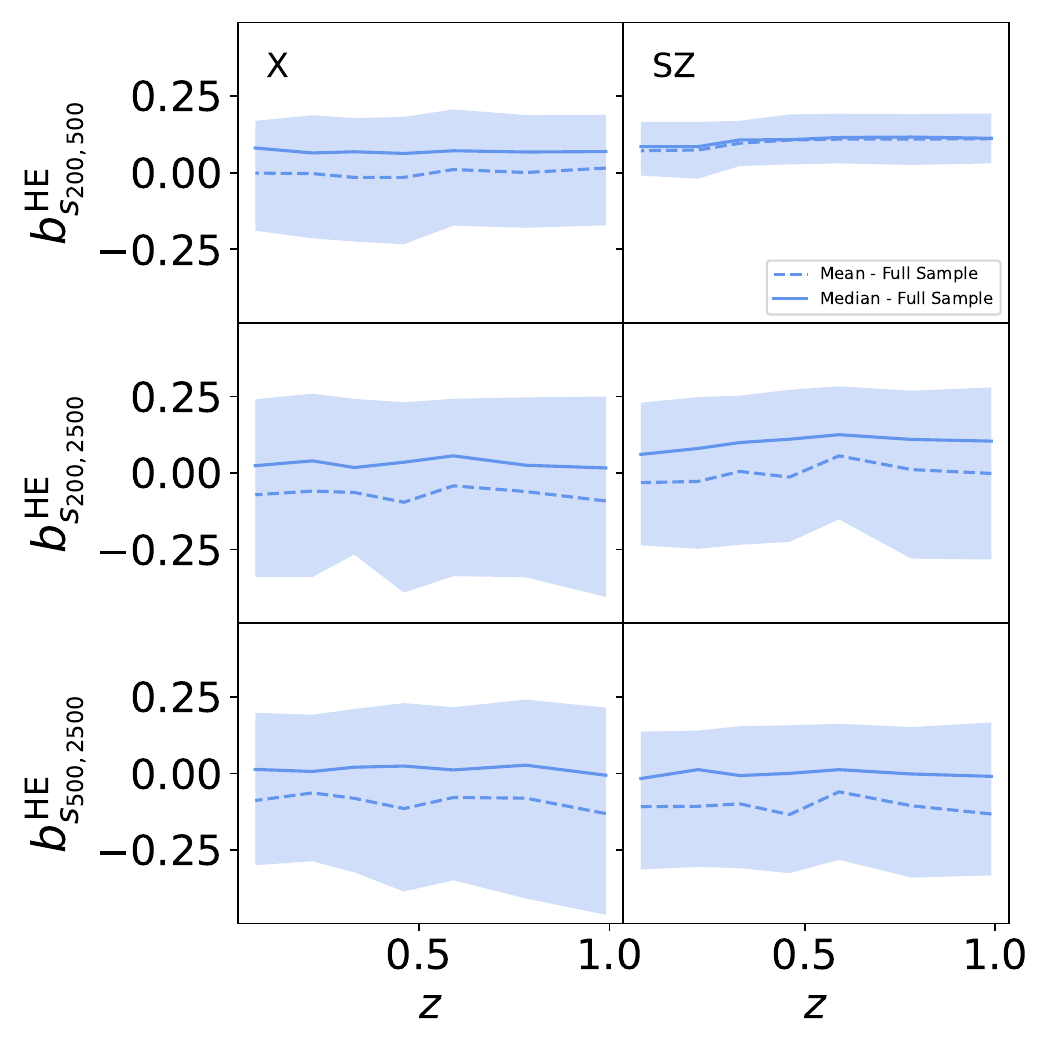}
    \caption{Summary statistics of the HE sparsity bias $b_{s_{200,500}}$ (top panels), $b_{s_{200,2500}}$ (mid panels) and $b_{s_{500,2500}}$ (bottom panels) as function of redshift for the X (left panels) and SZ (right panels) masses. The different lines correspond to the median (solid) and mean (dashed), while the shaded area corresponds to the region comprised between the 16-th and 84-th percentile.}
    \label{fig:he_bias_z}
\end{figure} 

\begin{table}
    \centering
    \caption{Mean and median (denoted as $\mu$) of the hydrostatic biases for three sparsities in the case of the X and SZ estimated masses respectively. The errors quoted on the means correspond to the standard error, $\sigma/\sqrt{N}$, while the errors on the medians correspond to the $16$-th and $84$-th percentile.}
    \begin{tabular}{c|ccc}
        \hline
         & $s_{200,500}$ & $s_{200,2500}$ & $s_{500,2500}$\\
        \hline
        $\langle b^{\rm X}_{\Delta_1,\Delta_2}\rangle$ & $-0.001 \pm 0.005$ & $-0.07 \pm 0.01$ & $-0.09 \pm 0.01$\\
        $\mu(b^{\rm X}_{\Delta_1,\Delta_2})$ & $0.07^{+0.11}_{-0.26}$ & $0.03^{+0.22}_{-0.37} $ & $0.01^{+0.20}_{0.37}$\\
        \hline
        $\langle b^{\rm SZ}_{\Delta_1,\Delta_2}\rangle$ &  $0.097 \pm 0.002$ & $-0.00 \pm 0.01$ & $-0.11 \pm 0.01$\\
        $\mu(b^{\rm SZ}_{\Delta_1,\Delta_2})$ & $0.10_{-0.09}^{+0.08}$ & $0.10_{-0.33}^{+0.16}$ & $-0.002_{-0.317}^{+0.156}$\\
        \hline
    \end{tabular}
    \label{tab:bias_HE_full}
\end{table}

\begin{figure*}
    \centering
    \includegraphics[width = \linewidth]{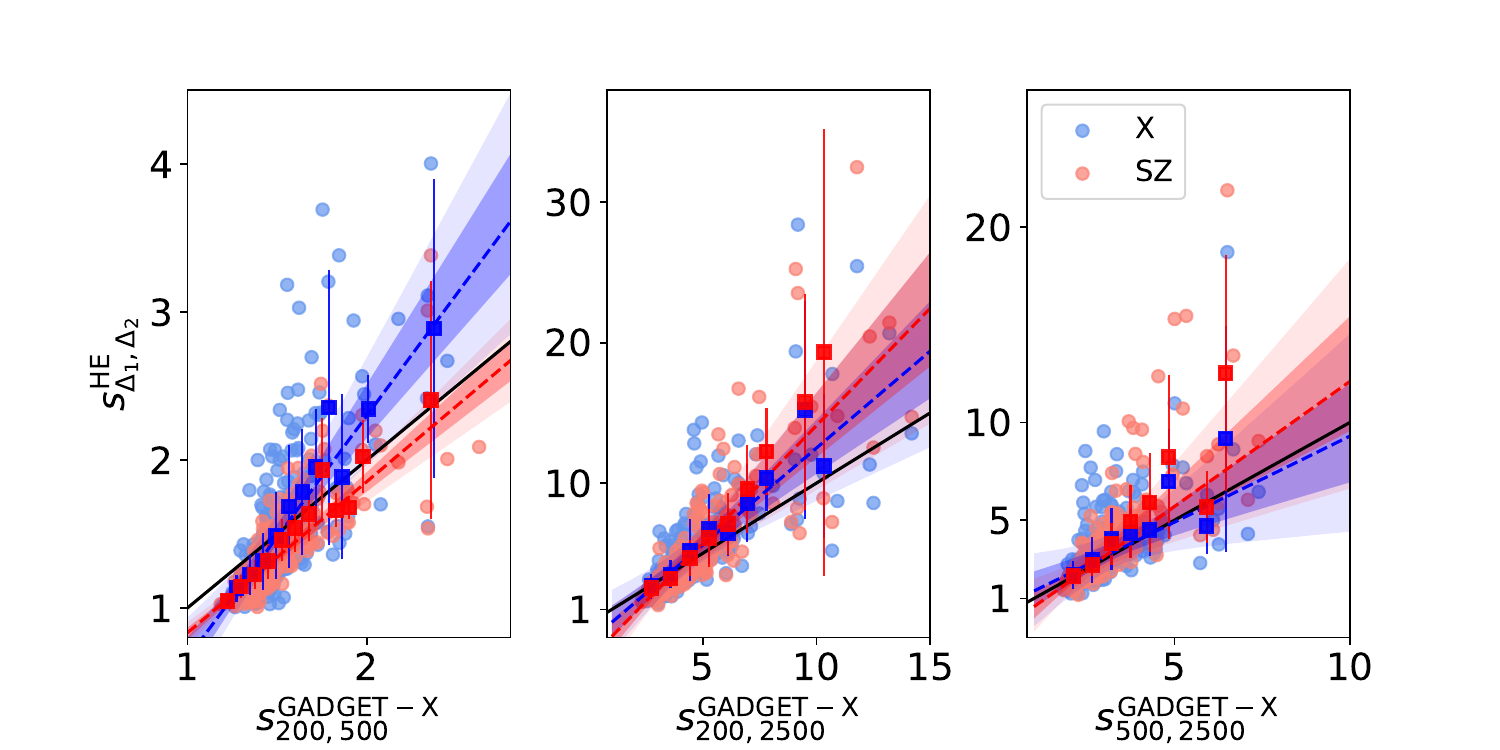}
    \caption{HE sparsities $s^{\rm HE}_{\Delta_1,\Delta_2}$ from the X (blue) and SZ (red) catalogs with respect to their \textsc{gadget-x} counterparts at $z=0$ in the case of $s_{200,500}$ (left panel), $s_{500,2500}$(central panel), and $s_{200,2500}$ (right panel). The coloured dashed line correspond to the average linear fit to the 25 equally spaced binned data points (squares) in bins of true sparsities containing at least two clusters. The shaded areas indicates that $1$ and $2\sigma$ credible regions. The black solid line corresponds to the unbiased case.}
    \label{fig:sparsity_bias_dependence}
\end{figure*}

\subsection{HE Sparsity vs True Sparsity}
Having discussed the summary statistics of the HE sparsity bias, it is worth investigating the relation between the true sparsities and the HE ones on a cluster-by-cluster basis. In Fig.~\ref{fig:sparsity_bias_dependence}, we plot the sparsities $s_{200,500}$ (left panel), $s_{200,2500}$ (central panel) and $s_{500,2500}$ (right panel) of the \textsc{gadget-x} clusters from the X-ray (light blue circles) and SZ (light red circles) samples at $z=0$ against the true sparsity values. For visual purposes, we plot the unbiased case as solid black lines. We also plot the mean of the X-ray (blue squares) and SZ (red squares) sparsities in 25 equally spaced bins of true sparsity values, where we have only retained the bins with more than two clusters. The errorbars represent the standard deviation. We find that a simple linear fit captures the relation between the true sparsities and HE ones, with a scatter that increases as function of true sparsity. We perform a Bayesian linear regression to infer the values of the slope $m_{\rm HE}$ and the intercept $c_{\rm HE}$. The average linear fits are shown in Fig.~\ref{fig:sparsity_bias_dependence} as dashed lines, while the shaded areas correspond the $1$ and $2\sigma$ credible regions. The results show that the steeper the slope the greater the bias as well as the level of scatter. For instance, in the case of $s_{200,500}$, we can see the the SZ sparsities lies nearly parallel to the diagonal line with a small scatter. This is consistent with the low level of bias and reduced scatter we have found in the case of the SZ bias summary statistics shown in the top right panels of Fig.~\ref{fig:he_bias_mass} and Fig.~\ref{fig:he_bias_z}. This is not the case of the X-ray sparsities, for which the relation to the true sparsities has a steeper slope and greater scatter.

We have performed a similar regression at other redshifts and found no statistically significant trend, consistently with the results of the analysis of the summary statistics. We quote in Table~\ref{tab:x_he_true_fit} (Table~\ref{tab:sz_he_true_fit}) the inferred values of the fitting coefficients for the X (SZ) sparsities respectively. These fits provide a model for the intrinsic HE sparsity bias that is necessary to take into account in the case of cosmological parameter inference analyses the make use of individual cluster sparsity measurements as proposed in \cite{2022arXiv221203233R}, an approach that we do not consider here.

\begin{table*}\label{tab:x_he_true_fit}
\centering 
\caption{Average and $1\sigma$ errors on the slope $m_X$ and intercept $c_X$ from the Bayesian linear regression fit of the HE sparsities vs the true sparsities for the different overdensity configurations using the samples at $z=0,0.2,0.5,0.8$ and $1$ respectively.}
\begin{tabular}{c|cccccc}
\hline
 & \multicolumn{2}{c}{$s_{200,500}$} & \multicolumn{2}{c}{$s_{200,2500}$} & \multicolumn{2}{c}{$s_{500,2500}$} \\
 \hline
$z$ & $m_X$ & $c_X$ & $m_X$ & $c_X$ &$m_X$ & $c_X$ \\
\hline
$0.0$ & $1.65\pm 0.30$ & $-0.99\pm 0.41$ &$1.40\pm 0.33$ & $-1.20\pm 1.40$ & $0.89\pm 0.37$ & $0.50\pm 1.30$ \\
$0.2$ & $0.98\pm 0.29$ & $-0.07\pm 0.48$ &$1.52\pm 0.35$ & $-1.80\pm 1.50$ & $1.77\pm 0.29$ & $-1.83\pm 0.84$ \\
$0.5$ & $1.46\pm 0.25$ & $-0.76\pm 0.36$ &$1.59\pm 0.29$ & $-2.07\pm 0.92$ & $1.64\pm 0.12$ & $-1.30\pm 0.45$ \\
$0.8$ & $1.41\pm 0.28$ & $-0.65\pm 0.36$ &$1.44\pm 0.20$ & $-1.71\pm 0.68$ & $1.88\pm 0.28$ & $-2.60\pm 0.63$ \\
$1.0$ & $0.87\pm 0.21$ & $0.11\pm 0.28$ &$1.37\pm 0.24$ & $-1.54\pm 0.87$ & $1.19\pm 0.35$ & $-0.49\pm 1.00$ \\
\hline
\end{tabular}
\end{table*}

\begin{table*}\label{tab:sz_he_true_fit}
\centering 
\caption{As in Table~\ref{tab:x_he_true_fit} for the SZ sparsities.}
\begin{tabular}{c|cccccc}
\hline
 & \multicolumn{2}{c}{$s_{200,500}$} & \multicolumn{2}{c}{$s_{200,2500}$} & \multicolumn{2}{c}{$s_{500,2500}$} \\
 \hline
$z$ & $m_{SZ}$ & $c_{SZ}$ & $m_{SZ}$ & $c_{SZ}$ &$m_{SZ}$ & $c_{SZ}$ \\
\hline
$0.0$ & $1.03\pm 0.10$ & $-0.20\pm 0.14$ &$1.63\pm 0.38$ & $-2.50\pm 1.50$ & $1.31\pm 0.40$ & $-0.70\pm 1.10$ \\
$0.2$ & $0.99\pm 0.24$ & $-0.11\pm 0.34$ &$1.58\pm 0.40$ & $-2.10\pm 1.40$ & $1.90\pm 0.33$ & $-2.20\pm 0.84$ \\
$0.5$ & $0.79\pm 0.10$ & $0.13\pm 0.13$ & $1.42\pm 0.27$ & $-1.70\pm 0.91$ & $1.86\pm 0.30$ & $-1.96\pm 0.72$ \\
$0.8$ & $0.85\pm 0.21$ & $0.07\pm 0.30$ &$1.38\pm 0.18$ & $-1.70\pm 0.55$ & $2.12\pm 0.28$ & $-2.63\pm 0.70$ \\
$1.0$ & $0.58\pm 0.13$ & $0.43\pm 0.19$ &$1.54\pm 0.19$ & $-2.35\pm 0.70$ & $1.62\pm 0.38$ & $-1.51\pm 0.99$ \\
\hline
\end{tabular}
\end{table*}

\section{Cosmological Implications}\label{sec:cosmo}
In light of the analyses presented in the previous sections, it is of interest to evaluate the implications of the sparsity bias on cosmological analyses performed using this type of measurement. For instance, \cite{Corasaniti2018}, \cite{Corasaniti2021}, and \cite{2022arXiv221203233R} propose different approaches to use the sparsities of galaxy clusters to constrain cosmology. These methodologies rely on either comparing the mean sparsity of a sample of clusters or the individual sparsities within the sample to a theoretical prediction computed from the halo mass function, and are therefore susceptible to the biases presented above. 

At the time of writing, sparsity constraints remain reliant on predictions from N-body simulations \citep[e.g.][]{2016MNRAS.456.2486D,Raygal2022} as the high-mass regime of the halo mass function remains poorly constrained in the presence of baryonic physics. This is primarily due to the computational cost of running simulations of a sufficiently large volume to accurately capture the cut-off at high masses with only a few hydrodynamical cosmological simulations being able to reach box side lengths of the order of a few $h^{-1}$Gpc, e.g Magneticum\footnote{\href{http://www.magneticum.org/}{http://www.magneticum.org/}} \citep{Magneticum2016}, Millenium TNG\footnote{\href{https://www.mtng-project.org/}{https://www.mtng-project.org/}} \citep{MilleniumTNG} and FLAMINGO\footnote{\href{https://flamingo.strw.leidenuniv.nl/}{https://flamingo.strw.leidenuniv.nl/}} \citep{Flamingo2023}. 

In the following, we perform a number of analyses using the approach of \citet{Corasaniti2018} to assess the impact of HE sparsity bias and selection effects on the cosmological parameter inferences. Before presenting the 
methods along with the additional corrections to account for such effects, we first present the synthetic datasets we have used for these analyses.

\subsection{Synthetic Sparsity Data \& Calibration Samples}
\label{subsec:synth_data}

We generate two distinct synthetic datasets consisting of different cluster sparsity catalogues, which we use to assess the impact of systematic biases. 

The first dataset consists of clusters from the \textsc{gadget-x} simulation. To avoid the repetition of clusters in the sample, we first select 277 \textsc{gadget-x} clusters with redshifts randomly drawn from the set of eight redshift snapshots of the \textsc{gadget-x} simulation for which the X-ray and SZ masses are computed. We subsequently assign to each cluster the mass estimates corresponding to that redshift. In doing so, we generate three different catalogues: one consisting of cluster sparsities obtained using the hydro masses from the \textsc{gadget-x} simulation: two others are generated using HE mass estimates from \citet{2023MNRAS.518.4238G}. We refer to these catalogues as the The300-GX, The300-X and The300-SZ respectively. 

The second dataset consists of haloes from the \textsc{uchuu} N-body simulation \citep{Ishiyama2021} described in Sec.~\ref{sec:uchuu}. In particular, we first identify all haloes in the catalogues with masses $M_{\rm 200} > 10^{14}h^{-1}{\rm M}_\odot$, in all snapshots at $z \leq 1.22$. Then, we randomly draw 277 haloes with redshift from the \textsc{uchuu} snapshots that are the closest to those of the The300 in terms of redshift. In building such a dataset, we verify that no halo is selected multiple times by checking that the haloes are sufficiently far away. For example, the two closest haloes are $40$ h$^{-1}$ Mpc apart (not accounting for the redshift difference). Indeed, the purpose of the \textsc{uchuu} synthetic dataset is twofold. Firstly, it provides a dataset with a minimal amount of selection bias since, as shown in \citet{Corasaniti2018, Corasaniti2021, 2022arXiv221203233R}, a random selection of haloes with a mass cut-off does not induce a considerable bias to the estimation of the sparsity statistics. Secondly, it provides a dataset for which the effect of selection bias is well constrained and as such it allows us to verify that the bias model correction we add to the likelihoods performs correctly. 

Having generated these datasets, we split each of the catalogues into two separate subsamples by randomly selecting $100$ clusters to form a data subsample, which is used for the cosmological parameter inference; while the remaining $177$ clusters constitute a calibration subsample, which we further bootstrap to calibrate statistical models of the different bias effects. 

\subsection{Cosmological Model Predictions}\label{subsec:cosmo_predictions}
\begin{figure}[ht]
    \centering
    \includegraphics[width = 0.95\linewidth]{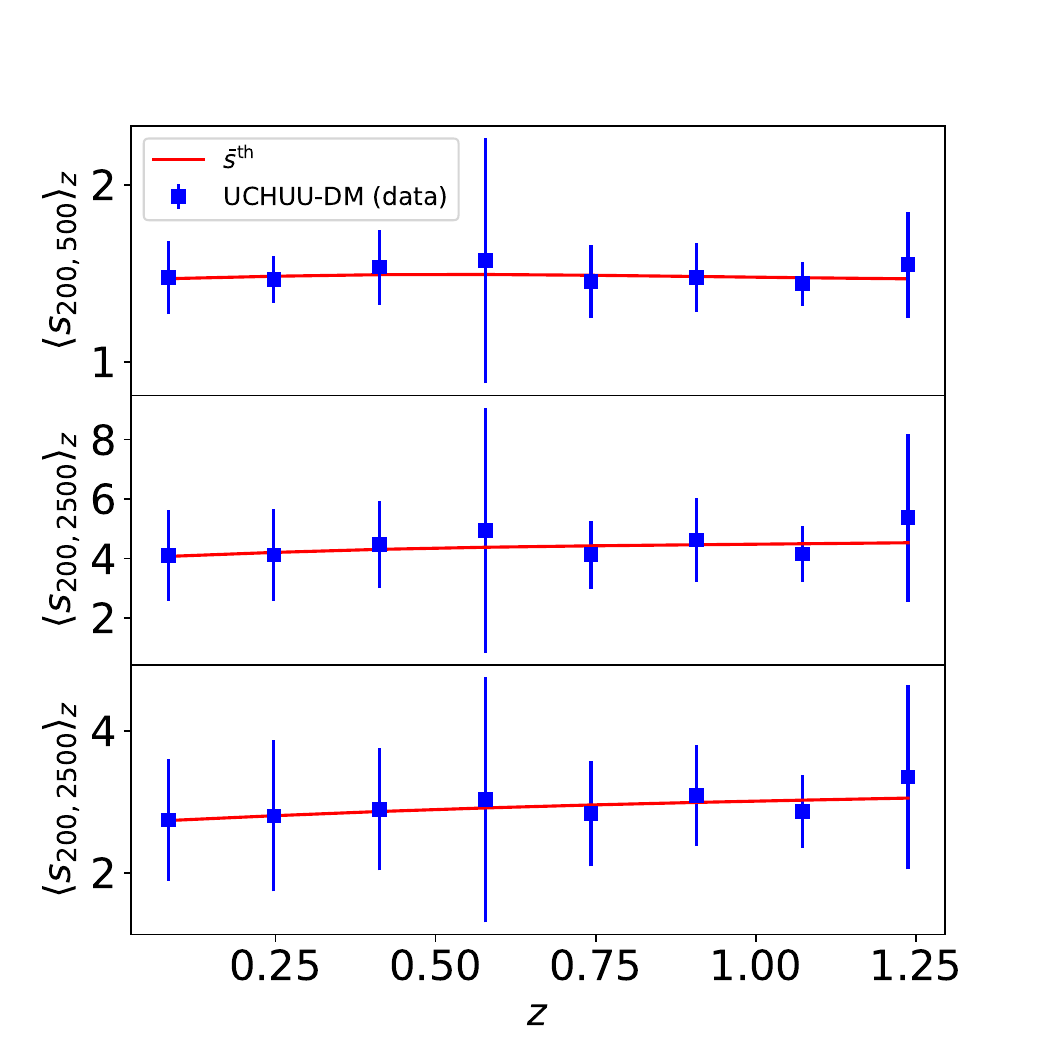}
    \caption{Mean sparsities predicted by Eq.~(\ref{eq:spars_hmf}) with the redshift dependent corrections (red solid lines) for the Uchuu fiducial cosmology against the estimates of the means from the \textsc{uchuu}-DM data sample in equally spaced redshift bins (blue squares).}
    \label{fig:predicted_sparsities}
\end{figure}

We predict the mean sparsity of the data for a given cosmological model using the integral relation \citep{Balmes2014,2022arXiv221203233R}:
\begin{equation}\label{eq:spars_hmf}
    \int^{M^{\rm max}_{\Delta_2}}_{M^{\rm min}_{\Delta_2}}\frac{\dd n}{\dd M_{\Delta_2}}\frac{\dd M_{\Delta_2}}{M_{\Delta_2}}=\langle s^{\rm HMF}_{\Delta_1,\Delta_2}\rangle
    \int^{\langle s_{\Delta_1,\Delta_2}\rangle M^{\rm max}_{\Delta_2}}_{\langle s_{\Delta_1,\Delta_2}\rangle M^{\rm min}_{\Delta_2}} \frac{\dd n}{\dd M_{\Delta_1}}\frac{\dd M_{\Delta_1}}{M_{\Delta_1}},
\end{equation}
which links the Halo Mass Function (HMF) at two distinct spherical overdensity contrasts, $\frac{\dd n}{\dd M_{\Delta_1}}$ and $\frac{\dd n}{\dd M_{\Delta_2}}$, to the mean sparsity of the sample, $\langle s_{\Delta_1, \Delta2}\rangle$. Given the functional form of the HMFs, this equation can be solved numerically with integration boundaries which reflect the data sample. In the case at hand, this means choosing an arbitrarily high upper bound, due to the exponential cut-off in the HMF, and a lower bound corresponding to the smallest cluster in the sample, $M_{\Delta_2}^{\rm min} \simeq 10^{14}h^{-1}{\rm M}_{\odot}$. For the HMF model, we choose to recalibrate the \cite{2016MNRAS.456.2486D} HMF on the large volume \textsc{uchuu} simulation \citep{Ishiyama2021}, see App.~\ref{hmf} for details. It is important to note that we also include a redshift dependent correction to the predicted sparsities to account for the HMF fitting inaccuracies. In particular, the correction to the mean sparsity takes the form of a multiplicative term,
\begin{align}
    \delta_{200,500}\; &= - 0.102 + 0.0038\cdot z,\label{corr_1}\\
    \delta_{200,2500} &= - 0.605 + 0.166\cdot z,\label{corr_2}\\
    \delta_{500,2500} &= - 0.414 + 0.075\cdot z,\label{corr_3}
\end{align}
such that $\bar{s}^{\rm th}_{\Delta_1, \Delta_2} = \langle s^{\rm HMF}_{\Delta_1,\Delta_2}\rangle\cdot (1 + \delta_{\Delta_1,\Delta_2})$, where $\langle s^{\rm HMF}_{\Delta_1,\Delta_2}\rangle$ is inferred from Eq.~(\ref{eq:spars_hmf}). These terms are required due to the rigidity of the \citet{2016MNRAS.456.2486D} fitting function at high masses which does not capture the high mass cut-off in the HMF with sufficient accuracy, which is particularly visible in Fig.~\ref{fig:hmf2500c}. These terms are obtained by fitting the residuals between the mean sparsity of the \textsc{uchuu}-DM calibration sample and that predicted by Eq.~(\ref{eq:spars_hmf}). For illustrative purposes, we plot in Fig.~\ref{fig:predicted_sparsities} the predicted $\bar{s}^{\rm th}_{\Delta_1, \Delta_2}$ for the Uchuu's fiducial cosmology against the \textsc{uchuu}-DM data.

\subsection{Bias Models}
\subsubsection{Selection \& Baryon Bias}
\label{subsec:selection}
As briefly mentioned, it has been shown that a random selection with a mass cut has a minimal impact on the cosmological parameter inferences using sparsity \citep{Corasaniti2018, Corasaniti2021, 2022arXiv221203233R}. While this type of selection is easily achievable when using data from a large volume cosmological simulation, this is generally not the case in most observational settings. For instance, the clusters from the Tier-1 CHEX-MATE observational program \citep{2021A&A...650A.104C}, while offering a random selection of clusters with masses in the range $2 < M_{\rm 500}\,[10^{14} M_\odot] <
9$ and redshift $0.05 < z < 0.2$, are still reliant on the Planck SZ selection function \citep{PlanckSZ2016}. Similarly, the Tier-2 CHEX-MATE observational program, which has observed the most massive systems with $M_{\rm 500} > 7.25 \cdot10^{14} M_\odot$, at low redshift, $z < 0.6$, is not only reliant on the Planck SZ selection function but also introduces its own selection.

The use of The300 sample poses a similar problem, since the selection of a subset of the $\sim$300 most massive clusters in the MDPL2 simulation at a random point in their mass assembly history is highly non-trivial. As already mentioned, to quantify the magnitude of this effect, we have generated a synthetic dataset to replicate the selection process described above using the halo catalogues of the {\sc uchuu} simulation \citep{Ishiyama2021}. This simulation has a volume eight times larger than the {\sc mdpl2} simulation, thus we can replicate the selection eight times without overlapping the sub-volumes. From each of the {\sc uchuu} sub-volumes, we select the 277 most massive objects to which, we randomly assign a redshift of observation, and recover the state of the halo at that redshift using the \textsc{rockstar} merger trees. Again, the redshifts are drawn from the set of the {\sc uchuu} snapshots whose redshifts are closest to those of The300. Finally, as already stated in Sec.~\ref{subsec:synth_data}, we randomly assign 100 haloes to a data sample and place the remaining haloes in a calibration sample. Since we aim to compare this selection to a purely random selection, we repeat this process by randomly selecting haloes with a mass $M_{200} > 10^{14}h^{-1}{\rm M}_{\odot}$ within each sub-volume. We bin the resulting datasets into eight redshift bins corresponding to the eight snapshots from which the cluster properties are drawn and compute the mean sparsities in each bin.

\begin{figure}[t]
    \centering
    \includegraphics[width = 0.95\linewidth]{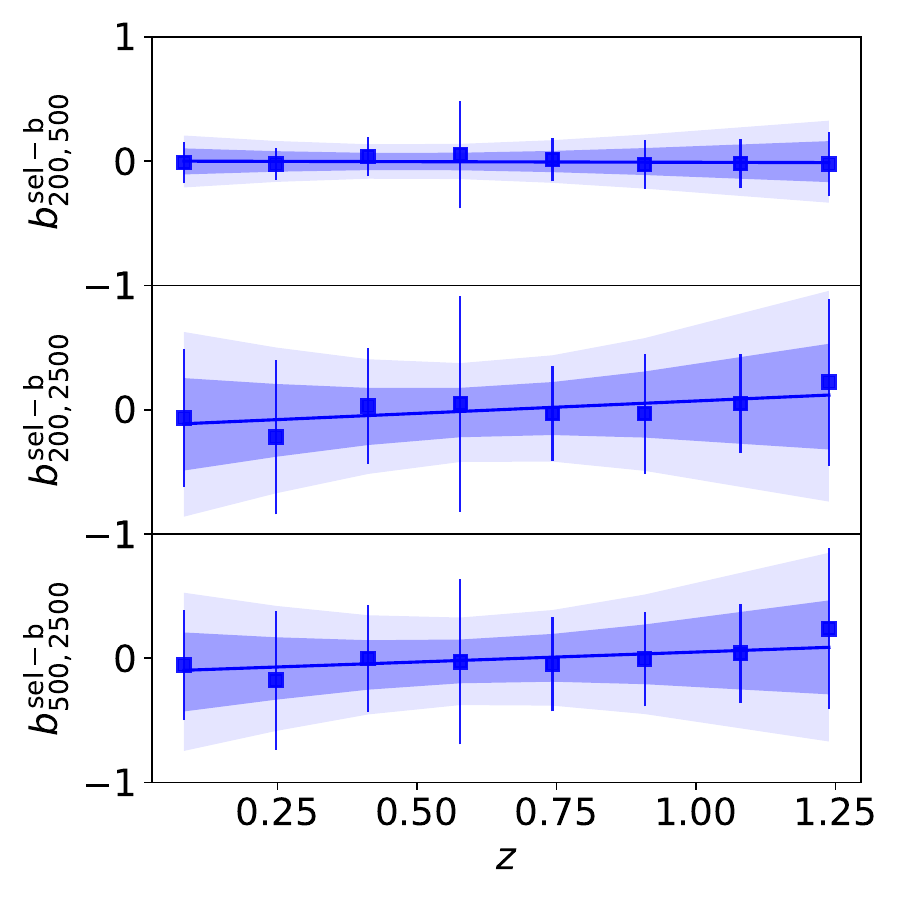}
    \caption{Estimates of the mean and standard deviation of selection-baryon bias for different sparsity configurations at different redshift bins (blue squares). The solid blue lines correspond to the average linear regression fits, while the shaded areas represents $1$ and $2\sigma$ credible regions.}
    \label{fig:selection_bias_ratio}
\end{figure}

\begin{table}[]
    \centering
    \caption{Average and $1\sigma$ error on the linear regression coefficients of the selection-baryon bias model.}
    \begin{tabular}{c|ccc}
        \hline
         & $b^{\rm sel-b}_{200,500}$ & $b^{\rm sel-b}_{200,2500}$ & $b^{\rm sel}_{500,2500}$\\
        \hline
        $b^{\rm sel-b}_0$ & $0.00 \pm 0.12$ & $-0.13 \pm 0.42$ & $-0.11 \pm 0.37$\\
        $b^{\rm sel-b}_1$ & $-0.01\pm 0.21$ & $0.20\pm 0.60$ & $0.16\pm 0.54$\\
        \hline
    \end{tabular}
    \label{tab:bias_sel_b}
\end{table}

We compare these estimates to the mean sparsities from the The300-GX sample at the same redshifts. Notice that as we are comparing results from a DM-only simulation against those from hydrodynamical runs, the comparison provides us with an estimate of bias due to selection as well as the effect of baryons on the cluster sparsities discussed in Section~\ref{baryonic_effects}. Specifically, we compute the selection-baryon bias: 
\begin{equation}
    b^{\rm sel-b}_{\Delta_1, \Delta_2} = 1 - \frac{\langle s_{\Delta_1, \Delta_2}^{\rm The300-GX} \rangle_{z_i}}{\langle s_{\Delta_1, \Delta_2}^{\rm UCHUU-DM} \rangle_{z_i}}.
\end{equation}
The mean and standard deviation for the different sparsity configurations at different redshifts are shown in Fig.~\ref{fig:selection_bias_ratio}.  
In all the cases the data points are consistent with a constant trend in redshift, though a linear model,
\begin{equation}\label{eq:sel-b}
b^{\rm sel-b}_{\Delta_1,\Delta_2} (z) = b^{\rm sel-b}_0 + b^{\rm sel-b}_1\cdot z, 
\end{equation}
provides a better fit to the bias estimates, especially in the case of $b^{\rm sel-b}_{200,2500}$ and $b^{\rm sel-b}_{500,2500}$. Henceforth, we infer the average and $1\sigma$ uncertainty of the fitting coefficients $b^{\rm sel-b}_0$ and $b^{\rm sel-b}_1$ through a Bayesian regression analysis for each sparsity configuration. The corresponding values are quoted in Table~\ref{tab:bias_sel_b}. Indeed, we may notice that slopes are consistent with zero within the $1\sigma$ errors. Nevertheless, to be as conservative as possible, we model the selection-baryon bias in terms of Eq.~(\ref{eq:sel-b}).

\subsubsection{HE Sparsity Bias}
\begin{figure}[ht]
    \centering
    \includegraphics[width = 0.95\linewidth]{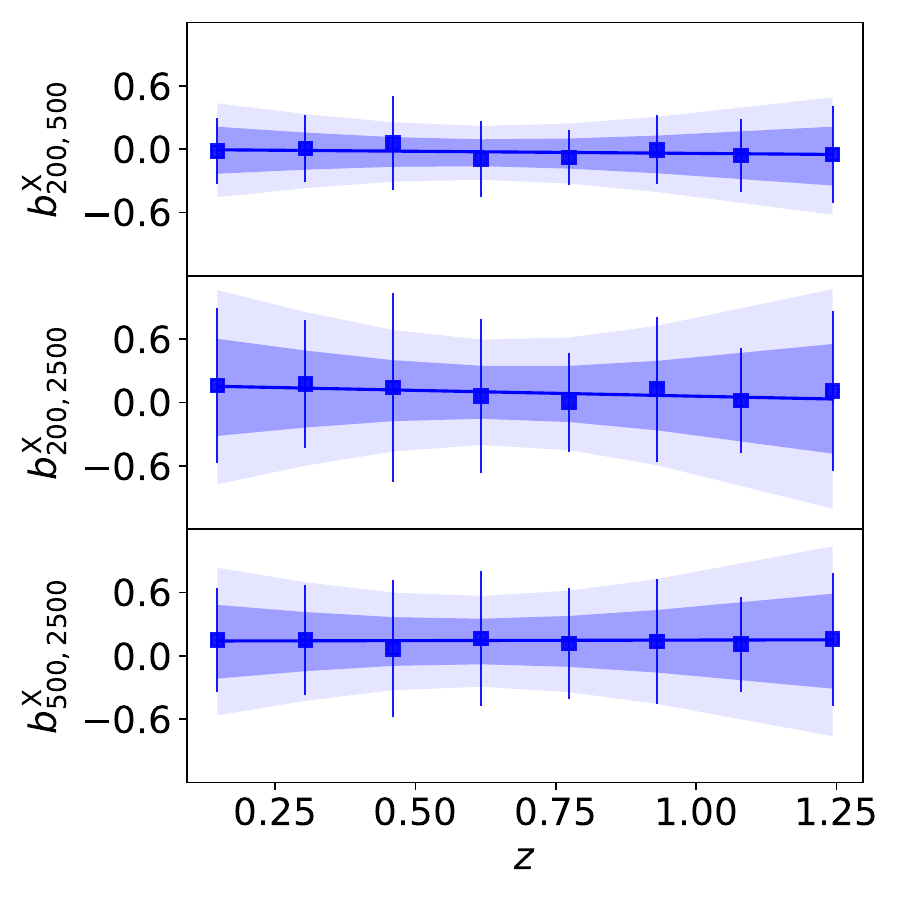}
    \caption{Estimates of the mean and standard deviation of the HE sparsity bias from The300-X calibration sample for the different sparsity configurations in different redshift bins (blue squares). The solid lines correspond to the average linear regression fits, while the shaded areas represents $1$ and $2\sigma$ credible regions.}
    \label{fig:he_x_bias_ratio}
    \includegraphics[width = 0.95\linewidth]{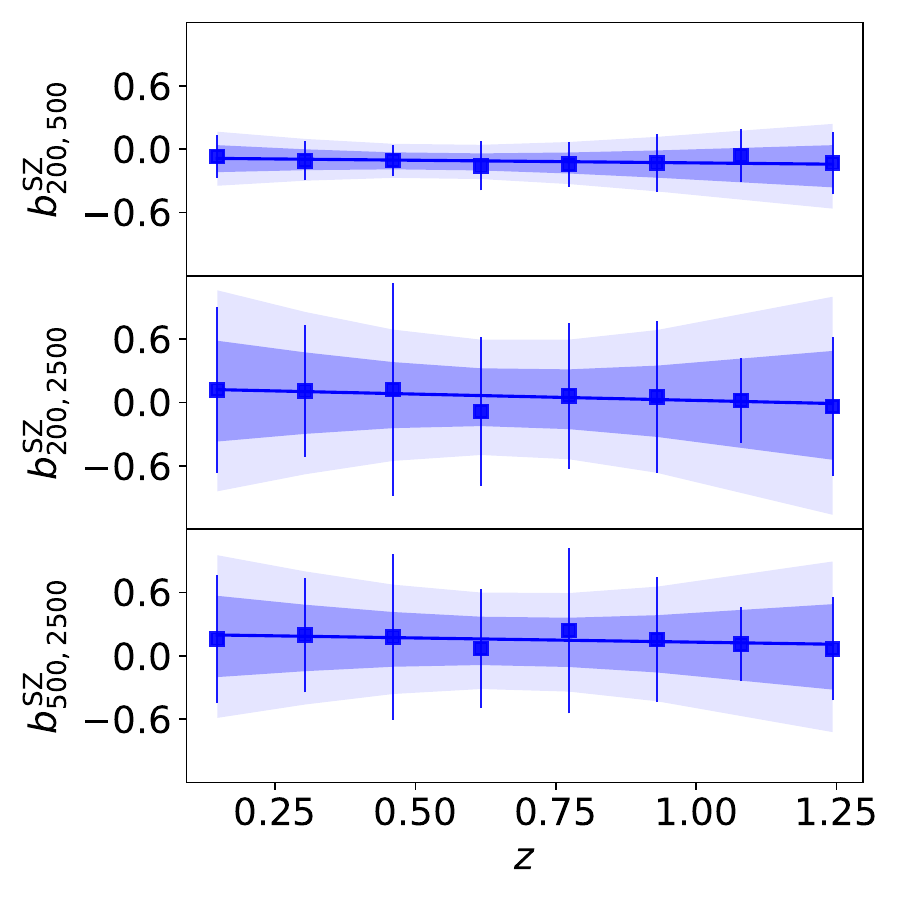}
    \caption{As in Fig.~\ref{fig:he_x_bias_ratio} from The300-SZ calibration sample.}
    \label{fig:he_sz_bias_ratio}
\end{figure}

We use the calibration samples from The300-X and The300-SZ datasets to estimate the HE bias by comparing the mean sparsities in different redshift bins against those estimated using The300-GX calibration sample. Namely, we compute
\begin{equation}
    b^{\rm HE}_{\Delta_1, \Delta_2} = 1 - \frac{\langle s_{\Delta_1, \Delta_2}^{\rm The300-HE} \rangle_{z_i}}{\langle s_{\Delta_1, \Delta_2}^{\rm The300-GX} \rangle_{z_i}},
\end{equation}
and plot the average and standard deviation for the different sparsity configurations at different redshift in Fig.~\ref{fig:he_x_bias_ratio} for the X-ray inferred masses and Fig.~\ref{fig:he_sz_bias_ratio} for the SZ ones. We can see that the estimates are consistent with a constant redshift bias, though a linear redshift dependence can provide a better fit. Again, as in the case of the selection-baryon bias, to be as conservative as possible, we model these trends with a linear regression model:
\begin{equation}
b^{\rm HE}_{\Delta_1,\Delta_2} (z) = b^{\rm HE}_0 + b^{\rm HE}_1\cdot z.
\end{equation}
In Table~\ref{tab:bias_he}, we quote the average and $1\sigma$ uncertainty on the fitting coefficients $b^{\rm HE}_0$ and $b^{\rm HE}_1$ from the Bayesian inference analysis for the The300-X and The300-SZ calibration sample respectively.

\begin{table}[]
    \centering
    \caption{Average and $1\sigma$ error on the linear regression coefficients of the HE bias model.}
    \begin{tabular}{c|ccc}
        \hline
         & $b^{\rm HE}_{200,500}$ & $b^{\rm HE}_{200,2500}$ & $b^{\rm HE}_{500,2500}$\\
        \hline
        $b^{\rm X}_0$ & $0.00 \pm 0.28$ & $0.17 \pm 0.58$ & $0.14 \pm 0.43$\\
        $b^{\rm X}_1$ & $-0.04\pm 0.40$ & $-0.11\pm 0.79$ & $0.01\pm 0.63$\\
        \hline
        $b^{\rm SZ}_0$ & $-0.08\pm 0.16$ & $0.14 \pm 0.58$ & $0.21 \pm 0.46$\\
        $b^{\rm SZ}_1$ & $-0.05\pm 0.28$ & $-0.12\pm 0.77$ & $-0.08\pm 0.61$\\  
        \hline
    \end{tabular}
    \label{tab:bias_he}
\end{table}

\begin{figure*}[ht]
    \centering
    \includegraphics[width = 0.9\linewidth]{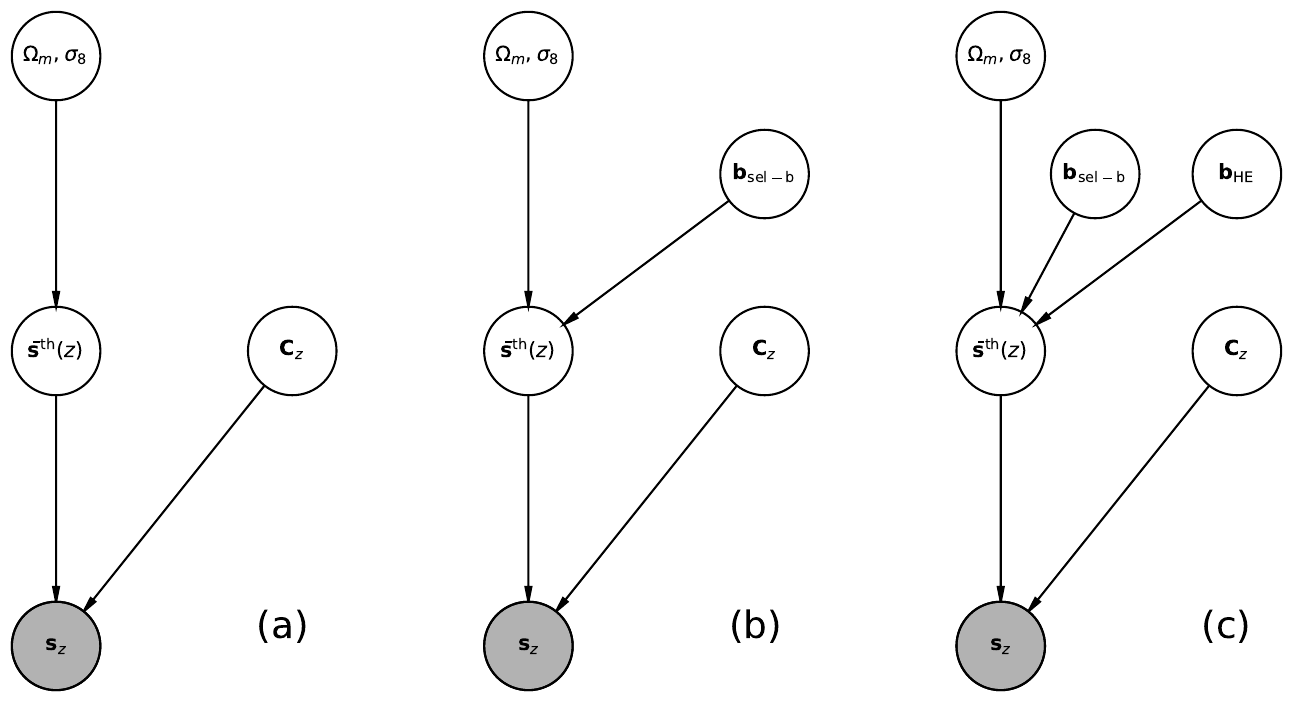}
    \caption{Bayesian graph models representing the relation between the cosmological parameters ($\Omega_m$, $S_8$), the mean sparsity model prediction at a given redshift ($\bar{\bf s}^{\rm th}$), the covariance among multiple sparsities at a given redshift (${\rm C}_z$, which reduces to the sparsity variance in the case of a single sparsity analysis) and the mean sparsity data (${\bf s}_z$). Panel (a) show the simplest data model with no bias included with the posterior sample as given by Eq.~(\ref{eq:posterior_basic}), while the data model shown in panel (b) accounts for the selection-baryon bias as in Eq.~(\ref{eq:posterior_selb}), and that shown in panel (c) extend the model to the HE bias with the posterior given by Eq.~(\ref{eq:posterior_he}).}
    \label{fig:bayes_graf}
\end{figure*}

\subsection{Cosmological Parameter Inference Analyses}
We are now able to investigate the impact of the different bias effects on the cosmological parameter constraints inferred from the analysis of mean sparsity measurements at different redshifts. Following the approach presented in \citet{Corasaniti2018}, we infer constraints using mean sparsity measurements from a single overdensity configuration at different redshifts, i.e. $\langle s_{200,500}\rangle$, as well as from the combination of multiple sparsity configurations, i.e. $\{\langle s_{200,500}\rangle,\langle s_{200,2500}\rangle,\langle s_{500,2500}\rangle\}$, as described in and \citet{2022MNRAS.516..437C}. To this purpose we use the synthetic data samples generated in Sec.~\ref{subsec:synth_data}. We summarise the results in terms of the constraints on the parameter $S_8=\sigma_8\sqrt{\Omega_m/0.3}$, which we infer by performing a Markov Monte Carlo Chain sampling of the posterior distribution assuming uniform priors on $\Omega_{\rm m} \in [0.2,0.6]$, and $\sigma_8 \in [0.2,1.2]$. We have set the remaining cosmological parameters to the fiducial values of the Planck cosmology. 

Next, we discuss the propagation of the bias model uncertainties and the priors assumed on the bias model parameters.

\subsection{Bayesian Graph Models}
A schematic summary of the computation of posterior distribution adopted under different bias model assumptions is given by the Bayesian graphs shown in Fig.~\ref{fig:bayes_graf}. 

Panel (a) corresponds to the simplest case, where we neglect any bias effect. Hence, the posterior reads simply as:
\begin{equation}\label{eq:posterior_basic}
P(\Omega_m,\sigma_8|{\bf s}_z)=\mathcal{L}({\bf s}_z|{\rm \bar{\bf s}}^{\rm th}(z|\Omega_m,\sigma_8))P(\Omega_m)P(\sigma_8),
\end{equation}
where $\mathcal{L}$ is the likelihood function with ${\bf s}_z$ being the data vector of mean sparsity measurements at different redshifts and ${\bf \bar{s}^{\rm th}}$ the vector of mean sparsity predictions, while $P(\Omega_m)$ and $P(\sigma_8)$ are the prior distributions of the sampled parameters. In the case of a multiple sparsity dataset, ${\bf s}_z=\{{\bf s}_{200,500},{\bf s}_{200,2500},{\bf s}_{500,200}\}$, while for a single sparsity this reduced to ${\bf s}_z=\{{\bf s}_{200,500}\}$. Assuming a Gaussian likelihood function, the log-likelihood reads as:
\begin{equation}
-2\log{\mathcal{L}} \propto \sum_{i=1}^{N}{\bf y}^{\top}_i {\bf\sf C}_{i}^{-1}{\bf y}_i,
\end{equation}
where $N$ is the number of redshift bins, and 
\begin{equation}
{\bf y}_{z_i} = {\bf s}_{z_i} - {\bf \bar{s}}^{\rm th}(z_i). 
\end{equation}
The matrix ${\bf C}_z$ is accounts for the covariance between the different sparsities at a given redshift. In the case of a single sparsity statistics, this reduces to the intrinsic variance of the estimated sparsity, i.e. $\sigma^2_{s_{z_i}}$. In the case of multiple sparsities, we compute the covariance matrix using the correlation matrix between different sparsity configurations, $r^{z}_{s_{\alpha},s_{\beta}}$, as given by the analytical formulae derived in \citet{2022MNRAS.516..437C} using the M2Csims simulations (see Appendix~\ref{appendix:corr}), such that $C^z_{s_{\alpha},s_{\beta}}=\sigma^z_{s_{\alpha}}\sigma^z_{s_{\beta}}r^z_{s_{\alpha},s_{\beta}}$.

Panel (b) summarises the sampling of the posterior marginalised over the selection-baryon bias model: 
\begin{eqnarray}\label{eq:posterior_selb}
P(\Omega_m,\sigma_8|{\bf s}_z) &= &\int d{\bf b}_{\rm sel-b}\mathcal{L}({\bf s}_z|{\rm \bar{s}}^{\rm th}(z|\Omega_m,\sigma_8),{\bf b}_{\rm sel-b})\nonumber \\
&\times &P({\bf b}_{\rm sel-b})P(\Omega_m)P(\sigma_8),
\end{eqnarray}
where $P({\bf b}_{\rm sel-b})$ is the prior distribution of the selection-baryon bias model parameters. In such a case, we assume the linear redshift dependent bias model Eq.~(\ref{eq:sel-b}), such that 
\begin{equation}
{\bf \bar{s}}^{\rm th}(z)\rightarrow \left[1-{\bf b}_{\rm sel-b}(z)\right]\cdot {\bf \bar{s}}^{\rm th}(z).
\end{equation}

In sampling the posterior distribution, Eq.~(\ref{eq:posterior_selb}), we assume Gaussian priors for the selection-baryon bias parameters $b_{0}^{\rm sel-b}$ and $b_{1}^{\rm sel-b}$ of each sparsity configuration, with mean and standard deviation specified by the values quoted in Table~\ref{tab:bias_sel_b}.

Finally, panel (c) shows the Bayesian graph for the sampling of the posterior in the presence of the selection-baryon bias and the HE bias:
\begin{eqnarray}\label{eq:posterior_he}
P(\Omega_m,\sigma_8|{\bf s}_z) &= &\int d{\bf b}_{\rm HE} d{\bf b}_{\rm sel-b}\mathcal{L}({\bf s}_z|{\rm \bar{s}}^{\rm th}(z|\Omega_m,\sigma_8),{\bf b}_{\rm sel-b},{\bf b}_{\rm HE})\nonumber \\
&\times &P({\bf b}_{\rm HE})P({\bf b}_{\rm sel-b})P(\Omega_m)P(\sigma_8),
\end{eqnarray}
where $P({\bf b}_{\rm HE})$ is the prior distribution of the HE bias model parameters. We assume the linear HE bias model given by Eq.~(\ref{eq:bias_he}), hence to leading order in the bias parameters we have 
\begin{equation}
{\bf \bar{s}}^{\rm th}(z)\rightarrow \left[1-{\bf b}_{\rm HE}(z)-{\bf b}_{\rm sel-b}(z)\right]\cdot {\bf \bar{s}}^{\rm th}(z).
\end{equation}

In sampling the posterior distribution, Eq.~(\ref{eq:posterior_he}), in addition to the Gaussian priors on the selection bias parameters, we also assume Gaussian priors for the HE bias parameters of each sparsity configuration with mean and standard deviation specified by the values given in Table~\ref{tab:bias_he}.

\subsection{Results}
\begin{figure}[ht]
    \centering
    \includegraphics[width = 0.9\linewidth]{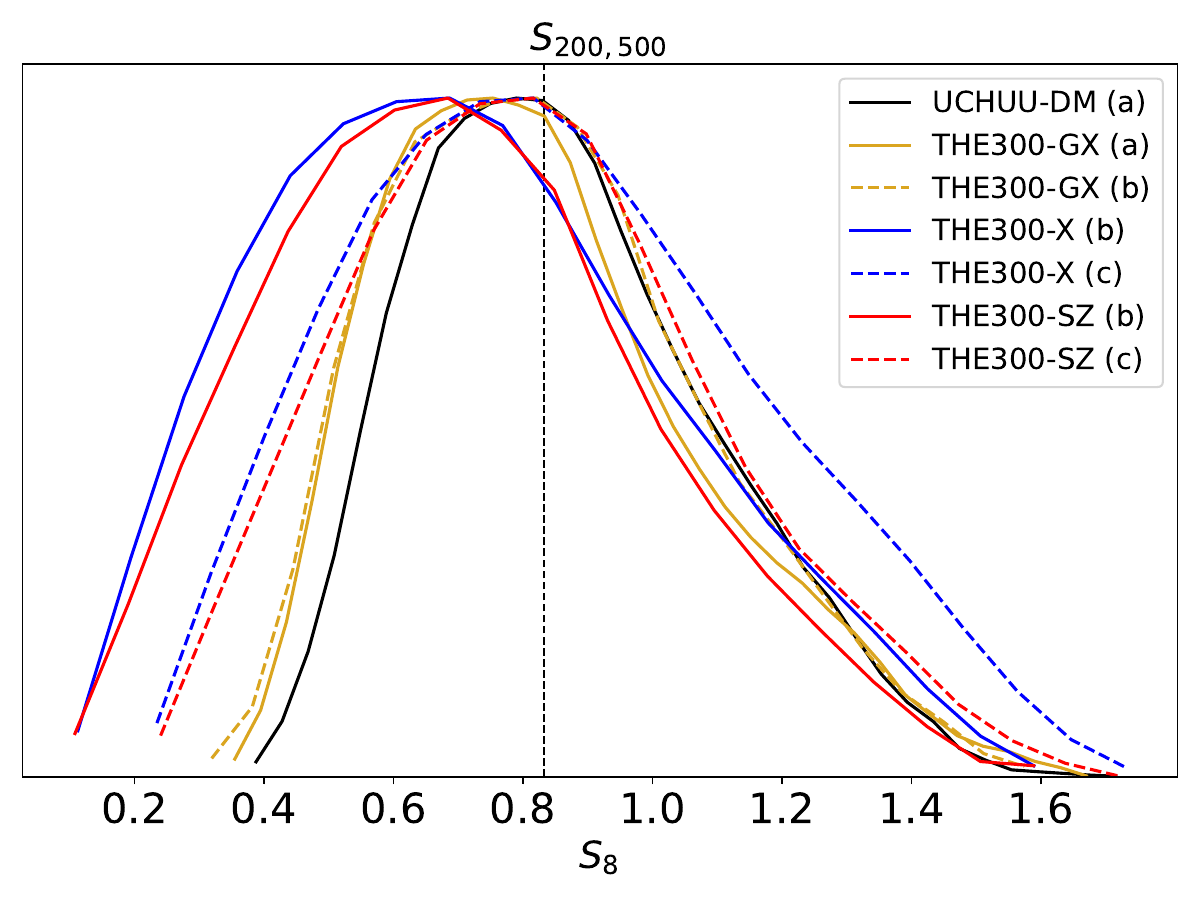}
    \caption{Marginalised posterior on $S_8$ from the analysis of single sparsity measurements for the \textsc{uchuu}-DM (black curve), The300-GX (yellow curves), The300-X (blue curves) and The300-SZ (red curves). The solid lines represent posterior inferred with the simplest data model, while the dashed lines corresponds to posteriors obtained marginalising over bias model parameters.}
    \label{fig:1S}
\end{figure}

\begin{figure}[ht]
    \centering
    \includegraphics[width = 0.9\linewidth]{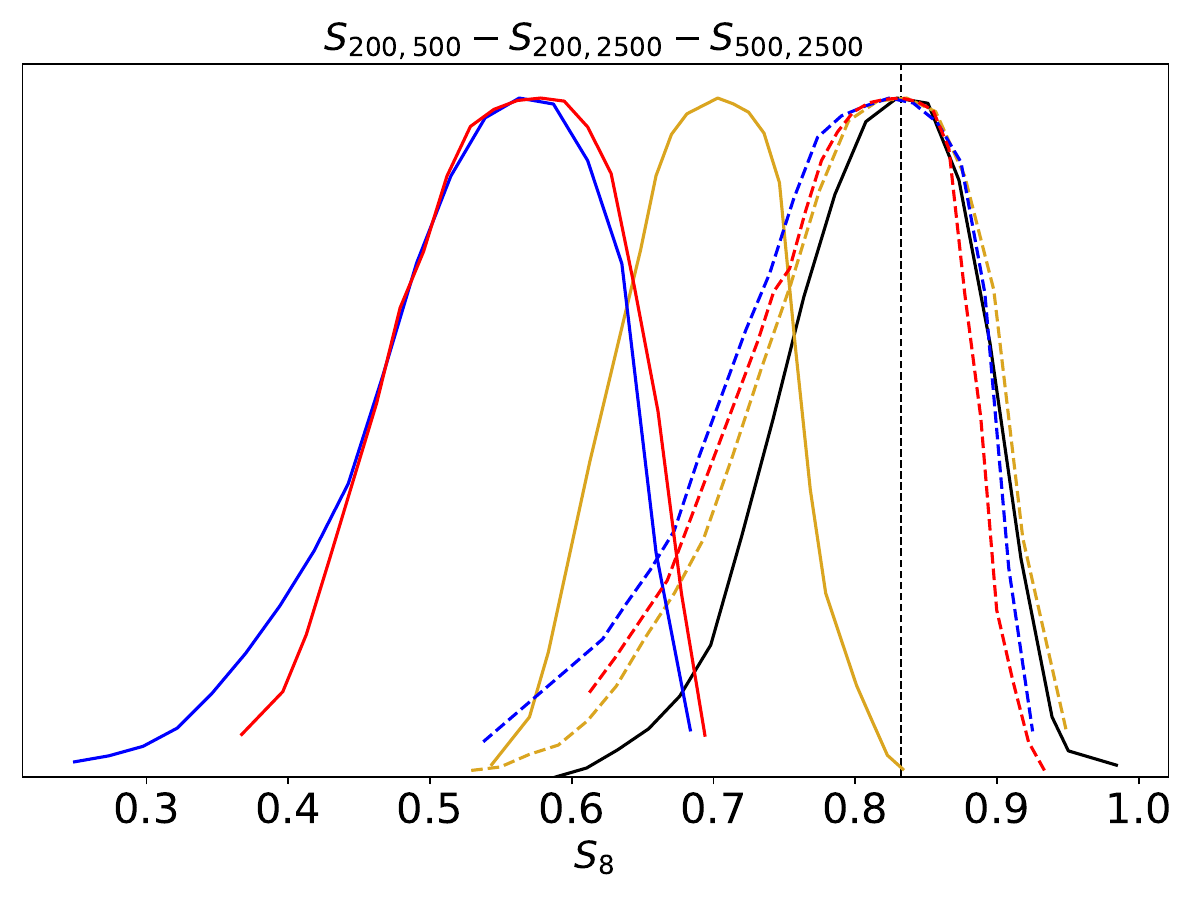}
    \caption{As in Fig.~\ref{fig:1S} in the case of multiple sparsity measurements.}
    \label{fig:3S}
\end{figure}

Here, we present the results of the MCMC likelihood analysis of the \textsc{uchuu}-DM, The300-GX, The300-X and The300-SZ synthetic data samples in the case of $s_{200,500}$ (1S) and multiple (3S) average sparsity measurements. We summarise the results in terms of the constraints on the parameter $S_8$ and remind the reader that the synthetic datasets have been generated with a fiducial value $S_8^{\rm f}=0.83$. The marginalised constraints are quoted in Table~\ref{tab:marge_S8}, where we report the mean and $1\sigma$ uncertainty, as well as the best-fit values of $S_8$ as inferred under the different data models described in the previous section. The marginalised posteriors are shown in Fig.~\ref{fig:1S} and Fig.~\ref{fig:3S} respectively.

The analysis of the \textsc{uchuu}-DM data sample provides us with a reference benchmark, since the cosmological model predictions obtained from Eq.~(\ref{eq:spars_hmf}) have been calibrated using the \textsc{uchuu} halo catalogues. In Fig.~\ref{fig:1S} and~\ref{fig:3S}, we plot the corresponding posterior distributions (solid black lines) inferred using the Bayesian model specified by Eq.~(\ref{eq:posterior_basic}) in the 1S and 3S case respectively. We can see that the peak of the distributions lies close to the fiducial value. Moreover, we may notice from the values quoted in Table~\ref{tab:marge_S8} that $S_8^{\rm f}=0.83$ lies well within the $1\sigma$ estimated errors. Quite importantly, we can see the constraint from the 3S analysis is nearly a factor of 2 more stringent than the 1S, consistently with the results of \citet{2022MNRAS.516..437C}.

The analysis of The300-GX data sample provides us with an estimate of the impact of the selection-baryon bias on the cosmological parameter inference. In particular, we can see from Fig.~\ref{fig:1S} that in the 1S case, the posterior distribution (solid yellow line) obtained using the basic data model Eq.~(\ref{eq:posterior_basic}) is only slightly shifted with respect to the benchmark posterior. This is consistent with the fact that the level of selection-baryon bias on $s_{200,500}$ is of the order of a few percent with respect to the DM-only case. We find $S^{\rm f}_8$ to be within the $1\sigma$ credible interval. This is not the case of the 3S analysis, for which where the effect of neglecting the selection-baryon bias is significantly larger, with $S^{\rm f}_8$ excluded at more than $2\sigma$. Instead, marginalising over the bias model, as specified by Eq.~(\ref{eq:posterior_selb}), results in unbiased constraints both in the 1S and 3S case. Though the inferred bounds on $S_8$ are weaker than those inferred in the benchmark case, a direct consequence of the marginalisation on the bias model parameters.

In Fig.~\ref{fig:1S} and Fig.~\ref{fig:3S}, we plot the marginalised posterior from the 1S and 3S analysis of The300-X (blue lines) and The300-SZ (red lines) data samples. We first perform a likelihood analysis only accounting for the selection-baryon bias model as in Eq.~(\ref{eq:posterior_selb}). We find posterior distributions that are systematically biased toward low $S_8$ values (solid curves). The effect is significantly more important for the 3S analysis than the 1S, as it can also be noticed from the values quoted in Table~\ref{tab:marge_S8}. In particular, in the 1S case the fiducial value remains within the $1\sigma$ errors, while it is excluded at more than $2\sigma$ for the 3S case. In contrast, accounting for the HE bias, as specified by Eq.~(\ref{eq:posterior_he}), results in unbiased constraints. We can see that the marginalised posterior distributions (dashed lines) peak close to the fiducial value and overlap with the benchmark distribution. We stress the fact that the use of multiple average sparsity measurements always lead to tighter bounds on $S_8$. After marginalising over the bias parameters, the joint analysis of $s_{200,500}$, $s_{200,2500}$ and $s_{500,2500}$ removes projection effects that limits the bound inferred using $s_{200,500}$ only. In particular, the joint analysis results in a constraint on $S_8$ that is a factor $\sim 2.6$ tighter in the case of the X-ray data sample and $\sim 3.6$ for the SZ one.


\begin{table*}
    \centering
    \caption{Marginalised constraints on $S_8$ from the \textsc{uchuu}-DM, The300-GX, The300-X and The300-SZ synthetic data samples with fiducial value $S^{\rm f}_8=0.83$. The different rows correspond to the different likelihood models considered, while the columns corresponds to the mean, $1\sigma$ error and best-fit value from the single (1S) and multiple (3S) sparsity analysis.} 
    \begin{tabular}{c|c|cc|cc}
        \hline
 Data & Bayes Graph & \multicolumn{2}{c}{\textsc{1S}} 
 & \multicolumn{2}{c}{\textsc{3S}} \\
        \hline
         \textsc{uchuu}-DM & (a) & $0.81\pm 0.27$ & $0.80$ & $0.80\pm 0.11$ & $0.82$ \\
         \hline
         The300-GX & (a) & $0.79\pm 0.27$ & $0.74$ & $0.70\pm 0.04$ & $0.70$\\
         The300-GX & (b) & $0.82\pm 0.30$& $0.80$ & $0.80\pm 0.13$ & $0.82$\\
         \hline
         The300-X & (b) & $0.71\pm 0.38$ & $0.68$ & $0.56\pm 0.10$ & $0.55$ \\
         The300-X & (c) & $0.84\pm 0.40$ & $0.81$ & $0.80\pm 0.15$ & $0.82$ \\
         \hline
         The300-SZ & (b) & $0.72\pm 0.35$ & $0.68$ & $0.56\pm 0.11$ & $0.57$ \\         
         The300-SZ & (c)& $0.81\pm 0.36$& $0.81$ & $0.80\pm 0.10$ & $0.82$ \\
         \hline
    \end{tabular}
    \label{tab:marge_S8}
\end{table*}

\section{Conclusions}\label{sec:conclu}

Halo sparsity is a simple, non-parametric proxy of the distribution of mass within haloes hosting galaxy groups and clusters \citep{Balmes2014}. Being a ratio of masses estimated at radii which enclose different overdensities, it provides an estimate of the local logarithmic slope of the halo mass profile that encodes cosmological information resulting from the halo mass assembly history. For a given cosmological setup, the statistical properties of the halo sparsity can be estimated using cosmological N-body simulations. 
These can in turn be used to calibrate analytical relations to predict the average sparsity at different redshifts and for different overdensity pairs for a given cosmological model \citep{Corasaniti2018,2019MNRAS.487.4382C,2022MNRAS.516..437C}. This has opened up the possibility to perform cosmological parameter inference analyses using sparsity measurements in clusters. However, with the increasing precision of cluster mass estimates, it has become critical to assess potential sources of systematic error. In particular, given that the prediction relies on DM-only simulations, it is important to assess the impact of baryons on the predicted sparsities and more generally the effects of selection and mass estimation biases. Here, we have investigated these potential sources of systematic uncertainty using the simulated clusters from the The300 zoom-in simulation suite \citep{2018MNRAS.480.2898C,2022MNRAS.514..977C}. 
Furthermore, we use hydrostatic equilibrium masses estimated from 3D ICM thermodynamic profiles of The300 clusters by \citet{2023MNRAS.518.4238G} to assess the effect of HE mass bias.

First, we find that the presence of baryons in simulated clusters alters the sparsity relative to DM-only predictions differently at different overdensities. In particular, the further two radii are from one another, the larger the differential change will be with respect to the DM-only sparsities. This is consistent with expectations from the baryon induced radial dependent mass bias found in simulated clusters \citep[e.g.][]{2016ApJ...827..112B,2017MNRAS.466.4442L}. Moreover, we find that the amplitude of this bias is directly correlated with the strength of the baryonic feedback implemented in the simulations, with stronger feedback models leading to a larger sparsity bias. Nevertheless, on average the effect remains minimal (at a few percent level) for sparsities probing the external cluster regions, such as $s_{200,500}$, while it is maximal at large overdensities, such as $s_{200,2500}$. Quite importantly, the scatter around the average bias strongly depends on the dynamical state of the simulated clusters.

Given the sensitivity of certain sparsity configurations to the astrophysical processes that shape the properties of the baryonic gas in clusters, we investigate the relation between sparsities and the gas fraction. We find a moderate anti-correlation ($\sim 40\%-60\%$) between the gas fraction within the inner cluster regions and the sparsities $s_{200,500}$ and $s_{500,2500}$ respectively. In particular, we find that the lower the value of the gas fraction inside the shell comprised between $R_{2500}$ and $R_{500}$, and the higher the value of the sparsity $s_{500,2500}$. This is a direct consequence of the effect of AGN on the gas and total mass distribution in clusters. On the other hand, we find that gas fraction measured within $R_{500}$ weakly correlate with the sparsity $s_{200,500}$ and $s_{500,2500}$, thus carrying complementary information that can be exploited for joint cosmological analyses. Furthermore, we find that ratios of the gas fraction estimated at different overdensities provide a bias tracer of the cluster sparsity, which is particularly sensitive to the underlying astrophysical feedback model implemented in the simulations.

Using the HE mass estimates of The300 cluster sample from \citet{2023MNRAS.518.4238G}, we study the impact of the HE mass bias on sparsity measurements. In particular, we find a median bias which ranges from a few to a maximum of $10\%$ level depending on the sparsity configuration. 
Moreover, we find that HE mass inferred from X-ray data leads to a larger scatter of the sparsity bias than SZ ones.

Finally, we generate synthetic datasets to estimate the joint effect of selection, baryon and HE bias on the cosmological parameter constraints inferred from single and multiple average sparsity measurements. By splitting the datasets into calibration and data samples, we use the former to calibrate linear redshift dependent models of the various sources of bias; whereas we use the data samples, consisting of sparsity measurements of $100$ clusters in the redshift range $0<z<1.2$, to perform a MCMC likelihood analysis and derive constraints on $S_8$ under different data model assumptions. We find that the effect of the sparsity biases is to systematically shift the posteriors toward lower $S_8$ values. In the case of single sparsity measurements ($s_{200,500}$), the effect is small and well within the statistical uncertainties. In contrast, the joint analysis of multiple sparsity measurements ($s_{200,500}$, $s_{200,2500}$ and $s_{500,2500}$) result in a larger systematic effect with the fiducial $S_8$ value excluded at more than $2\sigma$. We show that calibrated linear redshift dependent bias models can be used to account for such sparsity biases and marginalising over the sparsity bias model parameters is sufficient to recover the fiducial cosmology and infer unbiased constraints on $S_8$. The evaluation of the HE and selection mass biases presented in this study can be further improved through the analysis of a randomly selected cosmological samples of simulated clusters with 2D de-projected HE mass estimates.

In the case of application to observational data, one could calibrate the sparsity bias model parameters and propagate them through the cosmological parameter inference using the same methodology we have presented in this work. For instance, subsamples of data for which HE and lensing masses are measured at different overdensities could be used to calibrate the HE bias model parameters of different sparsity configurations, allowing to also capture projection effects. Similarly, one could estimate selection effects as in standard data analyses by comparing to random mocks.

\begin{acknowledgements}
This work has been made possible by The Three Hundred collaboration\footnote{\url{https://www.the300-project.org}}.
The authors acknowledge {\it The Red Espa\~nola de Supercomputaci\'on } for granting computing time for running most of the hydrodynamical simulations of {\sc The Three Hundred} galaxy cluster project in the Marenostrum supercomputer at the Barcelona Supercomputing Center. P.S.C., A.M.C.L.B. and Y.R. acknowledge support from CNES (Centre National d'Etudes Spatiales). T.R. acknowledge funding from the Spanish Goverment's grant program "Proyectos de Generaci\'on de Conocimiento" under grant  number PID2021-128338NB-I00. M.D.P. acknowledges financial support from PRIN-MIUR grant 20228B938N "Mass and selection biases of galaxy clusters: a multi-probe approach" funded by the European Union Next generation EU, Mission 4 Component 1  CUP C53D2300092 0006. S.E. acknowledges the financial contribution from the contracts Prin-MUR 2022 supported by Next Generation EU (M4.C2.1.1, n.20227RNLY3 {\it The concordance cosmological model: stress-tests with galaxy clusters}), and from the European Union’s Horizon 2020 Programme under the AHEAD2020 project (grant agreement n. 871158). W.C. and G.Y. would like to thank Ministerio de Ciencia e Innovaci\'on for financial support under project grant PID2021-122603NB-C21. W.C. is supported by the Atracci\'{o}n de Talento Contract no. 2020-T1/TIC-19882 granted by the Comunidad de Madrid in Spain. W.C. also thanks the ERC: HORIZON-TMA-MSCA-SE for supporting the LACEGAL-III project with grant number 101086388 and the China Manned Space Project for its research grants.

The CosmoSim database used in this paper is a service by the Leibniz-Institute for Astrophysics Potsdam (AIP). The MultiDark database was developed in cooperation with the Spanish Multi-Dark Consolider Project CSD2009-00064. The authors gratefully acknowledge the Gauss Centre for Supercomputing e.V. (www.gauss-centre.eu) and the Partnership for Advanced Supercomputing in Europe (PRACE, www.prace-ri.eu) for funding the MultiDark simulation project by providing computing time on the GCS Supercomputer SuperMUC at Leibniz Supercomputing Centre (LRZ, www.lrz.de). The authors also thank the Instituto de Astrofisica de Andalucia (IAA-CSIC), Centro de Supercomputacion de Galicia (CESGA) and the Spanish academic and research network (RedIRIS) in Spain for hosting \textsc{uchuu} DR1, DR2 and DR3 in the Skies \& Universes site for cosmological simulations. The \textsc{uchuu} simulations were carried out on Aterui II supercomputer at Center for Computational Astrophysics, CfCA, of National Astronomical Observatory of Japan, and the K computer at the RIKEN Advanced Institute for Computational Science. The \textsc{uchuu} Data Releases efforts have made use of the skun@IAA\_RedIRIS and skun6@IAA computer facilities managed by the IAA-CSIC in Spain (MICINN EU-Feder grant EQC2018-004366-P).
\end{acknowledgements}

%
%
\bibliographystyle{aa}
\bibliography{bibliography}

\begin{appendix} 
\section{Sparsity Mass Dependence}\label{sparsity_mass}
\begin{figure}[ht]
\centering
\includegraphics[width = 0.9\linewidth]{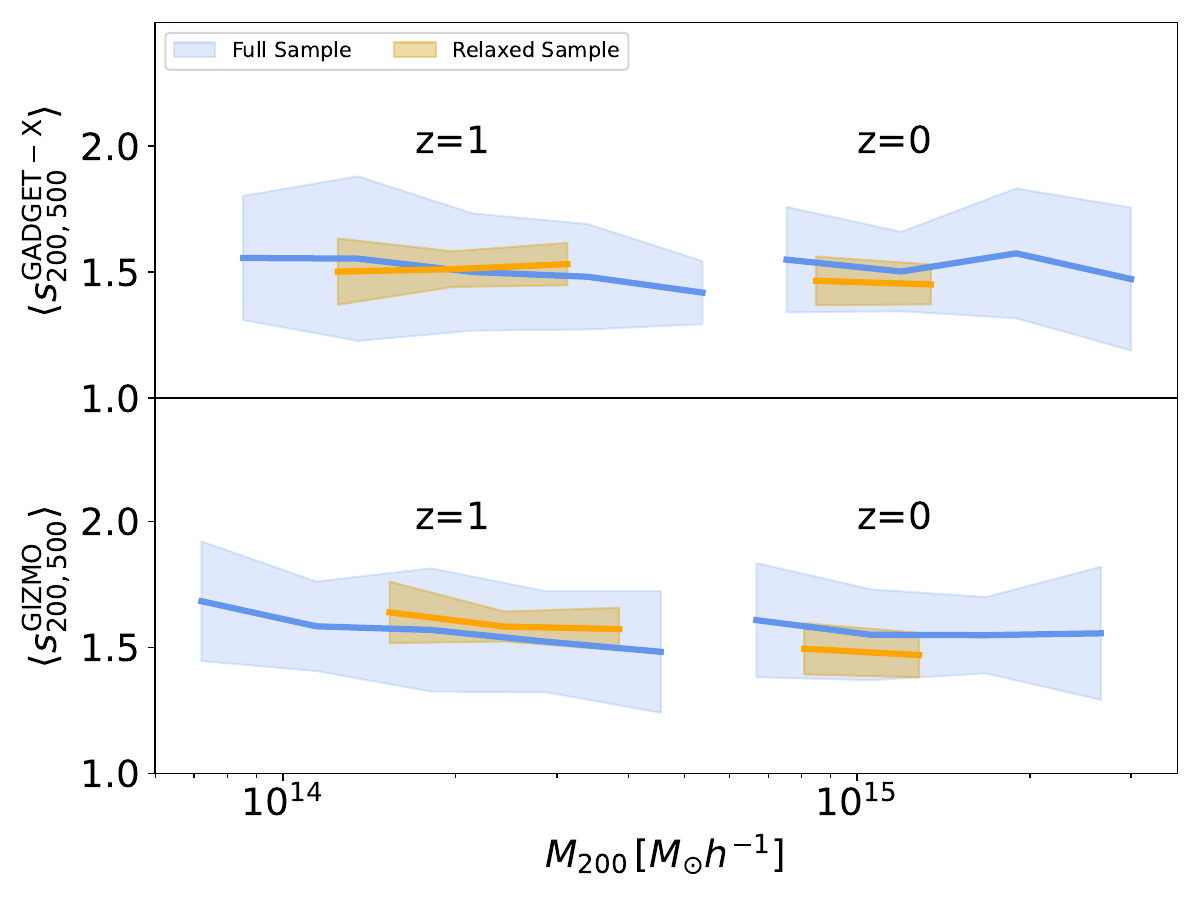}
\caption{Average sparsity $\langle s_{200,500}\rangle$ as function of $M_{200}$ for the clusters in the \textsc{gadget-x} (top panel) and \textsc{gizmo-simba} (bottom panel) catalogues at $z=0$ and $1$ respectively. The blue solid line corresponds to the case of the full cluster sample, while the goldenrod solid line corresponds to the relaxed one. The shaded areas correspond to the standard deviation in each mass bin.}\label{fig:s200500_vs_m200}
\end{figure}

\begin{figure}[ht]
\centering
\includegraphics[width = 0.9\linewidth]{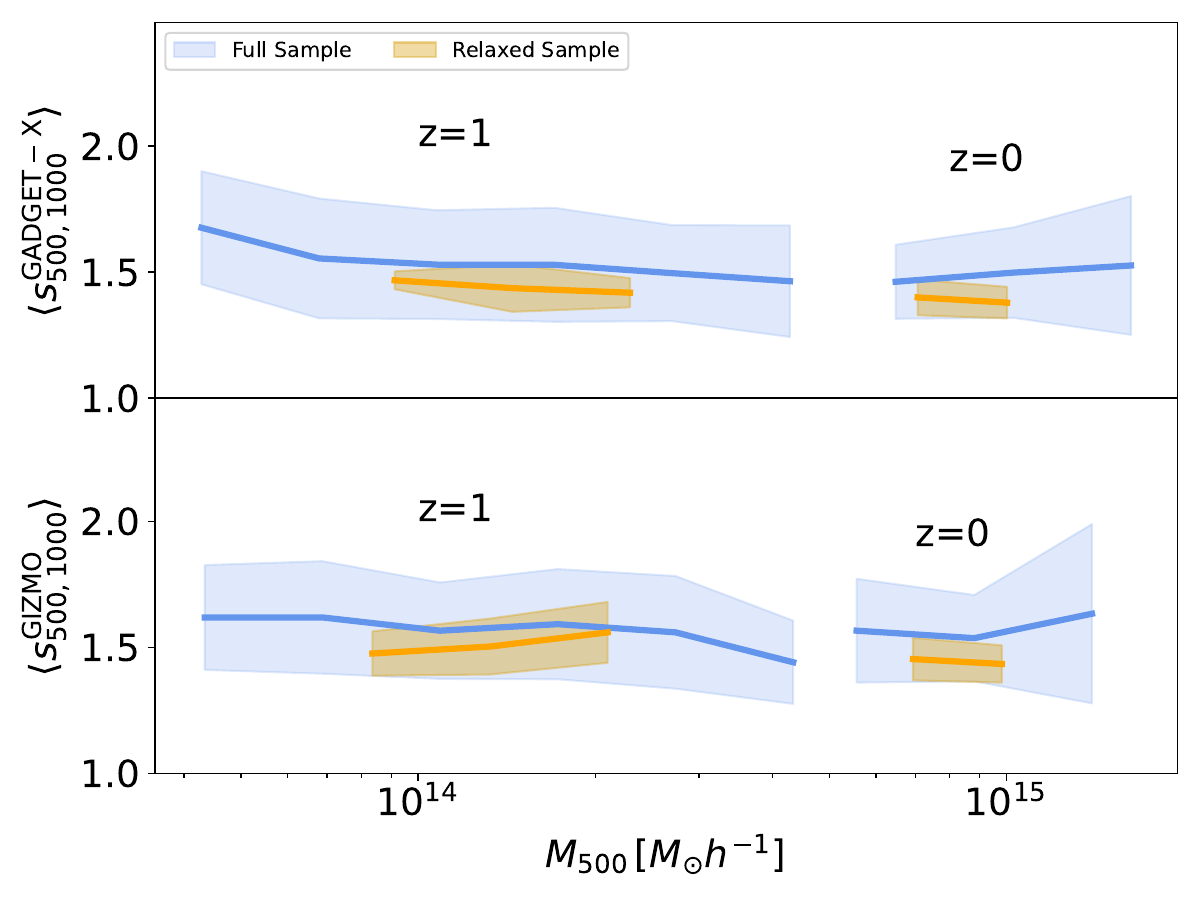}
\caption{As in Fig.~\ref{fig:s200500_vs_m200} for $\langle s_{500,1000}\rangle$ as function of $M_{500}$. }\label{fig:s5001000_vs_m500}
\end{figure}

\begin{figure}[ht]
\centering
\includegraphics[width = 0.9\linewidth]{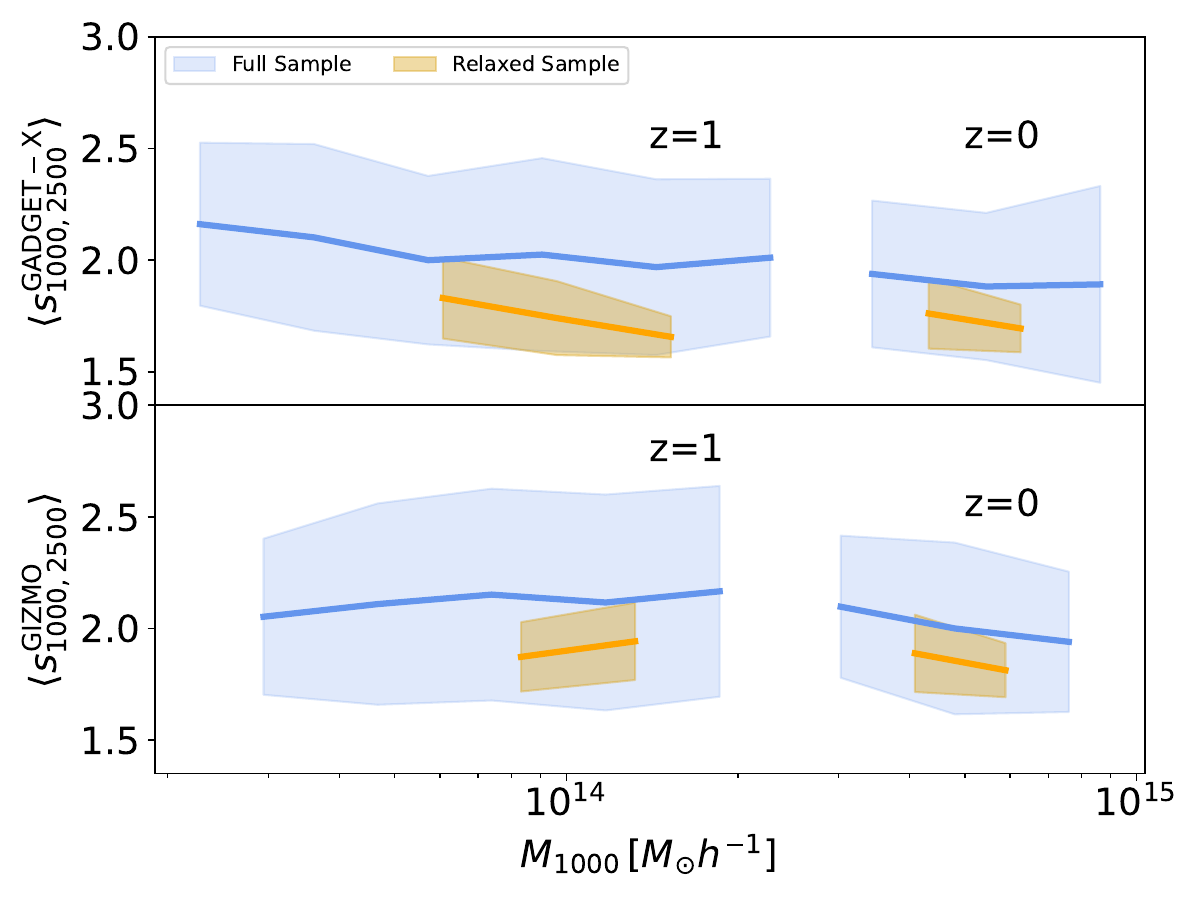}
\caption{As in Fig.~\ref{fig:s200500_vs_m200} for $\langle s_{1000,2000}\rangle$ as function of $M_{1000c}$.}\label{fig:s10002000_vs_m1000}
   \end{figure}

The analysis of halo catalogues from N-body simulations has shown that $s_{\Delta_1,\Delta_2}$ remains approximately constant as function of $M_{\Delta_1}$ \citep{Balmes2014,Corasaniti2018,2019MNRAS.487.4382C}. 
Using the data from the \textsc{gadget-x} and \textsc{gizmo-simba} simulated clusters, we show that this is the case also in the presence of baryons. In particular, using the total cluster masses at different overdensities we compute the average and standard deviation of the sparsities $s_{200,500}$, $s_{500,1000}$ and $s_{1000,2000}$ in equally spaced mass bins for the full cluster sample and the relaxed one at $z=0$ and $z=1$ respectively. These are shown in Fig.~\ref{fig:s200500_vs_m200}, Fig.~\ref{fig:s5001000_vs_m500} and Fig.~\ref{fig:s10002000_vs_m1000}, where the solid lines represent the average sparsity in a given mass bin for a given cluster sample at a given redshift, while the shaded areas represent the standard deviation. As we can see, in all the cases there is little variation of the sparsity with the cluster mass. We may also notice that in the case of relaxed clusters, the scatter as function of cluster mass is smaller than in the case of the full sample.


\section{Halo Mass Functions}\label{hmf}

In this work we make use of the theoretical prediction of the mean sparsity of dark matter haloes, as first introduced in \cite{Balmes2014}. To do so we require a description of the HMF that is consistent with the simulation data we are using for the sparsity analysis. Prescriptions commonly used in the literature \citep[e.g.][]{Tinker_2008, 2016MNRAS.456.2486D} being calibrated on smaller simulations, therefore lower mass ranges, tend to not be sufficiently accurate to be used in this context without additional corrective terms \citep[see e.g.][]{2019MNRAS.487.4382C}. As such we calibrate the HMF using the publicly available \textsc{rockstar} \citep{2013ApJ...762..109B} halo catalogues from the \textsc{uchuu} simulation \citep{Ishiyama2021}, which are consistent to prior parameterisations for low masses but allow a finer calibration of the high mass end thanks to the large volume of the simulation. 

\begin{figure}[ht]
    \centering
    \includegraphics[width = 0.9\linewidth]{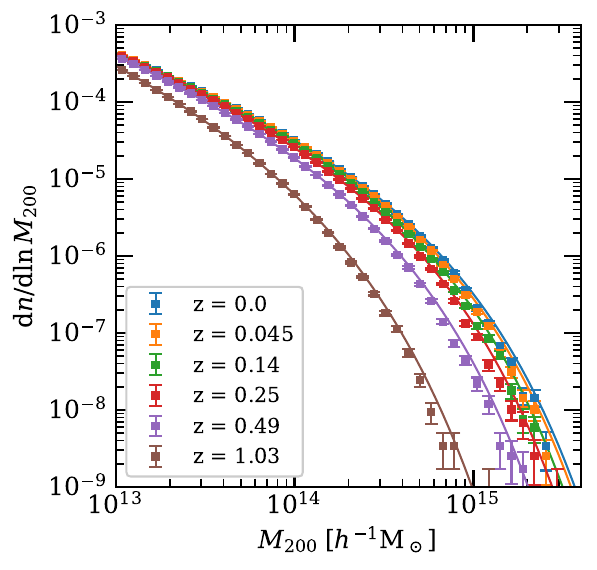}
    \caption{Halo mass function for halo masses $M_{200}$ at different redshift snapshots from the \textsc{uchuu} catalogues. The various lines correspond to the best-fit ST functions.}
    \label{fig:hmf200c}
\end{figure}

\begin{figure}[ht]
    \centering
    \includegraphics[width = 0.9\linewidth]{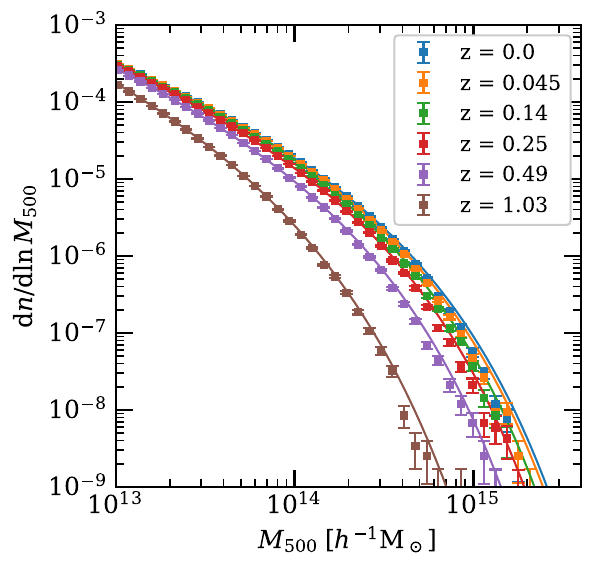}
    \caption{As in Fig.~\ref{fig:hmf200c} for halo masses $M_{500}$.}
    \label{fig:hmf500c}
\end{figure}

\begin{figure}[ht]
    \centering
    \includegraphics[width = 0.9\linewidth]{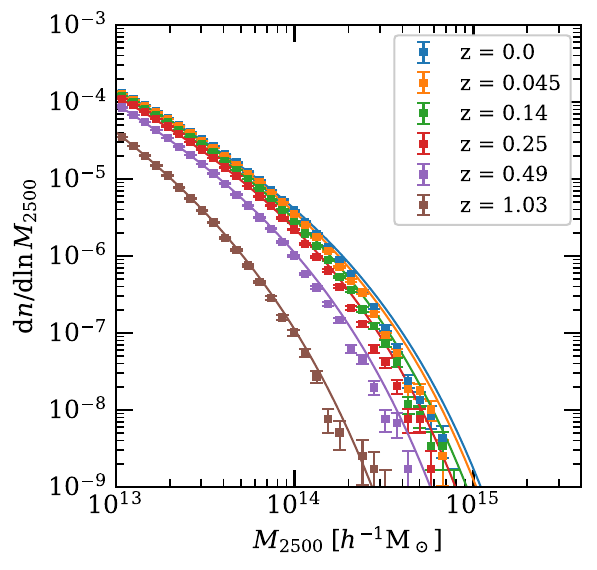}
    \caption{As in Fig.~\ref{fig:hmf200c} for halo masses $M_{2500}$.}
    \label{fig:hmf2500c}
\end{figure}

We recalibrate the HMF for the three spherical overdensity mass definitions, $\Delta=200,\ 500$, and $2500$, provided for each halo in the catalogues. we show the estimated HMF for $\Delta=200,\ 500$, and $2500$ in Fig.~\ref{fig:hmf200c}, Fig.~\ref{fig:hmf500c} and Fig.~\ref{fig:hmf2500c} respectively, where the error bars correspond to the Poisson error. We use these numerical estimates to calibrate an analytical parameterization of the halo mass function as given by the Sheth-Tormen (ST) formula \citep{1999MNRAS.308..119S}: 
\begin{equation}
    \frac{dn}{dM_{\Delta}}=\frac{\rho_m}{M_{\Delta}}\left(-\frac{1}{\sigma}\frac{d\sigma}{dM_{\Delta}}\right)f_{\rm ST}(\sigma),
\end{equation}
where $\rho_m$ is the cosmic matter density, $\sigma(M_\Delta)$ is the root-mean-square fluctuation of the linear density field smoothed on a scale enclosing the mass $M_{\Delta}$ and $f_{\rm ST}$ is the so-called multiplicity function parameterised as
\begin{equation}\label{st_multiplicity}
    f_{\rm ST}(\sigma) = A_{\Delta}\frac{\delta_c}{\sigma}\sqrt{\frac{2 a_{\Delta}}{\pi}}\left[1+\left(\frac{a_{\Delta}\delta_c^2}{\sigma^2}\right)^{-p_{\Delta}}\right]e^{-\frac{a_{\Delta}\delta_c^2}{2\sigma^2}},
\end{equation}
where $A_{\Delta}$, $a_{\Delta}$ and $p_{\Delta}$ are calibration parameters and $\delta_c$ is the linearly extrapolated spherical collapse threshold, which we estimate using the formula by \citet{KitayamaSuto1996}. We derive the best-fit values of the ST coefficients at different redshifts, the corresponding best-fit ST halo mass function are shown in Fig.~\ref{fig:hmf200c}, Fig.~\ref{fig:hmf500c} and Fig.~\ref{fig:hmf2500c} respectively. We note that as we are particularly interested in the high mass regime we truncate the catalogues such that, $M_{200} > 10^{13}h^{-1}{\rm M}_{\odot}$, this results in our fits being particularly ineffective at calibrating the $p$ parameter, which we set to the value given by \cite{2016MNRAS.456.2486D}. 

\begin{table}
\centering
\caption{Coefficients of the second-order polynomial expansion in $x$ parametrising the redshift evolution of the ST parameters for the \textsc{uchuu} halo mass functions.}
\begin{tabular}{c | c c c c c}
\hline
             & $c^{(0)}_{\Delta}$ & $c^{(1)}_{\Delta}$ & $c^{(2)}_{\Delta}$\\
\hline
$A_{200}$ & $0.39597325$ & $-0.56742399$ & $0.94041929$ \\
$a_{200}$ &  $0.75570691$ & $ 0.26064826$ & $0.79329127$ \\
\hline
$A_{500}$ & $0.33841282$ & $-0.18119179$ & $0.06966255$\\
$a_{500}$ & $0.20223081$ & $1.84052335$ & $-0.73711016$ \\
 \hline
 $A_{2500}$ & $0.12423476$ & $0.1211089$ & $-0.07360382$ \\
 $a_{2500}$ & $-4.94730559$ & $8.70280102$ & $-2.73112863$ \\
 \hline
 $p_{\Delta}$ & $0.2488$ & $0.2554$ & $-0.1151$ \\
\end{tabular}\label{tab:st_coeff}
\end{table}

In order to extrapolate to redshifts other than those probed by the simulation snapshots, following \citet{2016MNRAS.456.2486D} we parameterise the redshift dependence of the coefficients in terms of the variable $x_{\Delta}=\log_{10}(\Delta/\Delta_{vir}(z))$, where $\Delta_{\rm vir}$ is the virial overdensity, which we estimate from the formula of \citet{BryanNorman1998}. We approximate the evolution of the ST coefficient as function of $x_{\Delta}$ in terms of a second-order polynomial expansion:
\begin{equation}
    \Theta_{\Delta} =  c^{(0)}_{\Delta} + c^{(1)}_{\Delta}\cdot x_{\Delta} + c^{(2)}_{\Delta}\cdot x^2_{\Delta},
\end{equation}
where $\Theta_{\Delta}=\{A_{\Delta},a_{\Delta},p_{\Delta}\}$. We quote in Table~\ref{tab:st_coeff} the value of the best-fitting polynomial coefficients.

\section{Sparsity Correlations}\label{appendix:corr}
The use of sparsity mass ratios allow us to retrieve cosmological information encoded in the halo mass profile. In principle, the gravitational halo mass assembly process correlates the mass distribution across different radial shells thus leading to correlations among sparsities probing different regions of the mass profile. However, not all the sparsities are correlated to unity as one would expect if the halo density profile exactly follows the NFW profile. Indeed, as shown in \citet{2022MNRAS.516..437C}, certain sparsity combinations, such as those probing distant mass shells, exhibit a low level of correlations, thus allowing to infer cosmological constraints from multiple sparsity measurements. In \citet{2022MNRAS.516..437C}, the authors have evaluated the correlations among different sparsity combinations at different redshifts using the halo catalogues from large volume cosmological simulations (see their Fig.~3 and 4). 

\begin{figure}[ht]
\centering
\includegraphics[width = 0.9\linewidth]{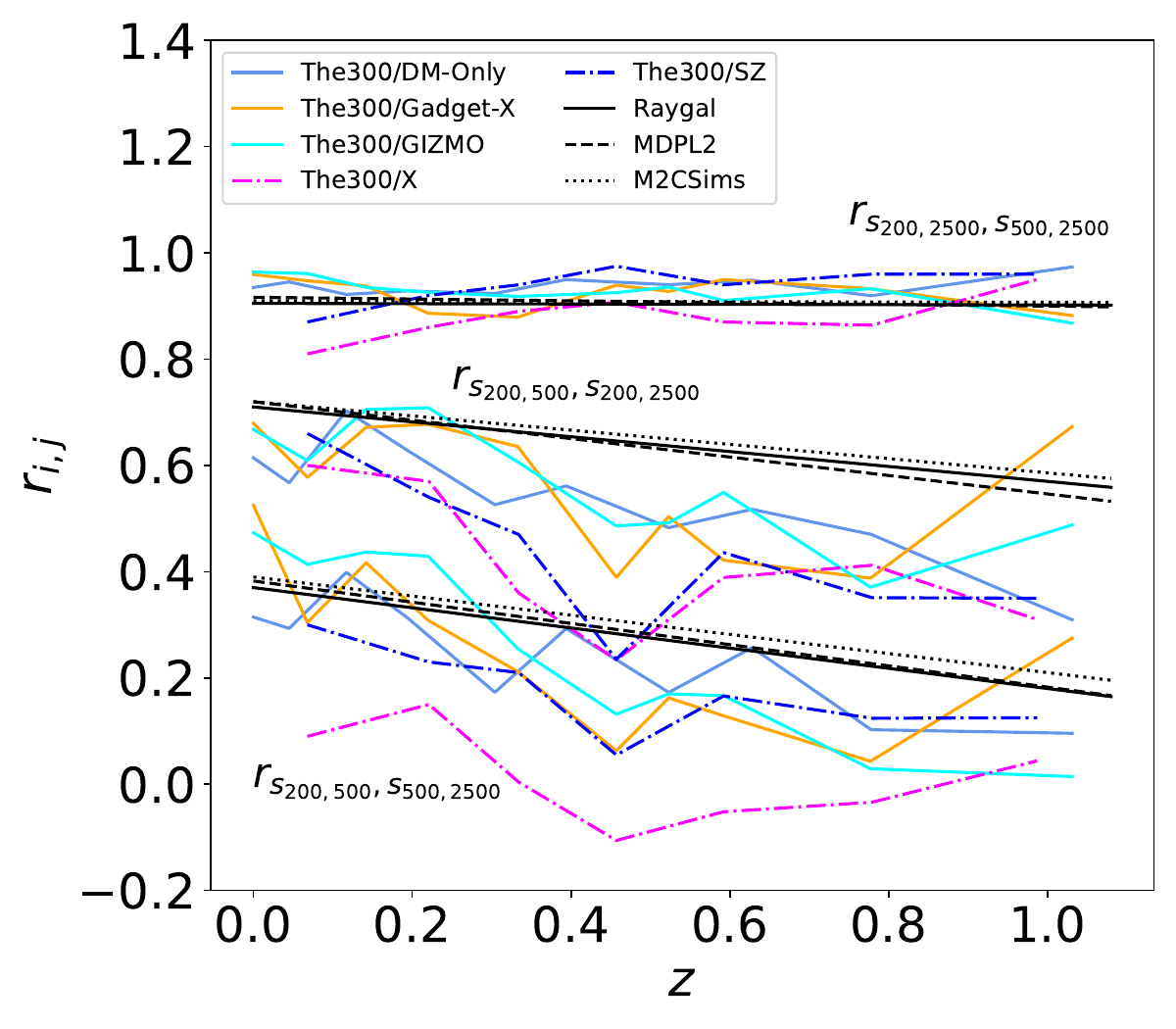}
\caption{Correlation coefficients of the different sparsity pairs $s_{200,500}-s_{500,2500}$ (lowest curves), $s_{200,500}-s_{200,2500}$ (middle curves), $s_{200,2500}-s_{500,2500}$ (top curves) as function of redshifts from the DM-Only (light-blue solid line), \textsc{gadget-x} (orange solid line), MDPL2 (red dashed line), Raygal (blue solid line) and M2Csims (green dotted line) samples.}\label{fig:corr}
\end{figure}

Here, we estimate the correlations among the sparsities $s_{200,500}$, $s_{200,2500}$ and $s_{500,2500}$ using the mass estimates at $M_{200}$, $M_{500}$ and $M_{2500}$ of the simulated clusters from the DM-only, \textsc{gadget-x} and \textsc{gizmo-simba} catalogues, as well as the HE masses from \citet{2023MNRAS.518.4238G}. In particular, for each sparsity pair we compute the correlation coefficient:
\begin{equation}\label{corr_coeff}
    r_{s_i,s_j}=\frac{\sum_{k=1}^{N_{\rm h}}\left(s_i^k-\langle s_i\rangle \right)\left(s_j^k-\langle s_j\rangle\right)}{\sqrt{\sum_{k=1}^{N_{\rm h}}\left(s_i^k-\langle s_i\rangle\right)^2\sum_{k=1}^{N_{\rm h}}\left(s_j^k-\langle s_j\rangle\right)^2}},
\end{equation}
where the index $i,j=\left\{(200,500),(200,2500),(500,2500)\right\}$ with $i\ne j$. These are shown as function of redshift in Fig.~\ref{fig:corr}. For comparison, we also plot the correlation coefficients obtained from the halo catalogues of large volume cosmological N-body simulations such as Raygal (black solid lines) and M2CSims (black dotted lines) presented in \citet{2022MNRAS.516..437C}, and the haloes from the MDPL2 simulation (black dashed line). As we can see, the correlations from the The300 simulations are very noisy. This is not surprising since they are estimated using only a few hundred clusters. Nonetheless, we can see a decreasing trend as function of redshift, consistent with the correlations from the large volume simulations. 

As already pointed out in \citet{2022MNRAS.516..437C}, the correlations among sparsities increase from high to low redshifts consistently with the fact that haloes grow from inside-out. As the halo mass is assembled over increasing time scales the distribution of mass within different mass shells becomes increasingly correlated. Hence, it is not surprising that such a trend remains unaltered even in the presence of baryons. We can see that sparsities such as $s_{200,500}$ and $s_{500,2500}$ probing different the mass distribution in different radial shells are less correlated than those probed by $s_{200,500}$ and $s_{500,2500}$. A larger sample of hydro simulated clusters is indeed necessary to obtain a more precise estimate of the correlations and test how they differ from those estimated from DM-only simulations.

\end{appendix}
\end{document}